\documentclass[12pt]{article}
\usepackage{amsmath}
\usepackage{amsthm}
\usepackage{amssymb}
\usepackage{amsfonts}
\usepackage{epic}
\usepackage{eepic}
 \usepackage{times}
\usepackage[matrix,arrow]{xy}
\usepackage{macros}
\usepackage{epsfig}

\theoremstyle{plain}

\theoremstyle{remark}

\newcommand{\sst}{\scriptscriptstyle}

\newcommand{\nn}{\nonumber}

\newcommand{\ka}{\kappa}

\newcommand{\phis}{{\phi_{\rm sing}}}
\newcommand{\beq}{\begin{equation}}
\newcommand{\eeq}{\end{equation}}

\newcommand{\pa}{\partial}
\newcommand{\ot}{\otimes}
\newcommand{\ra}{\to}

\newcommand{\paz}{\partial_z}
\newcommand{\paw}{\partial_w}

\newcommand{\bra}{\langle}

\newcommand{\ket}{\rangle}

\newcommand{\al}{\alpha}

\newcommand{\be}{\beta}
\newcommand{\ga}{\gamma}
\newcommand{\Ga}{\Gamma}
\newcommand{\de}{\delta}
\newcommand{\De}{\Delta}
\newcommand{\ep}{\epsilon}
\newcommand{\la}{\lambda}
\newcommand{\La}{\Lambda}

\newcommand{\Om}{\Omega}
\newcommand{\si}{\sigma}
\newcommand{\up}{\Upsilon}
\newcommand{\vf}{\varphi}

\newcommand{\bc}{{\mathbf c}}

\newcommand{\bz}{{\bar{z}}}

\newcommand{\CC}{{\mathcal C}}
\newcommand{\CD}{{\mathcal D}}
\newcommand{\CE}{{\mathcal E}}
\newcommand{\CF}{{\mathcal F}}
\newcommand{\CG}{{\mathcal G}}
\newcommand{\CH}{{\mathcal H}}

\newcommand{\CI}{{\mathcal I}}

\newcommand{\CL}{{\mathcal L}}
\newcommand{\CM}{{\mathcal M}}

\newcommand{\CO}{{\mathcal O}}
\newcommand{\CP}{{\mathcal P}}

\newcommand{\CR}{{\mathcal R}}
\newcommand{\CS}{{\mathcal S}}
\newcommand{\CT}{{\mathcal T}}

\newcommand{\CV}{{\mathcal V}}

\newcommand{\CZ}{{\mathcal Z}}

\newcommand{\SE}{{\mathsf E}}

\newcommand{\SL}{{\mathsf L}}

\newcommand{\SQ}{{\mathsf Q}}

\newcommand{\SV}{{\mathsf V}}

\newcommand{\fg}{{\mathfrak g}}

\newcommand{\fh}{{\mathfrak h}}

\newcommand{\sa}{{\mathsf a}}

\newcommand{\sfc}{{\mathsf c}}

\newcommand{\se}{{\mathsf e}}

\newcommand{\sq}{{\mathsf q}}
\newcommand{\spp}{{\mathsf p}}

\newcommand{\mst}{{\mathsf t}}

\newcommand{\BD}{{\mathbb D}}

\newcommand{\FB}{{\mathfrak B}}

\newcommand{\BL}{{\mathbb L}}

\newcommand{\BR}{{\mathbb R}}

\newcommand{\BC}{{\mathbb C}}

\newcommand{\BP}{{\mathbb P}}

\newcommand{\BZ}{{\mathbb Z}}

\newcommand{\Vect}{{\SV\se\sfc\mst}}

\DeclareMathOperator{\sgn}{sgn}

\newcommand{\rf}[1]{(\ref{#1})}

\newcommand{\aufz}
{\begin{list}{$\bullet$}{\topsep0cm \itemsep0cm \parsep0cm}}
\newcommand{\eaufz}{\end{list}}

\renewcommand{\nn}{\nonumber}
\begin{document}

\title{Irregular singularities in Liouville theory\\
and Argyres-Douglas type gauge theories, I}
\author{D. Gaiotto and J. Teschner}
\address{D.G.: Institute for Advanced Study, Einstein Dr., Princeton, NJ 08540, USA\\[2ex]
J.T.: DESY Theory, Notkestr. 85, 22603 Hamburg, Germany}
\maketitle

\begin{quote}
\centerline{\bf Abstract}
{\small 
Motivated by problems arising in the study of N=2 supersymmetric
gauge theories we introduce and study irregular singularities in
two-dimensional conformal field theory, here  Liouville theory.
Irregular singularities are associated to representations
of the Virasoro algebra in which a subset of the annihilation
part of the algebra act diagonally. In this paper we define 
natural bases for the space of conformal blocks in the presence
of irregular singularities, describe how to calculate their
series expansions, and how such conformal blocks can be constructed
by some delicate limiting procedure from ordinary conformal 
blocks. This leads us to a proposal for the structure functions
appearing in the decomposition of physical correlation functions
with irregular singularities into conformal blocks. Taken together,
we get a precise prediction for the partition functions of some
Argyres-Douglas type theories on $S^4$.}
\end{quote}

\section{Introduction}

\setcounter{equation}{0}

In this paper we answer a question which arises from the interplay \cite{Alday:2009aq} of four-dimensional ${\cal N}=2$ gauge theory and two-dimensional CFT. 
In the gauge-theory language, the question can be stated as: how can we compute the $S^4$ partition function \cite{Pestun:2007rz} of an Argyres-Douglas theory \cite{Argyres:1995jj,Argyres:1995xn}?
In the 2d CFT language, the question becomes: how can we compute Liouville theory correlation functions in the presence of an irregular vertex operator \cite{Gaiotto:2009ma}?
The equivalence of these two questions follows from two facts. First, the $S^4$ partition function of a certain class of $SU(2)$ gauge theories coincides with standard Liouville theory correlation functions.
Second, a limiting procedure which defines Argyres-Douglas theories from $SU(2)$ gauge theories \cite{Gaiotto:2009hg} has a simple interpretation as a collision limit in Liouville theory, 
which produces irregular vertex operators from the collision of several standard vertex operators.

The existence of a well-defined collision limit for Liouville theory correlation functions is far from obvious from a two-dimensional perspective. 
The connection to gauge theory is a crucial source of inspiration in defining the notion of irregular vertex operator in Liouville theory. 
Once we have a solid two-dimensional definition of our objective, the 2d CFT perspective is best suited for the actual calculation of the answer. 
The calculation proceeds in two stages. First, we define and compute a basis of conformal blocks with irregular singularities which has properties analogous to the standard BPZ conformal blocks \cite{BPZ}. 
Then we identify a measure which combines holomorphic and anti-holomorphic conformal blocks into a well-defined Liouville theory correlation function.
A posteriori, the various stages of the 2d CFT calculation can then be given an intuitive gauge-theory interpretation. 

The relations between N=2 supersymmetric gauge theories and Liouville
theory referred to above appear to be part of a larger story that 
has started unfolding, relating supersymmetric gauge theories, 
conformal field theories, (quantized) moduli spaces of flat 
connections, various integrable models and the 
geometric Langlands program, see \cite{KW,Gaiotto:2009hg,NS,AT,T10,NRS} for
an incomplete list of relevant references. A common theme in these 
developments are relations with the Hitchin integrable system and
with moduli spaces of flat connections on Riemann surfaces. The
consideration of irregular singularities appears to be 
a very natural generalization in this context. 
From this point of view it seemed overdue from this point of view to have
a Liouville theory with irregular singularities. 

This paper is meant to be the first of a series of papers on this subject.
While we here focus on more algebraic aspects like the construction of
the conformal blocks, subsequent publications will in particular
discuss analogs of the
modular transformations relating different bases for spaces of conformal
blocks, and relations with a generalization of the
quantum Teichm\"uller theory to cases with irregular singularities.

The structure of the paper mirrors this logical structure.
In section \ref{sec:irregular} we define the notion of an
irregular conformal block. In sections \ref{sec:vertex} and \ref{sec:degenerate} we describe two 
different, natural ways to define the same BPZ-like basis of irregular conformal blocks and clarify the nature of collision limits. 
In section \ref{sec:measure} we 
use the collision limit to derive the correct integration measure for a Liouville correlation function. Finally, in section \ref{sec:gauge} we provide the gauge theory interpretation of the various ingredients
of the 2d CFT answer. We refer the reader to the introductory part of each section for further details.  

While this paper was being written, reference \cite{BMT} appeared which has 
partial overlap with the discussion in Sections \ref{sec:irregular}
and \ref{sec:gauge}.

\noindent
{\bf Acknowledgements:} The work of DG is supported in part by 
NSF grant NSF PHY-0969448 and in part
by the Roger Dashen membership in 
the Institute for Advanced Study. Opinions and 
conclusions expressed here are those 
of the authors and do not
necessarily reflect the views of funding agencies.

JT would like to thank the Institute for Advanced Study for
where part of this work was carried out for hospitality.

\section{Irregular singularities in conformal field theory}
\label{sec:irregular}

\setcounter{equation}{0}

\subsection{Irregular vectors}

A primary field $\Psi_\De(z)$ in conformal field theory is defined by the
operator product expansion
\begin{equation}
T(y)\Psi_\De(z)\,\sim\,\frac{\De}{(y-z)^2}+\frac{1}{y-z}\frac{\pa}{\pa z}
\Psi_\De(z)\,,
\end{equation}
which is closely related to the statement the the state $|\De\rangle$
created as
\begin{equation}
|\De\rangle\,:=\,\lim_{z\ra 0}\Psi_\De(z)|0\rangle
\end{equation}
satisfies the highest weight property
\begin{equation}\label{hw}
L_n|\De\rangle\,=\,0\,,\quad n>0\,,\qquad
L_0|\De\rangle\,=\,\De\,|\De\rangle\,.
\end{equation}
Interpreting the Virasoro generators $L_n$ with $n>0$ as
generalizations of ``annihilation'' operators may lead one to
consider analogs of the coherent states where some subset of
the generators $L_n$ with $n>0$ acts diagonally as
\begin{equation}\label{Ldiag0}
L_k|\,I\,\rangle\,=\,\Lambda_k|\,I\,\rangle\,.
\end{equation}
From a mathematical perspective one may regard such vectors as
analogs of the so-called Whittaker vectors in the representation
theory of real reductive groups.

The Virasoro algebra, in particular the relations
$[L_k,L_{k'}]=(k-k')L_{k+k'}$ for $k,k'\geq 0$,
imply that $\La_{k+k'}=0$ if both $L_k$ and $L_{k'}$ are contained
in the set of generators which act diagonally as in
\rf{Ldiag0}. Based on this observation
it is easy to see that the values of the
indices $k$ for which the eigenvector property
\rf{Ldiag0} can hold with $\La_k\neq 0$
must be taken from one of the sets
$\{n,n+1,\dots,2n\}$, where $n$
is a positive integer. We will say that $|I_n\rangle$ is an
irregular vectors of order $n$ if it satisfies
\begin{subequations}\label{Ldiag}
\begin{align}\label{Ldiaga}
&L_k|\,I\,\rangle\,=\,\Lambda_k|\,I\,\rangle\,,\quad
k=n,\dots,2n\,,\\
&L_k|\,I\,\rangle\,=\,0\,,\quad
k>2n\,.
\label{Ldiagb}\end{align}
\end{subequations}
The collection $\Lambda=\{\Lambda_n,\dots,\Lambda_{2n}\}$
of eigenvalues parameterizes the irregular vectors of order $n$,
which may be expressed by using the notation $|I_n,\La\rangle$.

The representation of the generators $L_{k}$, $k=0,\dots,n-1$, is
severely restricted by the relations
$[L_k,L_{k'}]=(k-k')L_{k+k'}$ for $k,k'\geq 0$. A convenient way
to satisfy these relations can be introduced by using the parameterization
\begin{equation}\label{Lambdadef}
\Lambda_k\,=\,((k+1)Q-2\al)c_k-\sum_{l=1}^{k-1} c_{l}c_{k-l}\,,\qquad
k=n,\dots,2n\,.
\end{equation}
This expresses the $n+1$ parameters $\Lambda=\{\Lambda_n,\dots,\Lambda_{2n}\}$
in terms of the parameter $\al$ and the
collection of parameters $\bc=(c_1,\dots,c_n)$.
It is then elementary to check that the definitions
\begin{equation}\label{CLdef}
L_k|\,I_n\,\rangle\,:=\,\CL_k|\,I_n\,\rangle\,,
\qquad\begin{aligned}
&\CL_k\equiv\CL_k(\bc,\al):=\La_k
+\sum_{l=k+1}^{n-1} (l-k)c_{l}\frac{\pa}{\pa c_{l-k}}\,,\\
&\CL_0\equiv\CL_0(\bc,\al):=\al(Q-\al)+\sum_{k=1}^n k\, c_k
\frac{\pa}{\pa c_{k}}\,,
\end{aligned}
\end{equation}
are compatible with the algebraic relations $[L_k,L_{k'}]=(k-k')L_{k+k'}$,
where $k,k'\geq 0$.

It will often be convenient
to summarize the conditions \rf{Ldiag} and
\rf{CLdef} in the form
\begin{equation}\label{irrOPE}
T_>(y)|\,I_n\,\rangle\,=\,\Biggl[\;
\sum_{k=n}^{2n}\frac{\Lambda_k}{y^{k+2}}+
\sum_{k=0}^{n-1}\frac{\CL_k}{y^{k+2}}+\frac{1}{y}L_{-1}\,\Biggr]
|\,I_n\,\rangle\,,
\end{equation}
where $T_{>}(y):=\sum_{k\geq -1}y^{-n-2}L_n$. The formula
\rf{irrOPE} encodes
the singular behavior of the energy-momentum
tensor $T(y)$ in the vicinity of an
irregular singularity at $y=0$.

A more invariant point of view is to regard the conditions
\rf{Ldiagb} as natural generalization of the highest weight conditions
\rf{hw}. The highest weigth condition \rf{hw} says that
$|\Delta\rangle$ is fixed by the algebra of holomorphic vector
fields $\Vect_{\BD}$ on the unit disc $\BD=\{z\in\BC;|z|<1\}$
with generators $L_k\simeq z^k(z\pa_z+\De(k+1))$,
$k\geq 0$.
The space of all vectors which satisfy \rf{Ldiagb}
must then be a representation of the truncated
algebra $\Vect_{\BD}^{(n)}$ of
holomorphic vector fields on a disc which has
generators $l_k$, $k=0,\dots,2n$ and relations
\[
\begin{aligned}
&[\,l_k\,,\,l_{k'}\,]\,=\,(k-k')l_{k+k'}\,, \quad{\rm if}\;\;k+k'\leq 2n\,,\\
&[\,l_k\,,l_{k'}\,]\,=\,0\,,\quad{\rm if}\;\;k+k'> 2n\,,
\end{aligned}\qquad
0\leq k,k'\leq 2n\,.
\]
The equations \rf{Ldiaga}, \rf{Lambdadef} and \rf{CLdef}
define representations of $\Vect_{\BD}^{(n)}$ on spaces of
functions of the $n+1$ variables $\al$ and $\bc=(c_1,\dots,c_n)$.

\subsubsection{Comparison to free-field representation}

Let us introduce the (left-moving)
chiral free field $\vf(z)$, with mode-expansion given by
\begin{equation}
\phi(z)=q-\al_p \log z+\sum_{k\neq 0}\frac{i}{n}a_nz^{-n}\,,\qquad
\al_p:=ip+\frac{Q}{2}\,.
\end{equation}
The modes are postulated to have the following
commutation and hermiticity relations
\renewcommand{\sq}{q}
\renewcommand{\spp}{p}
\renewcommand{\sa}{a}
\renewcommand{\SL}{L}
\begin{equation}\label{ccr}
[\sq,\spp]=\frac{i}{2},\quad
\begin{aligned}\sq^{\dagger}=& \sq,\\
\spp^{\dagger}=&\spp,
\end{aligned}
\qquad [\sa_n,\sa_m]=\frac{n}{2}\de_{n+m},\quad
\sa_n^{\dagger}=\sa_{-n},
\end{equation}
which are realized in the
Hilbert-space
\begin{equation}
\CH^{\rm\sst F}_{\rm\sst L}\;\equiv \;L^2(\BR)\ot\CF,
\end{equation}
where $\CF$
is the Fock-space generated by acting with the modes $\sa_n$,
$n<0$ on the Fock-vacuum $\Om$
that satisfies $\sa_n\Om=0$,
$n>0$. We will mainly work in a representation 
where $\spp$ is diagonal.

The action of the Virasoro algebra on $\CH^{\rm\sst F}_{\rm\sst L}$ can
be defined in terms
of the generators $\SL_n\equiv\SL_n(\spp)$, where
\begin{equation}\label{FockVir}
\begin{aligned}\SL_n^{}(p)\;=\;&
(2p+inQ)\sa_n^{}+\sum_{k\neq 0,n}\sa_k^{} \sa_{n-k}^{},
\qquad n\neq 0, \\
\SL_0^{}(p)\;=\;&p^2 +\frac{Q^2}{4}+
2\sum_{k>0}\sa_{-k}^{}\sa_k^{}.
\end{aligned}
\end{equation}
Equations \rf{FockVir} yield a representation
of the Virasoro algebra with central charge
\begin{equation}
c\;=\; 1+6Q^2.
\end{equation}
Let us  consider coherent states $|\bc;\al\rangle^{(n)}$
that satisfy
\begin{equation}\begin{aligned}
&\sa_k\,|\bc;\al\rangle^{(n)}\,=\,-ic_k\,|\bc;\al\rangle^{(n)}\,,\\
&\sa_k\,|\bc;\al\rangle^{(n)}\,=\,0\,,
\end{aligned}\qquad
\begin{aligned} 
&{\rm for}\;\;
0<k\leq n\,,\\
&{\rm for}\;\;
k> n\,.
\end{aligned}\label{cohff}
\end{equation}
It follows directly \rf{cohff} that the 
coherent states $|\bc;\al\rangle^{(n)}$ represent a very special example of  irregular vector
of degree $n$ within the Fock space 
representation  \rf{FockVir}  of the Virasoro algebra. 
In a sense, the Ward identities for general irregular vectors are modeled on this specific 
example. 

We will discuss in a later section \ref{sec:screen} how to give a free-field description of more general 
irregular vectors by dressing such bare coherent state with screening charges. 

\subsubsection{Irregular modules} \label{sec:irrmod}

From a given irregular vector we may generate infinitely
many other vectors by acting with the Virasoro generators. 
It will be useful for us to formalize the point of view that
this leads to the definition of new types of Virasoro modules. 

To this aim, let us first note that the space 
${\rm DO}^{(n)}$
of algebraic 
differential operators in $n$ variables ${\mathbf c}=(c_1,\dots,c_n)$ 
is naturally
a module for the  subalgebra ${\rm Vir}_+$ isomorphic to $\Vect_{\BD}$
generated by $L_k$, $k\geq 0$. Identifying the trivial
differential operator $1$ with the irregular vector
$I^{(n)}_{{\mathbf c};\al}$ corresponds to defining the 
action of ${\rm Vir}_+$ on  ${\rm DO}^{(n)}$ via
\begin{equation}\label{L_konDO's}
L_k \cdot \CD\,=\,\CD \,\CL_k({\mathbf c};\al)\,,\qquad \forall \;\,
\CD\in{\rm DO}^{(n)}\,.
\end{equation} 
From this representation of  ${\rm Vir}_+$ one may then naturally
induce a representation $\CV_{\bc;\al}^{(n)}$ of the full Virasoro algebra.
As a vector space $\CV_{\bc;\al}^{(n)}$ is spanned by
expressions of the form
\begin{equation}
\BL_{-I}\,\CD\,:=\,
L_{-k}^{l_k}\,L_{-(k-1)}^{l_{k-1}}\,\cdots \,L_{-1}^{l_1}\,\CD\,,
\end{equation}
where $\CD$ is any element of a basis for ${\rm DO}^{(n)}$. The action
of the Virasoro algebra is defined in the usual way:
Writing
\begin{equation}
L_k\,\BL_{-I}\,=\,\sum_{k'=0}^k \sum_{I'} R_{kk}^{II'} \;\BL_{-I'}\,L_{k'}\,,
\end{equation}
with the help of the Virasoro algebra, we may apply \rf{L_konDO's}
with $k$ replaced by $k'$
to define the action of $L_k$ on any basis element of the form
$\BL_{-I}\,\CD$.

Irregular vectors and the associated modules were recently discussed
from a similar point of view in \cite{FJK}.

\subsection{Irregular singularities from collision of primary fields}
\label{collim}

Further motivation for the definitions above can be obtained
from the consideration
of certain collision limits of usual primary fields. Let us consider vectors
\[
|\,R_n(z)\,\rangle\,\,\equiv\,|R_n(z_1,\dots,z_n)\rangle:=\prod_{r=1}^{n+1} \Psi_{\De_r}(z_r)|\,0\,\rangle\,,
\]
that are created by
acting with a product of primary fields $\Psi_{\De_r}(z_r)$
on the vacuum $|0\rangle$.
The vectors $|\,R_n(z)\,\rangle$ satisfy the conditions
\begin{equation}\label{Rnconstr-k}
L_k|\,R_n(z)\,\rangle\,=\,\sum_{r=1}^{n+1}z_r^k
\left(z_r\frac{\pa}{\pa z_r}+\De_r(k+1)\right)|\,R_n(z)\,\rangle\,,\quad
k\geq -1\,,
\end{equation}
which are summarized in
\begin{equation}\label{Rnconstr}
T_{>}(y)|\,R_n(z)\,\rangle\,=\,\sum_{r=1}^{n+1}\left(\frac{\De_{\al_r}}{(y-z_r)^2}+
\frac{1}{y-z_r}
\frac{\pa}{\pa z_r}\right)|R_n(z)\rangle\,.
\end{equation}
We are going to argue that
the constraints \rf{irrOPE}
characteristic for irregular vectors follow from
\rf{Rnconstr} in a suitable limit which is defined by sending
$z_r\ra 0$ and $\De_r\ra\infty$ in a correlated way.

\subsubsection{Irregular puncture of degree $n=1$}\label{colln=1}

Let us now consider a limit which creates an irregular puncture of
degree $n=1$ in the collision of two regular punctures.
Let us study the behavior of $T(y)|\,R_1\,\rangle$
\begin{equation}
|\,R_1\,\rangle:=\,\Psi_{\De_z}(z)\,|\,\De_i\,\rangle
\end{equation}
in a suitable limit where $\De_z, \De_i \ra\infty$, 
$z \ra 0$, to be defined more precisely in the following.
We have
\begin{align*}
T_>(y)\,|\,R_1\,\rangle\,=\,
\bigg(& \frac{\De_z}{(y-z)^2}
+\frac{\De_i}{y^2}+\\
&\hspace{1cm}+\frac{1}{y-z}\frac{z}{y}\frac{\pa}{\pa z_1}
+\frac{1}{y}L_{-1}\bigg)|\,R_1\,\rangle
\end{align*}
It will be useful to rewrite this using $\De_{r}=\al_r(Q-\al_r)$ as
\begin{align}\notag
T_>(y)\,|\,R_1\,\rangle=\bigg[& T_{\rm sing}(y)
+\frac{2\al_z\al_i}{y(y-z)}\\
&
+\frac{1}{y(y-z)}{z}\frac{\pa}{\pa z}
+\frac{L_{-1}}{y}\bigg)\bigg]|\,R_1\,\rangle\,,
\label{R1constr}\end{align}
where we introduced $
T_{\rm sing}(y):=-(\pa_y\phis(y))^2+Q\pa_y^2\phis(y)$ with
\begin{equation}
\pa_y\phis(y)\,=\,-\frac{\al_z}{y-z}-
\frac{\al_i}{y}\,.
\end{equation}
In order to simplify the following discussions let us consider
the vector $|\,R_1'\,\rangle$ defined by
\begin{equation}
|\,R_1\,\rangle\,=\,z^{-2\al_z\al_i}\,|\,R_1'\,\rangle\,.
\end{equation}
In terms of $|\,R_1'\,\rangle$ the equations \rf{R1constr}
simplify to
\begin{align}
T(y)\,|\,R_1'\,\rangle=\bigg[T_{\rm sing}(y)
+\frac{1}{y(y-z)}{z}\frac{\pa}{\pa z}
+\frac{1}{y}L_{-1}\bigg]|\,R_1'\,\rangle
\label{R1'constr}\end{align}
Note that $\pa_y\phis(y)$ may be rewritten as
\begin{equation}
\pa_y\phis(y)\,=\,-
\frac{c_1+y \al'}{y(y-z_1)}\,,
\end{equation}
where
\begin{align*}
c_1:=-z \al_i\,,\qquad
\al':=\al_z+\al_i\,.
\end{align*}
In the limit to be taken, we will send $\al_z,\al_i\ra\infty$,
$z\ra 0$ keeping $c_1$ and $\al'$ finite.
This implies that $\phis(y)$ and
$T_{\rm sing}(y)$ have a finite limit.

In the limit of interest we reproduce the operator appearing on
the right hand side of
\begin{align*}
T_>(y)\,|\,I_1(c_1)\,\rangle\,=\,
\bigg[& \;\frac{\La_2}{y^4}+\frac{\La_1}{y^3}
+\frac{1}{y^2}
\bigg(
c_1\frac{\pa}{\pa c_1}+\De_{\al'}\bigg)
+\frac{1}{y}L_{-1}\bigg]|\,I_1(c_1)\,\rangle
\end{align*}
which are the constraints characterizing an irregular vector of order 1.

\subsubsection{Irregular puncture of degree $n=2$}\label{colln=2}

Let us now consider a limit which creates an irregular puncture of
degree $n=2$ in the collision of three regular punctures.
Let us study the behavior of $T(y)|\,R_2\,\rangle$
\begin{equation}
|\,R_2\,\rangle:=\,\Psi_{\De_1}(z_1)\Psi_{\De_2}(z_2)\,|\,\De_3\,\rangle
\end{equation}
in a suitable limit where $\De_i\ra\infty$, $i=1,2,3$,
$z_j\ra 0$, $j=1,2$ to be defined more precisely in the following.
We have
\begin{align*}
T_>(y)\,|\,R_2\,\rangle\,=\,
\bigg(& \frac{\De_1}{(y-z_1)^2}+\frac{\De_2}{(y-z_2)^2}
+\frac{\De_3}{y^2}+\\
&\hspace{1cm}+\frac{1}{y-z_1}\frac{z_1}{y}\frac{\pa}{\pa z_1}
+\frac{1}{y-z_2}\frac{z_2}{y}\frac{\pa}{\pa z_2}
+\frac{1}{y}L_{-1}\bigg)|\,R_2\,\rangle
\end{align*}
It will be useful to rewrite this using $\De_{r}=\al_r(Q-\al_r)$ as
\begin{align}\notag
T_>(y)\,|\,R_2\,\rangle=\bigg[& T_{\rm sing}(y)
+\frac{2\al_1\al_2}{(y-z_1)(y-z_2)}+
\frac{2\al_1\al_3}{y(y-z_1)}+\frac{2\al_2\al_3}{y(y-z_2)}\\
&
+\frac{1}{y(y-z_1)}{z_1}\frac{\pa}{\pa z_1}
+\frac{1}{y(y-z_2)}{z_2}\frac{\pa}{\pa z_2}
+\frac{L_{-1}}{y}\bigg)\bigg]|\,R_2\,\rangle\,,
\label{R2constr}\end{align}
where we introduced $
T_{\rm sing}(y):=-(\pa_y\phis(y))^2+Q\pa_y^2\phis(y)$ with
\begin{equation}
\pa_y\phis(y)\,=\,-\frac{\al_1}{y-z_1}-\frac{\al_2}{y-z_2}-
\frac{\al_3}{y}\,.
\end{equation}
In order to simplify the following discussions let us consider
the vector $|\,R_2'\,\rangle$ defined by
\begin{equation}
|\,R_2\,\rangle\,=\,z_1^{-2\al_1\al_3}z_2^{-2\al_2\al_3}
(z_1-z_2)^{-2\al_2\al_1}\,|\,R_2'\,\rangle\,.
\end{equation}
In terms of $|\,R_2'\,\rangle$ the equations \rf{R2constr}
simplify to
\begin{align}
T(y)\,|\,R_2'\,\rangle=\bigg[T_{\rm sing}(y)
+\frac{1}{y(y-z_1)}{z_1}\frac{\pa}{\pa z_1}
+\frac{1}{y(y-z_2)}{z_2}\frac{\pa}{\pa z_2}
+\frac{1}{y}L_{-1}\bigg]|\,R_2'\,\rangle
\label{R2'constr}\end{align}
Note that $\pa_y\phis(y)$ may be rewritten as
\begin{equation}
\pa_y\phis(y)\,=\,-
\frac{c_2+yc_1+y^2\al}{y(y-z_1)(y-z_2)}\,,
\end{equation}
where
\begin{align*}
c_2:=z_1z_2\al_3\,,\qquad
c_1:=-z_1(\al_2+\al_3)-z_2(\al_1+\al_3)\,,\qquad
\al:=\al_1+\al_2+\al_3\,.
\end{align*}
In the limit to be taken, we will send $\al_i\ra\infty$ for $i=1,2,3$,
$z_j\ra 0$ for $j=1,2$ keeping $c_2$, $c_1$ and $\al$ finite.
This implies that $\phis(y)$and
$T_{\rm sing}(y)$ have a finite limit.

The derivative terms in \rf{R2'constr}
may be rewritten using
\begin{align}
&z_i\frac{\pa}{\pa z_i}\,=\,
z_i\left(\frac{\pa c_1}{\pa z_i}\frac{\pa}{\pa c_1}+
\frac{\pa c_2}{\pa z_i}\frac{\pa}{\pa c_2}\right)\,=\,z_i(\al_i-\al)
\frac{\pa}{\pa c_1}
+c_2\frac{\pa}{\pa c_2}\,,
\end{align}
as
\begin{align}
\frac{1}{y(y-z_1)}{z_1}\frac{\pa}{\pa z_1}
&+\frac{1}{y(y-z_2)}{z_2}\frac{\pa}{\pa z_2}
=\\
&=\frac{2y-z_1-z_2}{y(y-z_1)(y-z_2)}c_2\frac{\pa}{\pa c_2}+
\frac{c_2+\al z_1z_2+yc_1}{y(y-z_1)(y-z_2)}\frac{\pa}{\pa c_1}\,.
\notag\end{align}
In the limit of interest we reproduce the operator appearing on
the right hand side of
\begin{align*}
T_>(y)\,|\,I_2(c_1,c_2)\,\rangle\,=\,
\bigg[& \;\frac{\La_4}{y^6}+\frac{\La_3}{y^5}+\frac{\La_2}{y^4}
+\frac{1}{y^3}\bigg(\La_1+c_2\frac{\pa}{\pa c_1}\bigg)
\\
&
+\frac{1}{y^2}
\bigg(2c_2\frac{\pa}{\pa c_1}+
c_1\frac{\pa}{\pa c_1}+\De_{\al}\bigg)
+\frac{1}{y}L_{-1}\bigg]|\,I_2(c_1,c_2)\,\rangle
\end{align*}
which are the constraints characterizing an irregular vector of order 2.

\subsubsection{Colliding one after the other}

It will sometimes be useful to decompose the limit above into
two steps: We may, for example, first
send $z_1\ra 0$ and $\al_1\ra\infty$, $\al_3\ra\infty$
such that $\al':=\al_1+\al_3$ and $c_1':=-z_1\al_3$ are kept fixed.
The constraints
reduce to
\begin{align}
T(y)\,|\,I_1(c_1',z_2)\,\rangle\,=\,
\bigg[& \;\frac{\La_2'}{y^4}+\frac{\La_1'}{y^3}+\frac{1}{y^2}\bigg(
c_1'\frac{\pa}{\pa c_1'}+\De_{\al'}\bigg)
\\
&\hspace{1cm}
+\frac{\De_{\al_2}}{(y-z_2)^2}
+\frac{1}{y(y-z_2)}{z_2}\frac{\pa}{\pa z_2}
+\frac{1}{y}L_{-1}\bigg]|\,I_1(c'_1,z_2)\,\rangle
\notag\end{align}
As before we may write
\begin{align}\label{I1'}
T(y)\,|\,I_1(c'_1,z_2)\,\rangle\,=\,
\bigg[& \;T_{\rm sing}(y)+\frac{2\al_2c_1'}{y^2(y-z_2)}
+\frac{2\al'\al_2}{y(y-z_2)}
\\
&\hspace{1cm}+\frac{1}{y^2}
d_1\frac{\pa}{\pa d_1}
+\frac{1}{y(y-z_2)}{z_2}\frac{\pa}{\pa z_2}
+\frac{1}{y}L_{-1}\bigg]|\,I_1(c'_1,z_2)\,\rangle
\notag\end{align}
using $
T_{\rm sing}(y):=-(\pa_z\phis(y))^2+Q\pa_z^2\phis(y)$ with
\begin{equation}
\pa_z\phis(y)\,=\,-\frac{c_1'}{y^2}-\frac{\al'}{y}-
\frac{\al_2}{y-z_2}\,.
\end{equation}
The part proportional to
$2\al_2(c_1'+y\al')$ in \rf{I1'} disappears in the constraints
characterizing
\begin{equation}
|\,I_1(c'_1,z_2)\,\rangle:= z_2^{-2\al_2\al'}e^{2\frac{\al_2c_1'}{z_2}}
|\,I_1'(c'_1,z_2)\,\rangle\,.
\end{equation}

The limit $z_2\ra 0$ is performed next.
$\phis(y)$ has a finite limit if we send
$z_2\ra 0$, $c_1'\ra\infty$ and $\al'\ra\infty$ such that
\begin{equation}
\al:=\al'+\al_2\,,\qquad c_1:=c_1'-z_2\al'\,,\qquad c_2:=-c_1'z_2\,.
\end{equation} are kept finite. We reproduce the
constraints characterizing $|\,I_2(c_1,c_2)\,\rangle$.

\subsection{Geometric interpretation}

In order to prepare for the more geometric interpretation of the
irregular vectors let us first revisit basic elements of the story in the
regular case from a convenient point of view.

\subsubsection{Conformal blocks}

Conformal blocks are the holomorphic building blocks for the correlation
functions in a conformal field theory. The correlation functions
of a conformal field theory can be defined as vacuum
expectation values $\big\langle 0|
\prod_{r=1}^{n} \Psi_{\De_r}(z_r)|0\big\rangle$
of a product of vertex operators.
They can be expanded as a sum of products of holomorphic and anti-holomorphic
building blocks called conformal blocks as
\begin{equation}\label{holofact1}
\bigg\langle\,0\,\Big|\,\prod_{r=1}^{n} \Psi_{\De_r}(z_r)\Big|\,0\bigg\rangle\,=\,
\int_{\CP}d\al(p)\;
|\CF_p(z_1,\dots,z_n)|^2.
\end{equation}
The integration is extended over tuples $p=(p_{1},\dots,p_{n-3})\in\CP:=\BR^{n-3}_+$.
More generally one may consider correlation function and conformal blocks associated to
Riemann surfaces $C_{P_1,\dots,P_n}$ with
$n$ punctures,
\begin{equation}\label{holofact2}
\bigg\langle
\prod_{r=1}^{n} \Psi_{\De_r}(P_r)\bigg\rangle_{C}\,=\,\int_{\CP}d\al(p)\;
|\CF_p(C_{P_1,\dots,P_n}^{})|^2.
\end{equation}
It is sometimes useful to fix a reference point $P_0$ on $C$,
and regard the conformal block as an overlap
\begin{equation}\label{overlap}
\langle\,V_C\,|\,R_n
\,\rangle
\end{equation}
between a vector $\langle\,V_C\,|$ characteristic for the
Riemann surface $C\setminus P_0$ with marked point $P_0$ and the vector
\begin{equation}\label{regvecdef}
|\,R_n(z)\rangle\,\,\equiv\,|R_n(z_1,\dots,z_n)\rangle:=\prod_{r=1}^{n+1} \Psi_{\De_r}(z_r)|\,0\,\rangle\,\in\CV_{\De}\,,
\end{equation}
created by acting with $n+1$ chiral vertex operators on the vacuum vector $|0\rangle$.
We will assume that the resulting vector is an element of a Verma module $\CV_{\De}$
of the Virasoro algebra. In the case $\langle\,V_C\,|=\langle 0|$ we must
assume $\De=0$, so that $\CV_{\De}$ is the representation generated
from the vacuum vector $|0\rangle$.

\subsubsection{Conformal Ward identities}

Let us briefly reformulate how conformal blocks are constrained by the
conformal Ward identities in a language that will be convenient for us.
Representing the conformal blocks
as an overlap \rf{overlap}
one may encode the conformal Ward identites for Riemann surfaces of
genus $0$ in the
statement that the vectors $|\,R_n(z)\,\rangle$
satisfy the equations \rf{Rnconstr}. The equations \rf{Rnconstr}
are equivalent to the conditions \rf{Rnconstr-k}
together with $L_{-1}|0\rangle=0$.
For genus zero one immediately gets the familiar formula
\begin{equation}\label{cfWard}
\big\bra \,T(y)\, \Psi_{n}(z_n)\dots \Psi_{1}(z_1)\,\big\ket=
\sum_{i=1}^{n+1}\left(\frac{\De_{\al_i}}{(y-z_i)^2}+
\frac{1}{y-z_i}
\frac{\pa}{\pa z_i}\right)
\big\bra\,\Psi_{n}(z_n)\dots \Psi_{1}(z_1)\,\big\ket
\end{equation}
from \rf{Rnconstr} if one bears in mind that $T(y):=\sum_{k\in\BZ}y^{-n-2}L_n$
and
\begin{equation}\label{vacconstr}
\langle 0|L_k=0\,,\quad {\rm for}\;\; k\leq 1.
\end{equation}
Equations \rf{Rnconstr}
can be read as an infinite set of
linear equations for the vectors $|R_n(z)\rangle$.
As will be discussed in more detail below, one finds an
infinite-dimensional set of solutions in general. Let us assume that
we have found a complete\footnote{The precise meaning of "complete" is subtle in
the case of infinite-dimensional vector spaces.
It will be clarified when it becomes relevant, which is not 
within this paper.}
 set of solutions
$\FB_n:=\{|R_{n,p}(z)\rangle;p\in\CP_n\}$. Each solution defines a
conformal block via
\begin{equation}
\CF_p(z):=\,\langle\, 0\,|\,R_{n,p}(z)\,\rangle\,.
\end{equation}

For surfaces $C$ of genus $g\geq 1$ with marked point $P_0$
one has to replace \rf{vacconstr} by the set of equations
\begin{equation}\label{CWIg>0}
\langle\,V_C\,|\,T[\eta]\,=\,0\,,\qquad T[\eta]:=\int_{\ga_{0}}dy\,\eta(y)T(y)\,,
\end{equation}
for any vector field $\eta=\eta(y)\pa_y$ that extend holomorphically
from a small circle $\ga_0$ surrounding $P_0$ to the rest
\footnote{The connected
component which is separated from $P_0$ by
$\CC_0$.} of the the Riemann surface $C$. We
may then consider
\begin{itemize}
\item[(i)]
a basis $\FB_n^p:=\{\,|R_{n,p''}^{\;p}(z)\rangle\,;\,p''\in\CP_n\,\}$
for the space
of solutions to the equations \rf{Rnconstr} within the same space
$\CV_{\De(p)}$,
\item[(ii)] a basis
$\FB_C^p:=\{\,\langle V_{C,p'}^{\,p}|\,;\,p'\in\CP_C\,\}$ for the space
of solutions to \rf{CWIg>0} within $\CV_{\De(p)}^{\dagger}$,
(we are using the notation $\CV_{\De(p)}^{\dagger}$ for
the hermitian dual to the space $\CV_{\De(p)}$),
\end{itemize}
and represent the conformal blocks as
\begin{equation}
\CF_{P}(C_{P_1\dots P_{n+1}})=
\langle\, V_{C,p'}^{\,p''}\,|\,R_{n,p'''}^{\;p''}\,\rangle \,,
\quad {P}:=(p',p'',p''')\in\CP\equiv\CP_C\times\BC\times\CP_n\,.
\end{equation}

The set of equations \rf{CWIg>0} which characterize the vector
$\langle\,V_C\,|$ is clearly dependent on the complex struture
of $C\setminus P_0$. We will next discuss how this dependence can
be described with the help of the Virasoro algebra.

\subsubsection{Complex structure dependence}

In order to see how the dependence on the complex structure of 
$C$ is represented
in this formulation let us temporarily consider the case $n=-1$.
We clearly have that
\begin{equation}\label{CWI}
\langle\,C\,|\,T[\eta]\,|\,0\,\rangle\,=\,0\,,
\end{equation}
for all vector fields $\eta$ that extend holomorphically
from the curve $\ga_0$ to the rest of the the Riemann surface $C$.
This simply follows by deforming the contour of
integration and using the residue theorem.
It is furthermore clear that \rf{CWI} holds for all
vector fields  $\eta$ that extend holomorphically
inside the disc $\BD_0$ bounded by $\ga_0$.
Such vector fields $\eta=\sum_{n}\eta_n y^{n+1}\pa_y$ 
have $\eta_n=0$ for $n< -1$,
so \rf{CWI} follows from $L_n|0\rangle=0$, $n\geq -1$.
Of particular interest are therefore the
vector fields  $\eta$ for which the left hand side of \rf{CWI} is
nonzero. The vector space of such vector fields may be represented
as the double quotient $\Vect(C\!\setminus\! P_0)\, \big\backslash \,\BC((y))\pa_y\,\big/\,\Vect(\BD_0)\,.$
It is a well-known mathematical result that this double
quotient is naturally isomorphic to the Teichm\"uller space
$\CT(C\!\setminus \!P_0)$ of deformations of complex structures
on the Riemann surface $C\setminus P_0$,
\begin{equation}\label{Viruni}
\CT(C\!\setminus \!P_0)\,\simeq\,
\Vect(C\!\setminus\! P_0)\, \big\backslash \,\BC(\! (y)\! )\pa_y\,\big/\,
\Vect(\BD_0)\,.
\end{equation}
Using the isomophism \rf{Viruni} one may associate to each
vector field $\eta$ an infinitesimal variation $\pa_{\eta}$ of the
complex structure on $C_{g,n}$. It is natural to require that
\begin{equation}\label{CWI2}
\pa_\eta\langle\,V_{C}\,|\,0\,\rangle\,=\,
\langle\,V_C\,|\,T[\eta]\,|\,0\,\rangle\,,\qquad \forall \eta\,.
\end{equation}

Turning to the case $n\geq 0$
one may note that the vector
$|R_n(z)\rangle$ is not annihilated by $T[\eta]$ for
all $\eta\in\Vect(\BD_0)$, but only by the subalgebra
$\Vect(\,\BD_0\setminus\{z_1,\dots,z_m\}\,)$ generated by
$\eta=\eta(y)\pa_y$ which vanish at $z_1,\dots,z_m$.
The vector fields $\eta$ for which
$\langle\,V_C\,|\,T[\eta]\,|\,R_n\,\rangle$ is non-vanishing
are naturally identified with the variations of complex structures of
surfaces $C_{g,n+1}$ obtained by gluing an $n+1$-punctured sphere $\CP^{1}\setminus \{z_1,\dots,z_{n+1},\infty\}$ to $C\setminus P_0$, the gluing being performed by
identifying annular neighborhoods of $P_0$ and $\infty$, respectively.
The variations of the positions $z_1,\dots,z_{n+1}$ become part of the
Teichm\"uller variations of $C_{g,n+1}$ in this way.
Note in particular that the case $n=0$ corresponds to the
insertion of a single vertex operator  at $P_0$
into $C$.

For our aims it is useful to observe that the additional deformations that
$C_{g,n}$ has
compared to $C$ can be characterized more abstractly as corresponding to
those vector fields $\eta\in\Vect(\BD_0)$ such that
$T[\eta]|R_n\rangle\neq 0$. The
action of these vector fields is represented explicitly in terms
of the derivatives $\pa_{z_r}$, $r=1,\dots,n+1$ via \rf{Rnconstr}.
Note furthermore that an overall translation of $z_1,\dots,z_n$ by the same
amount is equivalent to a variation of the marked point $P_0$ on $C$.
We may therefore without loss of generality assume that $z_{n+1}=0$.
The vector fields $\eta(y)\pa_y$
that preserve this condition must vanish at $y=0$. The set of all
such vector fields will be denoted as $\Vect^{(0)}(\BD_0)$.
The remaining parameters $z_1,\dots,z_{n}$ can be considered as
variables that represent explicitly the part of the complex structure
dependence of the conformal blocks coming from $|R_n(z)\rangle$. Variations of these
parameters correspond to  vector fields $\eta\in\Vect^{(0)}(\BD_0)$ such that
$T[\eta]\,|\,R_n(z)\,\rangle\neq 0$.

\subsubsection{Moduli of the irregular vectors}

Let us finally return to the discussion of irregular vectors $|I_n(c)\rangle$.
It is natural to interpret
\begin{equation}\label{irrblock}
\langle \,V_C\,|\,I_n(c)\,\rangle
\end{equation}
as a conformal block obtained by inserting into $C$ a vertex operator which creates
an irregular singularity at position $P_0$. We note that
$T[\eta]\,|\,I_n(c)\,\rangle$ are non-vanishing for $\eta\in\Vect^{(n)}(\BD_0)$.
The action of $T[\eta]$ on $|I_n\rangle$
is represented by differential operators with respect to $c_1,\dots,c_n$.
Having followed the discussion above it is clearly natural to regard the
parameters $c_1,\dots,c_n$ as generalizations of the complex structure moduli
associated to an irregular singularity of order $n$ at $P_0$.

It may also be helpful to compare conformal blocks with insertion of an irregular vector
to the conformal blocks constructed as
\begin{equation}
\langle\,V_C\,|\,I_\infty\,\rangle\,,\qquad |\,I_{\infty}(\chi)\,\rangle:=
e^{T[\chi]}|\,\Delta\,\rangle\,,
\end{equation}
where $T[\chi]:=\sum_{k\in\BZ}\chi_kL_k$.
It is clear that
\begin{equation}
L_k\,|\,I_{\infty}(\chi)\,\rangle\,=\,\CD_{\chi,k}\,|\,I_{\infty}(\chi)\,\rangle\,.
\end{equation}
where $\CD_{\chi,k}$ is a linear combination of derivatives
$\pa_{\chi_k}$ for $k>0$. The vector $|\,I_{\infty}(\chi)\,\rangle$ behaves
formally as an irregular vector of infinite order.

One may, on the other hand, regard $e^{T[\chi]}$ as the operator which represents
a re-parametrization of the local coordinate around $P_0$. Conformal blocks
as $\langle\,V_C\,|\,I_\infty\,\rangle$ therefore represent functions on open
subsets of an infinite dimensional generalization of the moduli space $\CT(C\setminus P_0)$
which parameterizes tuples $(C,P_0,y)$, where $y$ denotes the choice of a
local
coordinate around $P_0$. The moduli space of all such tuples $(C,P_0,y)$ is
closely related to the moduli space of Riemann surfaces with a hole which has a
parameterized boundary: To given $(C,P_0,y)$ one may consider the surface
$C\setminus \BD_{\ep}$, where $\BD_{\ep}$ is a disc with radius $\ep$ around $P_0$, defined using the local coordinate $y$ by the condition $|y|<\ep$.
Changes of coordinate $y$ induce reparameterizations of the 
boundary of $C\setminus \BD_{\ep}$.

As a finite, but arbitrarily large part
of the reparameterizations of $y$ acts 
nontrivially on the irregular vectors, we may regard such vectors
as an approximation to the insertion of  a hole with 
parameterized boundary. We will make this point of view 
more precise in the second part of our paper.
It will be shown that the irregular vectors can be
used as a useful regularization in the study of the 
infinite-dimensional moduli spaces associated to 
surfaces with holes.

\section{Algebraic construction of bases 
for spaces of irregular vectors} \label{sec:vertex}

\setcounter{equation}{0}

In order to construct physical correlation functions in a
holomorphically
factorized form like
\rf{holofact1} or \rf{holofact2} one first needs to find
useful bases for the spaces of conformal blocks. It is our
next aim to define such bases in the case of conformal blocks
constructed from irregular vectors as in \rf{irrblock}. This
is equivalent to defining bases for the space of solution
to the Virasoro constraints summarized in \rf{irrOPE}.

\subsection{The problem}

It will again be useful to compare with the case of regular vectors 
$|\,R_n(z)\rangle$ defined in \rf{regvecdef}.
For this case it is well-known how to construct useful bases for the
space of solutions to the constraints
characterizing the vectors $|\,R_n(z)\rangle$.
One may, for example, introduce vertex operators 
$\Psi^{\al}_{\al_f,\al_i}(z)$
that map from the Virasoro module $\CV_{\al_i}$ to $\CV_{\al_f}$.
Such vertex operators are defined uniquely up to a constant by the
intertwining property
\begin{equation}\label{Psinconstr}
L_n \cdot\Psi^{\al}_{\al_f,\al_i}(z)\,=\, z^n(z\pa_z+\al(n+1))
\Psi^{\al}_{\al_f,\al_i}(z)+ 
\Psi^{\al}_{\al_f,\al_i}(z) \cdot L_n \,.
\end{equation}
Out of these vertex operators one may then construct families of
regular vectors defined by expression such as
\begin{equation} \label{eq:regvec}
|\,R^{(n)}(z,\be)\,\rangle \equiv \, \Psi^{\al_1}_{\al_0,\be_1}(z_1)\Psi^{\al_2}_{\be_1,\be_2}(z_2)\cdots
\Psi^{\al_{n}}_{\be_{n-1},\al_{n+1}}(z_n)
\,|\,\al_{n+1}\,\rangle\,.
\end{equation}
\begin{figure}\begin{center}
\includegraphics[width=7.5cm]{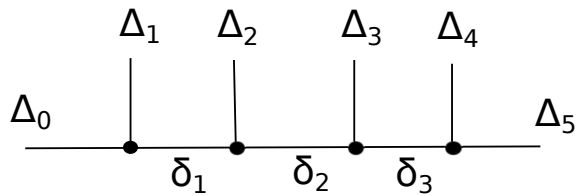} 
\caption{\it The standard graphical representation of a conformal block. In the figure, $\De_i$ denotes $\De_{\al_i}$ and $\de_i$ denotes $\De_{\be_i}$.}
\label{Fig:reg}\end{center}
\end{figure}
The elements of this family are labelled by the tuple $\be$ of
intermediate dimensions, $\be=(\be_1,\dots,\be_{n-1})$. 
The same tuple may therefore be taken as  
label for a basis $|\,R^{(n)}(z,\be)\,\rangle$ for the space of
solutions to the constraints \rf{Rnconstr} 
inside the Verma module $\CV_{\al_0}$. A diagrammatical representation
for the conformal blocks $\langle\al_0|R^{(n)}(z,\be)\rangle$
is given in Figure \ref{Fig:reg}.

We have seen first evidence for the claim
that irregular vectors can be constructed from the
collision of ordinary primary fields. This suggests that
one may define bases for the space of irregular
vectors by taking a suitable limit of the family of
vectors $|\,R^{(n)}(z,\be)\,\rangle$. This also suggests 
that the set of parameters labelling 
bases of irregular vectors is related to the one
appearing in the case of regular vectors: There will
be $n-1$ parameters $\be=(\be_1,\dots,\be_{n-1})$ 
labelling elements 
$|\,I^{(n)}(z,\be)\,\rangle$ of a basis for the space
of irregular vectors of n-th order.
Alternatively,
one may look for more direct ways of defining such bases, 
for example by generalizing the construction \rf{eq:regvec}.
We'll propose ways to realize both options, but it will turn
out that none of them will be straightforward to realize.

One may begin looking for a generalization of the construction 
\rf{eq:regvec} by recalling that 
this definition produces a representation of the regular 
vector $|\,R^{(n)}(z,\be)\,\rangle$ as a power series
in $z_1$, $z_{2}/z_1$, $z_3/z_2$, etc., with a leading powers 
$(z_k/z_{k-1})^{\De_{\be_{k-1}}-\De_{\be_k}-\De_{\al_k}}$ 
controlled by 
the intermediate conformal dimensions. This basis therefore
has a simple behavior at the boundary of the complex structure moduli 
space where the punctures are colliding in a specific pattern, 
$z_n \ll z_{n-1}\ll \cdots \ll z_2 \ll z_1$. 
The role of the component  of the boundary of the complex structure
moduli space considered above would in the 
case of irregular vectors 
naturally be taken by  
regimes in which the parameters $\bc=(c_1,\dots c_n)$ tend to 
zero in a specific hierarchical order.
This suggests that part of the characterization of 
irregular counterparts of the vectors \rf{eq:regvec} 
will be a specification of their asymptotic behavior
in such regimes. 

It is important to notice, however, that the leading asymptotic behavior 
alone does 
not suffice to define 
the basis of conformal blocks uniquely: 
Adding arbitrary 
linear combinations of the vectors 
$|R^{(n)}(z,\be)\rangle$ with 
intermediate dimensions $\De_{\be_i}$ replaced by $\De_{\be_i}+k_i$, 
$k_i\in\BZ^{\geq 0}$, would yield 
vectors which have the same asymptotic behavior.
Very similar problems will arise in overly naive 
attempts to  characterize bases for spaces of irregular vectors in 
terms of their asymptotic behavior when $\bc=(c_1,\dots c_n)$ degenerates. 
One therefore needs additional requirements to characterize the
elements $|I^{(n)}(z,\be)\rangle$ of a basis for the irregular
vectors uniquely. 

\subsection{The proposed solution}\label{sec:vertintro}

To begin with, let us note that
it is easy to find inside a generic Verma module 
$\CV_{\al_0}$ a unique solution $|\,I^{(1)}(c_1)\,\rangle$ 
to the Ward identities for a rank 1 irregular vector. 
As we review in Appendix \ref{sec:rk1}, this can be done 
either by direct solution of the Ward identities, or from the 
collision limit of $|\,R^{(1)}(w)\,\rangle$ \cite{Gaiotto:2009ma}. 
Thus there is no problem defining a basis of conformal block 
with one (or more) rank $1$ punctures. 

On the other hand, it is easy to see that 
we can find infinitely many solution $|\,I^{(2)}(c_1,c_2)\,\rangle$ 
to the Ward identities for a rank 2 irregular vector, simply by picking an arbitrary 
$c_1$ functional dependence for the coefficient of the highest weight vector in 
$\CV_{\al_0}$.

The solution we are going to propose for the problems
arising when $n>1$ may again be 
motivated by reconsidering the regular case. Let us look at the
simplest nontrivial case, $n=2$, for example. 
The vector $\Psi^{\al_{2}}_{\be,\al_{3}}(z_2)|\al_{3}\rangle$
appearing in the definition \rf{eq:regvec} can be expanded
as sum over Virasoro descendants of $|\,\be\,\rangle$,
\begin{equation}
\Psi^{\al_{2}}_{\be,\al_{3}}(z_2)\,|\al_{3}\,\rangle\,=\,
z_2^{\De_\be-\De_{\al_2}-\De_{\al_3}}
\sum_{I}z_2^{|I|}\,C_I\,{\BL}_{-{I}}\,|\,\be\,\rangle\,.
\end{equation} 
with ${\BL}_{-{I}}$ being monomials in Virasoro generators,
and $|{I}|$ being the $L_0$-weight of $L_{-{I}}$. Moving 
${\BL}_{-{I}}$ through $\Psi^{\al_{1}}_{\al_0,\be}(z_1)$
by means of \rf{Psinconstr} will yield for $|R^{(n)}(z,\be)\rangle$
an expression of the 
form
\begin{align}\notag
|\,R^{(2)}(z,\be)\,\rangle\,&=\,z_2^{\De_\be-\De_{\al_2}-\De_{\al_3}}
\sum_{I}z_2^{|I|}\,C_I\,{\CL}_{-{I}}\Psi^{\al_{1}}_{\al_0,\be}(z_1)
\,|\,\be\,\rangle\\
& =: z_2^{\De_\be-\De_{\al_2}-\De_{\al_3}}
\sum_{I}z_2^{|I|}\,C_I\,\CL_{-I}|\,R^{(1)}(z_1,\be)\,\rangle\,,
\label{eq:Regexp}\end{align}
where ${\CL}_{-{I}}$ is obtained from ${\BL}_{-{I}}$ by
replacing every Virasoro generator $L_{-k}$ in ${\BL}_{-{I}}$
by $L_{-k}-z_1^{-k}(z_1\pa_{z_1}+(1-k)\De_{\al_1})$. We see that the
vector $|R^{(2)}(z,\be)\rangle$ can be expanded as 
a sum over vectors that may be called
generalized descendants of the vector
$|R^{(1)}(z,\be)\rangle$. 

This recursive structure can be 
used to characterize the vectors $|R^{(2)}(z,\be)\rangle$ uniquely.
Indeed, imposing the compatibility of the 
expansion \rf{eq:Regexp} with the 
constraints \rf{Rnconstr-k} characterizing the vectors $|R^{(2)}(z,\be)\rangle$
and $|\,R^{(1)}(z_1,\be)\,\rangle$ yields an infinite set of equations
on the coefficients $C_I$ in \rf{eq:Regexp}
which turns out (see Appendix \ref{app:CVO}) 
to fix them uniquely up to an overall normalization.

Anticipating that the elements $|I^{(2)}(c,\be)\rangle$
of a basis for the space of irregular vectors can be obtained
from $|R^{(2)}(z,\be)\rangle$ in a suitable limit, suggests that
the vectors $|I^{(2)}(c,\be)\rangle$ may be characterized
by a recursive relation to the vectors $|I^{(1)}(c)\rangle$
that is similar to \rf{eq:Regexp}. Indeed, we will
propose that an analog of \rf{eq:Regexp} will be given by an expansion
of the form
\begin{equation}\label{eq:rk2ansatz}
|\,I^{(2)}(c,\al'')\,\rangle\,=\,c_2^{\nu_2}c_1^{\nu_1}\,
e^{(\al''-\beta') \frac{c_1^2}{c_2}}\,
\sum_{k=0}^{\infty}c_2^k\,|\,I^{(1)}_{2k}(c_1,\beta')\,\rangle\,,
\end{equation}
where the vectors $|I^{(1)}_{2k}(c_1,\beta')\rangle$ are 
generalized descendants of the rank 1 irregular vector 
$|I^{(1)}(c_1,\beta')\rangle$. 
With the term ``generalized descendant'' we mean 
linear combinations of vectors obtained from 
$|I^{(1)}(c_1,\beta')\rangle$ by acting on it with Virasoro generators or 
derivatives with respect to $c_1$. The coefficients in this 
expansion are strongly constrained by
the equations following from the 
consistency of \rf{eq:rk2ansatz} with the 
constraints characterizing the irregular vectors
$|I^{(2)}(c,\al'')\rangle$ and $|I^{(1)}(c_1,\beta')\rangle$,
respectively.

We conjecture that there exists a 
solution of the resulting 
equations which determines the vectors 
$|I^{(1)}_{2k}(c_1,\beta')\rangle$ uniquely
in terms of $|I^{(1)}(c_1,\beta')\rangle$.
We have performed extensive checks of this conjecture
by calculating low orders in the expansion above. 
A more detailed discussion is given in  Appendix \ref{app:CVO}.
This conjecture is furthermore supported by our discussion of the
collision limits which indicate that bases of irregular vectors 
characterized by expansions of the form  \rf{eq:rk2ansatz}
can be constructed by taking certain limits of regular vectors.

\subsection{Generalization to higher rank irregular vectors}

\begin{figure}\begin{center}
\includegraphics[width=6cm]{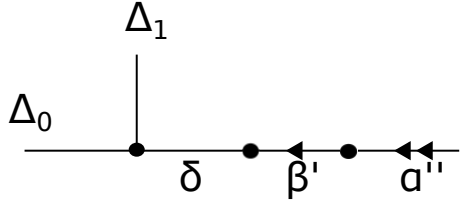} 
\caption{\it The graphical representation of a conformal block with two regular punctures, fused in a channel of dimension $\delta$, and a rank $2$ puncture of momentum $\al''$, realized inside a rank $1$ channel of momentum $\beta'$
We denote rank $2$ channels with a double arrow. Black dots denote standard or generalized chiral vertex operators.}\label{Fig:Irr}
\end{center}
\end{figure}

We furthermore 
conjecture that such bases of solutions can be built 
recursively for irregular vectors of any rank. 
We find it natural to denote the basis with a notation 
which resembles the regular case, as 
\begin{equation}\label{eq:genVO}
|\,I^{(n)\,} \rangle \,=\, \Psi^{r,1}_{\al_0,\beta'}(c_1)
\Psi^{1,2}_{\beta',\beta''}(\bc^{(2)})\cdots
\Psi^{n-1,n}_{\beta^{(n-1)},\al^{(n)}}(\bc^{(n)})\,
|\,I_n\,\rangle\,.
\end{equation}
Here $\Psi^{k-1,k}_{\beta^{(k-1)},\beta^{(k)}}(\bc^{(k)})$ denotes the linear operation of expanding (any descendant of) 
a rank $k$ irregular vector of momentum $\beta^{(k)}$ 
as the appropriate sum over descendants 
of a rank $k-1$ irregular vector of momentum $\beta^{(k-1)}$, 
and $\Psi^{r,1}_{\al_0,\beta'}(c_1)$ the realization of (any descendant of) 
a rank $1$ irregular vector of momentum $\beta'$ inside of the Verma module 
$\CV_{\al_0}$. 
The elements of such a basis are labelled by the tuple of
momenta $(\beta',\dots,\beta^{(n-1)})\in\BC^{n-1}$ which label the intermediate
irregular vectors used in the expansion. A diagrammatical
representation for the elements of such a basis is depicted in 
Figure \ref{Fig:Irr}.

More formally one may consider the maps 
$\Psi^{k-1,k}_{\beta^{(k-1)},\beta^{(k)}}(\bc^{(k)})$ 
as intertwining operators between the irregular modules 
$\CV_{\bc;\al}^{(k)}$
introduced in Subsection \ref{sec:irrmod} as follows:
We may consider $\Psi^{1,2}_{\beta,\al}(\bc)$, with $\bc=(c_1,c_2)$, 
for example, 
as an operators between the spaces
\begin{equation}
\Psi^{1,2}_{\beta,\al}(\bc)\,:\,\CV_{\bc;\al}^{(2)}
\;\ra\;\CV_{c_1;\beta}^{(1)}\ot \BC[\![\,c_2/c_1^2\,]\!]'\,
c_2^{\nu_2}c_{1}^{\nu_{1}}e^{(\al-\be)\frac{c_1^2}{c_2}}\,,
\end{equation}
where $\BC[\![z]\!]'$, the algebraic dual of the polynomial 
ring $\BC[\![z]\!]$, is the space for formal Taylor series 
in the variable $z$. The operator $\Psi^{1,2}_{\beta,\al}(\bc)$
is supposed to satisfy the
intertwining property
\begin{equation}\label{eq:intertw}
L_k\cdot\Psi^{1,2}_{\beta,\al}(\bc)\,=\,
\Psi^{1,2}_{\beta,\al}(\bc)\cdot
L_k\,.
\end{equation}
In order to describe the image of $\CV_{\bc;\al}^{(2)}$
within $\CV_{\bc;\be}^{(1)} \ot \BC[\![\,c_2/c_1^2\,]\!]'\,
c_2^{\nu_2}c_{1}^{\nu_{1}}e^{(\al-\be)\frac{c_1^2}{c_2}}$, 
it clearly suffices to find the vector 
\begin{equation}
\Psi^{1,2}_{\beta,\al}(\bc) \, |\,I_{\bc;\al}^{(2)}\,\rangle
\,\in\, \CV_{c_1;\beta}^{(1)}\ot \BC[\![\,c_2/c_1^2\,]\!]'\,
c_2^{\nu_2}c_{1}^{\nu_{1}}e^{(\al-\be)\frac{c_1^2}{c_2}}\,,
\end{equation}
the rest being determined by \rf{eq:intertw}. This vector must satisfy
the equations following from the combination of 
$L_k |I_{\bc;\al}^{(n)}\rangle=
\CL_k(\bc;\al)|I_{\bc;\al}^{(n)}\rangle$ with \rf{eq:intertw}.  But these 
equations are easily seen to be equivalent to the equations
determining the generalized descendants 
$|I^{(1)}_{2k}(c_1,\beta')\rangle$ in \rf{eq:rk2ansatz} above.

\subsection{Other types of bases in the presence of irregular singularities}

The constructions above do not exhaust the family of bases 
for irregular vectors that may be of interest. One may wish to study
conformal blocks of mixed type 
containing both regular and irregular singularities like,
for example  $\langle\al_0|RI^{(1)}(\be)\rangle$, where
\begin{equation} \label{eq:regrk11}
|\,RI^{(1)}(\be)\,\rangle\,=\, 
\Psi^{\al_2}_{\al_0,\be}(z)
\Psi^{r,1}_{\be,\al'}(c_1)|\,I_1\,\rangle \,.
\end{equation}
The constructions above give a representation
as a power series in $c_1/z$ which characterizes the 
conformal blocks $\langle\al_0|RI^{(1)}(\be)\rangle$ 
near $c_1/z=0$. It is natural to ask if there
exist alternative bases for the solutions to the
constraints characterizing $|RI^{(1)}(\be)\rangle$
which have a simple behavior in the opposite limit where 
$z/c_1\ra 0$.

And indeed, we are going to propose that there exist 
solutions of the Ward identities which admit an expansion over
generalized descendants of 
a rank $1$ irregular vector of momentum $\beta'$,
\begin{equation}\label{eq:regrk1exp}
|\,IR^{(1)}(\be)\,\rangle =\,z^{\mu_z}c_1^{\mu_1}\,
e^{(\al'-\beta') \frac{2c_1}{z}}\,
\sum_{k=0}^{\infty}z^k\,|\,I^{(1)}_k(c_1,\beta')\,\rangle\,.
\end{equation}
The vectors $|I^{(1)}_k(c_1,\beta')\rangle$ are 
generalized descendants of $|I^{(1)}(c_1,\beta')\rangle$ 
as introduced in \rf{eq:rk2ansatz} above,
with coefficients which only depend on $\al_2$, $\al'$ and $\beta'$.
We have again found ample evidence for the conjecture that
a solution to the constraints for $|\,IR^{(1)}\,\rangle$ of the form 
\rf{eq:regrk1exp} exists and is unique. 

\begin{figure}\begin{center}
\includegraphics[width=4cm]{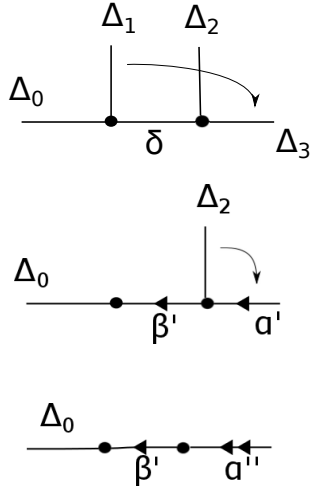} 
\caption{\it The sequence of collision limits which give the conformal blocks of rank $2$.}\label{fig:coll}
\end{center}
\end{figure}

In order to represent the resulting new 
basis for the conformal blocks in a way analogous to \rf{eq:genVO}
it may be convenient to introduce a generalization 
of the vertex operator $\Psi^{\al}_{\al_f,\al_i}(z)$ that
is defined in the usual way in terms of the intertwining
property \rf{Psinconstr}, but which is now mapping the 
irregular module $\CV^{(1)}_{c_1;\al_i}$ to $\CV^{(1)}_{c_1;\al_f}$.
We will denote  the resulting object as 
$\Psi^{(1)\al}_{\al_f,\al_i}(z)$. The basis 
$|IR^{(1)}(\be)\rangle$ defined by means of  
the expansions \rf{eq:regrk1exp}
could then be represented as
\begin{equation} \label{eq:regrk1not}
|\,IR^{(1)}(\be)\,\rangle= \Psi^{r,1}_{\al_0,\beta'}(c_1) 
\Psi^{(1)\al_2}_{\beta',\al'}(z)|\,I_1\,\rangle \,.
\end{equation} 
We have given a diagrammatical representation
of the basis $|IR^{(1)}(\be)\rangle$ in Figure \ref{fig:coll}. 
\begin{figure}\begin{center}
\includegraphics[width=4.5cm]{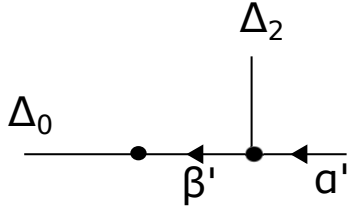} 
\caption{\it The graphical representation of a conformal block where a regular puncture and a rank $1$ puncture are realized inside a rank $1$ channel of momentum $\beta'$.}
\end{center}
\end{figure}\

This basis will also turn out to be 
useful as an intermediate step in the analysis of relations 
between the basis $|R^{(2)}(\be)\rangle$ for regular vectors,
and the basis $|I^{(2)}(\be)\rangle$. We will indicate how 
vectors $|IR^{(1)}(\be)\rangle$ can be constructed 
in a simple, careful collision limit from the usual 
$|R^{(2)}(\be)\rangle$. Furthermore, 
the power series defining the vectors $|IR^{(1)}(\be)\rangle$
is now adequate to reproduce the vectors $|I^{(2)}(\be)\rangle$
in a careful collision limit $z_2 \to 0$.
A diagrammatical representation
of this sequence of operations is given in Figure \ref{fig:coll}. 

The study of the collision limits
gives another way to argue that 
vectors $|IR^{(1)}(\be)\rangle$ and
$|I^{(2)}(\be)\rangle$ with series expansions \rf{eq:rk2ansatz}
and \rf{eq:regrk1exp}, respectively,
really exist, as will be discussed from 
two points of view in Appendix \ref{nullcoll}.

\subsection{Further generalizations}

 In the case of regular conformal blocks, 
 there are several other useful bases of solutions for the Ward identities with $n$ punctures. 
 Indeed, the chiral vertex operator $\Psi^\al_{\al_f,\al_i}(z)$ can be readily promoted 
 to a map $\CV_{\al_i} \to \CV_{\al}\ot
\CV_{\al_f}$, using Virasoro Ward identities 
 in order to place descendants of the primary of dimension $\al$ at z. 
Then one can fuse the punctures in any order, forming a basis labeled by a rooted binary tree
\begin{equation}
|R_n\rangle = \Psi_{\al_f,\be_1}^{\be_2}\left[\Psi_{\be_1, \be_3}^{\be_4}[\cdots], \Psi_{\be_2, \be_5}^{\be_6}[\cdots] \right]
\end{equation} 
 In the case with irregular punctures, one can similarly promote  
$\Psi_{\beta^{(n)},\al^{(n)}}^{\al_z}(z)$ to a map 
$\CV_\al \otimes \CI_n \to \CI'_n$. Exchanging the role of $0$ and $z$, 
one can thus define a map  $\Psi_{\beta^{(n)},\al}^{\al^{(n)}}$ which fuses a rank $n$ irregular vector at $z$ and a regular vector at the origin 
into a rank $n$ irregular module. Then starting from 
 \begin{equation}
 \Psi_{\gamma^{(n)},\beta^{(n)}}^{\al_w}(w) \Psi_{\beta^{(n)},\al}^{\al^{(n)}} |\al \rangle
 \end{equation}
and colliding $w \to 0$ it may be possible to define a formal power series for 
$\Psi_{\beta^{(n)},\al'}^{\al^{(n)}}(z) |I_1\rangle$ which fuses an irregular vector of rank $n$ and an irregular vector of rank $1$ 
into an irregular module of rank $n+1$.  Iterating this procedure, 
one may arrive to the most general map $\Psi_{\beta^{(n+m)},\al^{(n)}}^{\al^{(m)}}$, fusing irregular punctures of rank $n$ and $m$ 
into an irregular module of rank $n+m$. These maps could be combined to produce very general bases of conformal blocks with irregular singularities,
which explore more general boundary components of Teichm\"uller space for several irregular punctures. We leave a more detailed discussion of such possibilities to
the future.



\section{Free field construction}\label{sec:screen}

\setcounter{equation}{0}

As an alternative approach to the construction of bases for
spaces of irregular vectors we will now describe constructions
based on the free field representation of the Virasoro algebra.
This will give strong additional support for our
previous claims about existence of irregular vectors 
with a certain structure of their expansions around 
the degeneration limit. It will furthermore give 
strong hints towards the existence of Stokes phenomena 
in such limits. 

\subsection{Primary fields}

At first, we can review the free field construction of chiral vertex operators. 
We will mostly consider the case that $Q>2$ in the following, corresponding
to central charge $c>25$. It turns out, however, that the results
that we obtain for this regime have
an analytic continuation w.r.t. the parameter $Q$
which allows one to cover the case $c>1$ as well.
The basic building blocks of all constructions will be the following
objects:\\[1ex]
\noindent{\it Normal ordered exponentials : }
\begin{equation}
\SE^{\al}(z)\,
\equiv\,\exp\Bigg(2\al\sum_{k< 0}\frac{i}{k}a_kz^{-k}\Bigg)\,
e^{2\al(\sq -\al_{\spp}\log z)}\,
\exp\Bigg(2\al\sum_{k> 0}\frac{i}{k}a_kz^{-k}\Bigg)\,.
\end{equation}
{\it Screening charges:}
\begin{equation}
\SQ(z)\;\equiv\;
\lim_{\ep\downarrow 0}\int_{\CC_{z,\ep}}dw\;\SE^b(w)\;,
\end{equation}
with integration contour being the circle $\CC_{z,\ep}=\{w\in\BC;
|w|=e^{\ep}|z|\}$.

Out of these building blocks 
we may now construct an important class of chiral primary fields,
\begin{equation}\label{CVO}
\SV_s^{\al}(z)\;=\;
\big(\SQ(z)\big)^s\,\SE^{\al}(z)
\;.\end{equation}
These objects are a priori only defined under suitable restrictions
on the parameters $\al$, $s$ and $b$ which ensure that the short-distance
singularities arising from the operator product expansions of the 
fields in \rf{CVO} are all integrable.
Similar objects can be defined for more
general values of the parameters $\al$, $b$ {\it and} $s$ 
by analytic continuation \cite{TL}. For explicit calculations
it may also be useful to 
replace $(\SQ(z))^s$ in \rf{CVO} by expressions of the form
$\int_{\Ga_1}dt_1\dots\int_{\Ga_s}dt_s \,\SE^b(t_1)\cdots\SE^b(t_s)
\SE^{\al}(z)$ for a suitable collection of 
contours $\Ga_1,\dots,\Ga_n$. 

The covariant transformation law under conformal transformations,
\begin{equation}
[L_k, \SV_s^{\al}(z)]\,=\,z^k(z\pa_z+\De_\al(k+1))\SV_s^{\al}(z)\,,
\end{equation}
follows from the well-known facts that the fields 
$E^{\al}(z)\equiv \SV_0^{\al}(z)$ satisfy this transformation law,
and that the fields $\SE^b(w)$ transform as
total derivatives due to $\De_b=1$.

\subsection{Irregular vectors}

To begin with, let us introduce coherent states 
$|\bc;\al\rangle^{(n)}$ as before, defined by the properties
\begin{equation}\begin{aligned}
&\sa_k\,|\,\bc\,;\al\,\rangle^{(n)}\,=\,-ic_k\,|c;\al\,\rangle^{(n)}\,,\\
&\sa_k\,|\,\bc\,;\al\,\rangle^{(n)}\,=\,0\,,
\end{aligned}\qquad
\begin{aligned} 
&{\rm for}\;\;
0<k\leq n\,,\\
&{\rm for}\;\;
k> n\,.
\end{aligned}
\end{equation}
and thus satisfy the Ward identities for irregular vectors
of degree $n$. The vectors $|\bc\,;\al\rangle^{(n)}$ can be 
considered as coherent states created 
from the Fock vacuum $|\al\rangle$ as
\begin{equation}
|\,\bc\,;\al\,\rangle^{(n)}\,=\,\exp\Bigg(
\sum_{k=1}^n \frac{1}{n}c_na_{-n}\Bigg)|\al\rangle\,.
\end{equation}

More general irregular vectors may then be constructed by
acting on the vectors $|\bc\,;\al\rangle^{(n)}$ with powers of the
operators $\SQ_\ga$
\begin{equation}
\SQ_\ga\;\equiv\;
\int_{\ga}dw\;\SE^b(w)\;,
\end{equation}
where $\ga$ is any contour that starts and ends at $w=0$ in 
sectors for which ${\rm Re}(c_n/w^n)<0$. There are $n$ such sectors,
explicitly given by
\begin{equation}
\CS_k^{(n)}:=\left\{\,w\in\BC\,;\,-\frac{\pi}{2}+2\pi k <n\arg(w)-(\ga_n-\pi)<
-\frac{\pi}{2}+2\pi k\,\right\}\,,
\end{equation}
for $k=0,\dots,n$. Vectors like 
\begin{equation}\label{ffirrvecdef}
\left(\SQ_{\ga_1}\right)^{s_1}\,\cdots
\,\left(\SQ_{\ga_m}\right)^{s_m}\,|\,\bc\,;\al\,\rangle^{(n)}
\end{equation}
will then be well-defined for collections of non-intersecting
contours $\ga_1,\dots,\ga_m$ of the type introduced above.
Moreover, the operators $\SQ_\ga$ are easily seen to 
commute with the Virasoro generators. This implies that 
the vectors defined in \rf{ffirrvecdef} behave under
conformal transformation in the same way as the 
vectors $|\bc\,;\al\rangle^{(n)}$.
One can consider a basis of $n$ non-intersecting contours $\ga_l$ which
start {\it and} end at $w=0$ in the sectors $\CS_k^{(n)}$, and 
thereby generate families of irregular vectors which depend
on $n$ additional positive-integer valued parameters $s_1,\dots,s_n$.

\begin{figure}\begin{center}
\includegraphics[width=3.5cm]{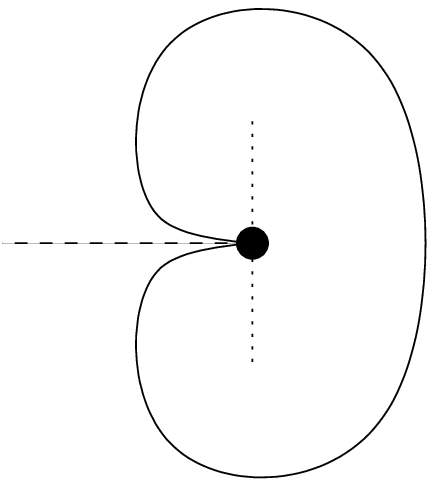} \quad \quad
\includegraphics[width=3.5cm]{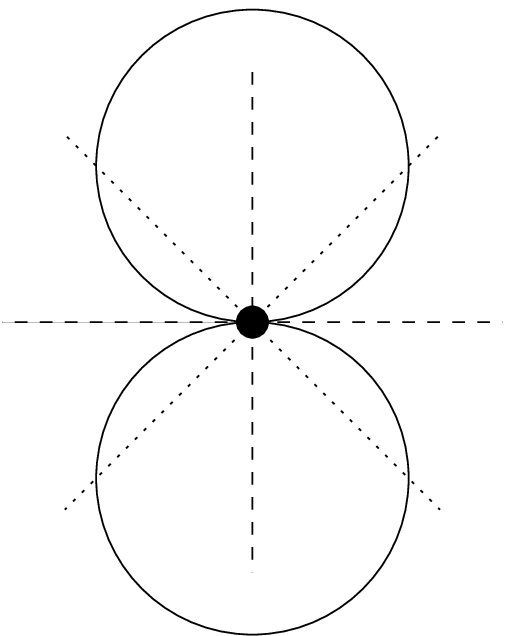} \quad \quad
\includegraphics[width=3.5cm]{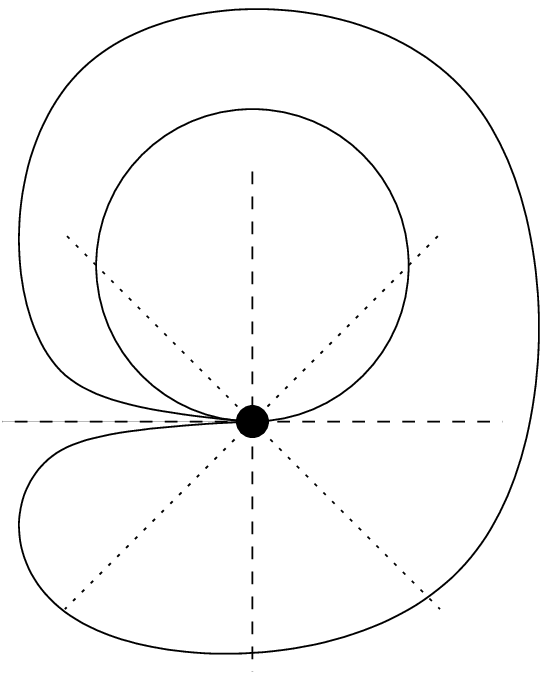} \quad \quad
\caption{\it Examples of screening paths. From left to right, the unique rank $1$ example, a rank $2$ example with two short paths, another rank $2$ example}
\label{rk1rk2}
\end{center}
\end{figure}

Notice that there are several inequivalent choices of set of $n$ contours $\ga_l$, which give distinct bases of irregular vectors. 
For example, one could consider the $n$ ``shortest'' contours, joining consecutive sectors. 
Alternatively, one could use nested sets of longer contours. Some examples
are shown in Figure \ref{rk1rk2}.

We can now consider collision and degeneration limits in the free field setup, in order to match the free-field bases of irregular vectors with the
formal power series built from expansion over descendants of irregular vectors. 

\subsection{Degeneration limits}

\subsubsection{Saddle point analysis}

In a degeneration limit  the singularity of $\partial \phi(y)$ goes from $n+1$ to $n$. Correspondingly, 
there must be a zero $y_{\bc}$ of $\partial \phi(y_{\bc})$ which moves to the origin. The approximate position of the zero is $- c_n/c_{n-1}$
and it is easy to see that the value of $\phi(y_{\bc})$ diverges as $c_n \to 0$. 
This has an interesting implication: the zero $y_{\bc}$ of $\partial \phi(y)$ is a saddle for the screening charge contour integral,
and the saddle point approximation is increasingly good in the degeneration limit for an integration contour corresponding to the 
steepest descent contour of $y_{\bc}$. 

This means that any screening contours which can be deformed to the steepest descent contour 
will collapse in the degeneration limit, and their 
contribution can be computed in the saddle point approximation.
The position of the saddle point and the value of 
$\partial \phi(y)$ on the saddle are 
not affected much by the presence of other screening charges. 
The value on the saddle is controlled by the value 
$\phi_{\mathrm{sing}}(y_{\bc})$ of
\begin{equation}
\phi_{\mathrm{sing}}(y) = \frac{c_n}{ny^n} + \frac{c_{n-1}}{(n-1)y^{n-1}} + 
\cdots + \frac{c_1}{y}
\end{equation}
plus logarithmic terms which are affected by the other screening charges. 

Let us apply these observations to the study of the behavior of
an irregular vector of rank $n$ of the form
\[
\left(\SQ_{\ga_1}\right)^{s_1}\,\cdots
\,\left(\SQ_{\ga_n}\right)^{s_n}\,|\,\bc\,;\al_n\rangle^{(n)}
\]
in a degeneration 
limit where $c_n\ra 0$. Let us assume that $\ga_n$ can be deformed to the steepest descent
contour for the saddle point which is collapsing to the origin, while the $\ga_1,\dots,\ga_{n-1}$ are chosen so that they do not 
receive contributions from that saddle point, i.e. do not intersect the path of steepest ascent from that saddle. 
When $c_n\ra 0$ we will then get an irregular vector
of order $n-1$ proportional to
\[
\left(\SQ_{\ga_1'}\right)^{s_1}\,\cdots
\,\big(\SQ_{\ga_{n-1}'}\big)^{s_{n-1}}\,|\,\bc^{(n-1)}\,;
\al_{n-1}\rangle^{(n-1)}\,,
\]
multiplied with 
a prefactor which contains
\begin{equation}
e^{2bs_n\phi_{\rm sing}(y_{\bc})}\,=\,
e^{-2(\al_n - \al_{n-1}) \phi_{\rm sing}(y_{\bc})}\,.
\end{equation}
The contours $\ga_1',\dots,\ga_{n-1}'$ are obtained from 
$\ga_1,\dots,\ga_{n-1}$ by deforming these contours such that
they start and end in the sectors $\CS_k^{(n-1)}$.
The logarithmic terms give important powers of $c_n$, which 
are harder to compute from the free field analysis.

It may be instructive to observe that the free field 
construction gives a rather concrete realization of the
intertwining operators $\Psi^{n-1,n}_{\beta^{(n-1)},\beta^{(n)}}(\bc)$
introduced in Section \ref{sec:vertex}.
Indeed, by expanding 
\begin{equation}
|\,\bc\,;\al_n\rangle^{(n)}\,=\,\sum_{m=0}^\infty \frac{c_n^m}{m!}
(a_{-n})^m\,|\,\bc^{(n-1)}\,;\al_{n-1}\rangle^{(n-1)}\,,
\end{equation}
combined with an application of the saddle-point method as outlined
above one will get a representation
for $\left(\SQ_{\ga_n}\right)^{s_n}|\bc\,;\al_n\rangle^{(n)}$
as a formal series in powers of $c_n$ of the form
\begin{equation}\label{FFseries}
\left(\SQ_{\ga_n}\right)^{s_n}\,|\,\bc\,;\al_n\rangle^{(n)}
\,=\,e^{-2(\al_n - \al_{n-1}) \phi_{\rm sing}(y_{\bc})}\prod_{k=1}^{n}
c_k^{\nu_k}\sum_{m=0}^{\infty}c_n^m\,|\,\bc^{(n-1)}\,;\al_{n-1};m\,\rangle^{(n-1)},
\end{equation}
where the $|\bc^{(n-1)}\,;\al_{n-1};m\,\rangle^{(n-1)}$ are generalized descendants 
of $|\bc^{(n-1)}\,;\al_{n-1}\,\rangle^{(n-1)}$. It follows that
the formal expansion in powers of $c_n$ of 
$\left(\SQ_{\ga_n}\right)^{s_n}|\bc\,;\al_n\rangle^{(n)}$
represents the intertwiner $\Psi^{n-1,n}_{\beta^{(n-1)},\beta^{(n)}}
(\bc)$ within the free field representation.

\subsubsection{Stokes phenomena}

In a degeneration limit where $c_n/c_{n-1}\ra 0$, there will be
a unique steepest descent
contour $\ga_n$ for the saddle point which tends to the origin in this
limit. The remaining contours can always be assumed to 
have zero intersection with the contour of steepest ascent. 
Using such contours in the construction above will define
irregular vectors which have an asymptotic behavior for 
$c_n/c_{n-1}\ra 0$ that is well-approximated by the formal
series \rf{FFseries} only in a certain sector of the complex
plane parameterized by $c_n/c_{n-1}$. Indeed, assuming for 
example 
that for a given initial value of the parameters $c$ 
the steepest descent contour
$\ga_n$ ends in the sector $\CS_n^{(n)}$, a variation of the 
phase of $c_{n-1}$ may move the phase of the saddle point 
$-c_n/c_{n-1}$ too far away from the sector $\CS_n^{(n)}$
for having a steepest descent contour that would still 
end in  $\CS_n^{(n)}$. We conclude that there are Stokes
phenomena in the asymptotic behavior of irregular vectors
in the degeneration limit.

\begin{figure}\begin{center}
\includegraphics[width=5cm]{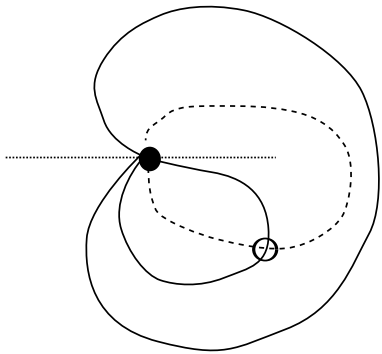}
\caption{\it A choice of screening paths with good degeneration limit. The open circle denotes the saddle point, the dashed path is the steepest ascent path. }\label{rk1rk2'}
\end{center}
\end{figure}

We may observe an analogy with the classification of different natural
bases for regular conformal blocks: Different natural bases are labelled
by the boundary components of the Teich\-m\"uller space of the 
Riemann surface one is working on. The elements of a basis associated
to a given boundary component 
are characterized by having a simple form of the expansion in powers
of the gluing parameters in the given boundary component only. 
They may be analytically continued to other boundary components, 
but will have a much more complicated behavior there. There
exist, however, linear transformations between the bases 
associated to different boundary components that can be
decomposed into the so-called fusion-, braiding- and modular
transformation moves.

Considering conformal blocks in cases
with irregular singularities, the considerations above 
strongly suggest that the data
classifying boundary components of the Teichm\"uller spaces 
may include the choices of Stokes sectors. A given basis
for the space of conformal blocks is characterized by having
a simple asymptotic expansion only in one particular
Stokes sector. The analytic continuation of the
elements of a given
basis associated to one Stokes 
sector into another sector 
will be shown in the 
second part of this series to be representable as
linear combinations of the elements of the basis 
associated to the other sector. The chiral bootstrap
is in the irregular case therefore characterized by an 
enlarged set of data containing analogs of the Stokes
matrices in addition to the fusion-, braiding- and modular
transformation matrices (or integral kernels). The second
part in this series will in particular contain
explicit calculations of these data.

Let us finally stress one important observation: Vectors
like $\left(\SQ_{\ga_n}\right)^{s_n}|c;\al_n\rangle^{(n)}$
which can be expanded as in \rf{FFseries} are perfectly 
well-defined objects. Identifying the expansion on the
right hand side of \rf{FFseries} with the
formal expansion of the intertwiner 
$\Psi^{k-1,k}_{\beta^{(k-1)},\beta^{(k)}}(c_k)$
suggests that the conformal blocks constructed
using this intertwiner do not only exist in the
sense of formal series, but that there exist 
actual functions for which the
algebraic constructions discussed above give 
the asymptotic series expansions in suitable Stokes sectors.

\subsection{Collision limits}

The free field representation is in many respects particularly well-suited for 
discussing the production of irregular vectors in collision 
limits. In order to illustrate some important qualitative 
features we will restrict attention to the case $n=2$, leaving
the discussion of more general cases to the future. 
Let us start from
\begin{equation}\label{start}
\SV^{\al_1}_{s_1}(z_1)\SV^{\al_2}_{s_2}(z_2)|\,\al_3\,\rangle\,.
\end{equation}
The normal ordered exponential fields satisfy exchange relations
of the form
\begin{equation}\label{ex}
\SE^{\al_1}(z_1)\SE^{\al_2}(z_2)\,=\,e^{-2\pi i\al_1\al_2}
\SE^{\al_2}(z_2)\SE^{\al_1}(z_1)\,,
\end{equation}
valid for $|z_1|=|z_2|$, $\arg(z_1)>\arg(z_2)$. Introduce the partial
screening charges 
\begin{equation}
\SQ_I:=\int_{z_2}^{z_1}dw\;\SQ^b(w)\,,
\qquad
\SQ_{I'}:=\int_{z_1}^{\hat{z}_2}dw\;\SQ^b(w)\,,
\end{equation}
where $\hat{z}_2:=e^{2\pi i}z_2$. The exchange relation 
\rf{ex} allows us to 
move the normal ordered exponentials in the definition of \rf{start}
to the right of the screening charges. Assuming $\arg(z_1)>\arg(z_2)$ we
thereby find
\begin{align}\label{reorder}
\big[\SQ(z_1)\big]^{s_1}\SE^{\al_1}(z_1)\,&   
\big[\SQ_{I}+\SQ_{I'}\big]^{s_2}\SE^{\al_1}(z_2)\,=\\
=\,& \big[\SQ(z_1)\big]^{s_1}\,\big[e^{-2\pi ib\al_1}\SQ_{I}+e^{2\pi ib\al_1}\SQ_{I'}\big]^{s_2}\,
\SE^{\al_1}(z_1)\SE^{\al_2}(z_2)\,.
\nn\end{align}
In this form it becomes straightforward to take the collision limits
producing irregular vectors of degree $2$. 
The collision limits of $\SE^{\al_1}(z_1)\SE^{\al_2}(z_2)|\al_3\rangle$
produce vectors $|c;\al''\rangle^{(2)}$ with $\al''=\al_1+\al_2+\al_3$.
It is furthermore easy to show that the partial screening charge
$\SQ_{I}$ becomes the operator $\SQ_{\ga}$
with contour $\ga$ which connects sector $\CS_2$ with $\CS_1$,
while $\SQ_{I'}$ turns into the operator $\SQ_{\ga'}$ associated
to the contour $\ga'$ which connects $\CS_1$ with $e^{2\pi i}\CS_2$.
The operator $\SQ(z_1)$ becomes an operator $\SQ_{\xi}$ associated to
a contour $\xi$ which starts in $\CS_1$, encircles $w=0$ and ends in
$e^{2\pi i}\CS_1$.

What is interesting to observe is the fact that there are
two different ways to take the limit of \rf{reorder} which yield  
either 
\begin{equation} \label{eq:choice}
\big[\SQ_\xi\big]^{s_1}\big[\SQ_{\ga}\big]^{s_2} |\,\bc;\al''\,\rangle^{(2)}\,,
\qquad
{\rm or} 
\qquad
\big[\SQ_\xi\big]^{s_1}\big[\SQ_{\ga'}\big]^{s_2} 
|\,\bc;\al''\,\rangle^{(2)}\,,
\end{equation}
as a result, depending on whether ${\rm Im}(\al_1)$ tends to $\pm\infty$
in the limit.

One of the main points to be observed here is the fact that after 
multiplying with some simple numerical factors we get vectors
that stay finite in the collision limits.

 \section{Conformal blocks from solutions of null vector equations} 
\label{sec:degenerate}

\setcounter{equation}{0}

Degenerate fields in Liouville theory satisfy differential equations.
We will use these differential equations in order to get an alternative
approach to the definition of bases in the space of conformal blocks,
the calculation of their series expansions,
and for the study of the collision limits producing 
irregular singularities.

This is necessarily somewhat tricky, as generic conformal blocks do not 
satisfy a closed system of partial differential equations. The idea 
may be informally described as follows:
Inserting {\it additional} degenerate fields into the conformal block
gives modified conformal blocks that satisfy the differential equations
following from null vector decoupling. These differential equations
can be used to obtain power series expansions for the modified 
conformal blocks. The original conformal block can be recovered in 
a certain limit where the additional degenerate fields fuse with some
of the primary fields inserted into the modified conformal blocks. 
In order to calculate this limit it suffices to know the braiding
transformations involving degenerate fields, which is explicitly known.
We will show how this idea can be used to calculate series expansions
for conformal blocks involving both regular and irregular vertex
operators.

 \subsection{Degenerate chiral vertex operators} \label{sec:degvec}

Let us consider the special chiral vertex operators 
\begin{equation}
V_{\pm}(y)\,\equiv\, \Psi^{-b/2}_{\be',\be}(y)\,.
\end{equation}
It is well-known that these  vertex operators
satisfies the
operator differential equation
\begin{equation}\label{NVop}
\frac{1}{b^2}\frac{\pa^2}{\pa y^2}\,V_\pm(y)\,+:T(y)
V_{\pm}(y):\,=\,0\,,
\end{equation}
with normal 
ordering $:T(y)
V_{\pm}(y):$ defined as 
\begin{equation}
:T(y)
V_{\pm}(y):\,\equiv\sum_{k<-1}y^{-k-2}L_k
V_{\pm}(y)+\sum_{k\geq -1}y^{-k-2}
V_{\pm}(y)L_k\,.
\end{equation}
The operator differential equation \rf{NVop} is equivalent to the
decoupling of null vectors in the Verma module of descendants of 
$V_{\pm}(y)$. We will therefore call the equations following from 
\rf{NVop} null vector equations. Note in particular that
conformal blocks containing insertions of $V_{\pm}(y)$
like
\begin{align}\label{CF:def}
&\CF^{}(y)\equiv\CF^{}(y;z_1,z_2):=
\big\langle\,\al_0\,|\,\Psi^{\al_1}_{\al_0,\be'}(z_1)\,
V_+(y)\,
\Psi^{\al_2}_{\be,\al_3}(z_2)\,|\,\al_3\,\big\rangle\,,
\end{align}
will satisfy a partial differential equation of 
second order in $\frac{\pa}{\pa y}$ which is obtained 
by moving the Virasoro generators in $:T(y)
V_{\pm}(y):$ to the left or right until they hit the highest weight
vectors. The resulting differential equation is of the generic form
\begin{equation}\label{PDEreg0}
\left(\frac{1}{b^2}\frac{\pa^2}{\pa y^2}+\CT(y)\right)
\CF^{}(y;z_1,z_2)\,=\,0\,,
\end{equation}
where $\CT(y)$ is a first order differential operator which for 
the case \rf{CF:def} above is explicitly given as
\begin{align*}
\CT^{}(y):=&\frac{\De_1}{(y-z_1)^2}+\frac{\De_2}{(y-z_2)^2}
+\frac{\De_3}{y^2}+\notag 
\frac{1}{y-z_1}\frac{z_1}{y}\frac{\pa}{\pa z_1}
+\frac{1}{y-z_2}\frac{z_2}{y}\frac{\pa}{\pa z_2}-\frac{1}{y}
\frac{\pa}{\pa y}\,.
\end{align*}
We are using the notations $\De_{\al_r}=\al_r(Q-\al_r)$, $r=0,1,2,3$,
$\de_b=-\frac{1}{2}-\frac{3}{4}b^2$. 

\subsubsection{Fusion rules}

The null decoupling condition can only hold if one restricts the 
Liouville momentum $\be$ 
to jump across the degenerate field as $\be'=\be\pm b/2$. 
Indeed, if we apply this relation to the leading term in the expansion of 
$V_\pm(z) |\De_\be \rangle$ and denote
$\delta = \De_{\be'} -\De_\be -\De_{-b/2}$, 
we get the constraint $\delta (\delta-1-b^2) + b^2 \De_\be =0$. 
This is solved by $\de = b \be$ or $\de = b (Q-\be)$,
which means $\be' = \be -b/2$ or $\be' = \be + b/2$,
respectively. It is known that this condition is also sufficient for 
$V(z)$ to satisfy the null vector equation. 
We will also denote the solutions with $\be' = \be \mp b/2$ as
\begin{equation}
V_{\pm}(y)\,\equiv\, V_{\be;\pm}(y)\,\equiv\,\Psi^{-b/2}_{\be\mp b/2,\be}(y)\,.
\end{equation} 

Notice that $\de$ controls the monodromy of $V(y)$ around the origin. 
This fact is of crucial importance, as it represents a link
between the characterization of bases for the spaces of conformal
blocks in terms of the series expansions for solutions of the 
null vector equations on the one hand, 
to the characterization in terms
of Verlinde line operators \cite{AGGTV,DGOT} on the other hand. 
The latter is closely connected to the characterization 
of bases for the spaces of conformal blocks by means of 
the geodesic length operators in quantum Teichm\"uller
theory. 

We can easily repeat the analysis to find the constraints appropriate 
for the vertex operator $V^{(1)}(y)$ 
which represents the insertion of a degenerate field near a rank $1$ irregular vector as a sum over generalized descendants
of the rank $1$ irregular vector. 
Thus in order for  $V^{(1)}(y)$ 
to satisfy the null vector decoupling, it must shift the Liouville momentum of the irregular puncture by $\pm b/2$.
We can thus define 
\begin{equation}
V^{(1)}_{\pm}(y)\,\equiv\, V_{\be';\pm}^{(1)}(y):=
\Psi^{-b/2(1)}_{\be'\mp b/2,\be'}(y)\,.
\end{equation} 
This reasoning readily extends to the vertex operators 
$V^{(n)}_\pm(y)$.

\subsubsection{Monodromy and formal monodromy}

Notice an important fact. Our expansion \rf{eq:regrk1exp}
for $V^{(1)}_\pm(z)$ has a prefactor
$y^{\nu}c_1^{\nu_1}\, e^{(\beta'-\al') \frac{2c_1}{y}}$
where 
\begin{equation}
\nu = - 2 \Delta_{- b/2}- 2 (\beta' - Q)(\beta' - \al') = 1 + 3/2 b^2 \mp b(\beta'-Q)
\end{equation}
which means either $\nu = b \al'$ or $\nu = b( 2 Q - \al')$.
The parameter $\nu$ controls the formal monodromy of the asymptotic expansion. 
This is a rather intuitive result. If the irregular puncture arises from 
the collision of two regular punctures, of Liouville momenta which 
add to $\al'$, it appears that the formal monodromy around the 
irregular puncture 
is simply the sum of the monodromy eigenvalues 
around each individual puncture. 

The formal monodromy is  a very important piece of information. 
We expect it to provide a link between the bases for the irregular 
conformal blocks constructed in this paper
and the irregular generalization of the quantum Teichm\"uller theory. 

Furthermore, this confirms the expectation that the solutions built from 
$V^{(1)}_\pm(y)$
may be well-defined beyond the formal power series definition, but depend on a choice of Stokes sector 
where the expansion would be valid. In the case of the degenerate insertion, 
the exponential prefactor $\exp \pm \frac{b c_1}{y}$ suggests that there are two possible choices of Stokes sector:
sectors including either of the two half-lines  
$\frac{y}{c_1} \in \pm i {\mathbb R}^+$.

Finally, we can extend the formal monodromy statement to any rank $k$. We expect that for an irregular puncture 
of rank $k$ and momentum $\al_k$ at the origin, the degenerate insertion will shift the Liouville momentum 
by $\pm b/2$, and will have formal monodromy $\nu = b \al_k$ or $\nu = b( (k+1) Q - \al_k)$.
We can test this at the leading order of the $y$ expansion 
for $V^{(k)}_\pm(y)|I_k \rangle$. The null decoupling constraint 
takes the form
\begin{equation}
\left( \partial_y^2 - b^2 y^{-1} \partial_y\right)+ b^2 T_{\gg}(y)
\end{equation}
where with $T_{\gg}(y)$ we denote the singular part of $T(y)$ near the irregular vector, excluding the $L_{-1}$ piece.
This constraint can be patiently applied to the ansatz 
\begin{equation}
y^\nu \exp \pm \sum_{n=1}^k \frac{c_n}{n y^n}
\end{equation}
to verify that this ansatz can be the starting point of a systematic 
asymptotic expansion of the solution. We refer the reader to section \ref{sec:n2} 
for an example of such systematic expansion, at rank $2$. 

\subsection{Construction of bases with the help of null vector equations - the regular case}

We now want to explain in some more detail how to use the null vector 
equations in order to construct certain bases for the space 
of conformal blocks. 
As indicated above, the basic idea is to first consider conformal 
blocks which contain degenerate fields $V_{\pm}(y)$, exploit the
information given by the differential equations that 
such conformal blocks satisfy, and finally remove the degenerate
fields by taking some limit which produces conformal blocks 
without degenerate fields. We'll here give an outline of this 
procedure, leaving several details to Appendix \ref{app:PDE}.


Let us explain the basic idea a bit more precisely
in the case $n=0$. The object of our interest is the 
expansion of the conformal block
\begin{equation}
F(z_1,z_2):=\,\big\langle\,\al\,|\,\Psi^{\al_1}_{\al,\be}(z_1)
\Psi^{\al_2}_{\be,\al_3}(z_2)\,|\,\al_3\,\big\rangle
\end{equation}
in powers of $z_2$. The scaling properties of this conformal block
imply the general form 
\begin{equation}
F(z_1,z_2)\,=\,z_1^{\De_\al-\De_{\al_1}-\De_{\al_2}-\De_{\al_3}}
\sum_{k=0}^{\infty}
\left(\frac{z_2}{z_1}\right)^{\chi+k}F_k\,.
\end{equation}

As a technical tool for its study we shall
modify the conformal blocks by additional insertions of the
special chiral vertex operators $V_{\pm}(y)$.


The differential equations satisfied by $\CF^{}(y;z_1,z_2)$
will turn out to have a unique solution in the form 
of a double power series in $z_2/y$ and $y/z_1$ such that
\begin{equation}\label{F0series}
\CF^{}(y;z_2,z_1)\,=\,
\sum_{k=0}^\infty\left(\frac{z_2}{y}\right)^{\chi+k}\!\!
\CF^{}_k(y;z_1)\,,
\quad\CF^{}_k(y;z_1)\,=\,z_1^\ka
\sum_{l=0}^\infty\left(\frac{y}{z_1}\right)^{\eta+l}\!\!
\CF^{}_{k,l}\,,
\end{equation}
where
\begin{equation}
\ka\,=\,\De_0-\De_1-\De_2-\De_3-\de_b\,,\qquad
\chi\,=\,\De_{\be}-\De_{\al_2}-\De_{\al_3}\,,\qquad
\eta\,=\,b\be\,.
\end{equation}
This form of the series as specified in \rf{F0series} is necessary
for the solution to be identified
with the conformal block
$\CF^{}(y;z_1,z_2)$. It follows from the representation theoretic
construction of the conformal blocks by summing over 
states from fixed intermediate representations.

To avoid confusions, let us note that it is not at all straightforward
to find series expansions in $z_1/y$ that could be identified 
with the conformal blocks $\CE^{}(y;z_2,z_1):=\big\langle\al_0|V_{\pm}(y)
\Psi^{\al_1}_{\al_0\pm b/2,\be}(z_1)
\Psi^{\al_2}_{\be,\al_3}(z_2)\,|\,\al_3\,\big\rangle$. The differential
equation does not give any constraints on the leading
coefficients $\CE_{k,0}$ of an expansion like
\begin{equation}
\CE^{}(y;z_2,z_1)\,=\,
\sum_{k=0}^\infty\left(\frac{z_2}{y}\right)^{\chi+k}\!\!
\CE^{}_k(y;z_1)\,,
\quad\CE^{}_k(y;z_1)\,=\,z_1^\ka
\sum_{l=0}^\infty\left(\frac{z_1}{y}\right)^{\eta'+l}\!\!
\CE^{}_{k,l}\,.
\end{equation}

It is therefore necessary for us to start from an 
expansion of the form \rf{F0series}, and continue
analytically to $y\ra \infty$ afterwards to recover the 
conformal block $F(z_1,z_2)$ we are after.
As a tool for carrying out this analytic continuation we may use the
exchange relation
\begin{align}\label{degbraid}
 & \Psi^{\al_1}_{\al_0,\be-b/2}(z_1)\, 
V_{\be;+}^{}(y)\,=\\
&\quad\,=\,B_+(s)\,
V_{\al_0+b/2;+}^{}(y)\Psi^{\al_1}_{\al_0+b/2,\be}(y)
+B_-(s)\,
V_{\al_0-b/2;-}^{}(y)\Psi^{\al_1}_{\al_0-b/2,\be}(y)\,,
\notag\end{align}
valid for $|z_1|=|y|$ 
furthermore allows us perform the analytic continuation to $|y|>|z_1|$.
The coefficients
$B_{\pm}(s)$ in \rf{degbraid}
depend on $s:=\sgn(\arg(z_1/y))$, as usual. 
One may then study 
the limit $y\ra\infty$ using the OPE
\begin{equation}\label{infOPE}
\big\langle\,\al_0\,|\,V^{}_{\al_0+\ep b/2;\ep}(y)\,=\,
\big\langle\,\al_0+\ep b/2\,|\;y^{\frac{1}{2}(2b^2+1)+\ep b(\al_0-Q/2)}\,
\big(C_{\ep}+\CO(y^{-1})\big)\,.
\end{equation}
Assuming that $\Re(2\al_0-Q)>0$, the term with $\ep=1$ in \rf{infOPE}
dominates for $y\ra\infty$. Let us assume that $s=1$ and set $B_+\equiv
B_{+}(1)$. Assuming furthermore that $\al=\al_0+b/2$, 
we may then calculate the 
sought-for coefficients $F_k$ as
\begin{equation}\label{Fkreconstr}
F_k\,=\,\frac{1}{C_+B_+}
\lim_{y/z_1\ra\infty}y^{-b(\al_0+b/2)}
z_1^{k-\ka}\CF_{k}(y;z_1)\,.
\end{equation}
We want to show that \rf{Fkreconstr} leads to a purely algebraic
procedure for the calculation of the expansion
coefficients $F_k$. To this aim let us note that 
the representation theoretic definition of 
$\CF^{}(y;z_1,z_2)$ via \rf{CF:def}
yields power series in $y/z_1$, $z_2/z_1$ 
convergent for $|z_1|>|y|>|z_2|$. The expansion
in powers of $z_2$ can be obtained from \rf{eq:Regexp}.
We thereby get an expansion for $\CF^{}(y;z_1,z_2)$ of the form \rf{F0series}.
The coefficient functions $\CF^{}_k(y;z_1)$
 are proportional to the conformal blocks
\begin{equation}\label{desccontr}
\CF^{}_k(y;z_1)\,=\,y^{k} 
\,\langle\,\al_0\,|\,\Psi_{\al_0,\be-b/2}^{\al_1}(z_1)\,
V_+(y)\,
|\,\be,v_k\,\rangle\,, 
\end{equation}
where $|\,\be,v_k\,\rangle$ is the descendant
$\sum_{{I};|{I}|=k}C_{{I}}{\BL}_{-{I}}|\,\be\,\rangle$. 
By moving ${\BL}_{-{I}}$ to the left in \rf{desccontr}
above one may rewrite $\CF^{}_k(y;z_1)$ in the
form
\begin{equation}\label{GkDG0}
\CF^{}_k(y;z_1):=\,y^{k}\CD_k(y,z_1)\CF^{}_0(y;z_1)\,,
\end{equation}
where the differential operator $\CD_k(y,z_1):=
\sum_{{I};|{I}|=k}
C_{{I}}{\mathbf \CL}_{-{I}}(y,z_1)$
creating
$\CG^{}_k$ from $\CG^{}_0$ is
obtained from $\sum_{{I};|{I}|=k}
C_{{I}}{\BL}_{-{I}}$ by replacing
\[
L_{-k}\ra
-y^{-k}(y\pa_y+\De_{-b/2}(1-k))-z_1^{-k}(z_1\pa_{z_1}+\De_1(1-k))
\]
from
the left to the right. The differential operator
$\CD_k(y,z_1)$
is of the form
\begin{equation}\label{CDexpCP}
\CD_k(y,z_1)\,=\,\sum_{l=0}^k\left(\frac{z_1}{y}\right)^{l}\,
\sum_{m=0}^k \CD_{k;l,m}\left(y\frac{\pa}{\pa y}\right)^m\,,
\end{equation}
as follows easily from its scaling behavior. The relation \rf{GkDG0}
will allow us to calculate the asymptotics $\CF^{}(y;z_1,z_2)$ 
in terms of the asymptotics of the lowest order term $\CF^{}_0(y;z_1)$
as soon as we have determined the differential operator
$\CD_k$. Having constructed the full 
power series expansion \rf{F0series} of $\CF^{}(y;z_1,z_2)$
one can view  the relation \rf{GkDG0} as a linear equation for
the differential operator $\CD_k(y,z_1)$.
It is equivalent to the linear system
\begin{equation}\label{CDklm}
\sum_{m=0}^k\,\sum_{\substack{l',l''=0\\l'+l''=l}}^{k}\,\CD_{k;l',m}\,
(\eta+l'')^m
\CF_{0,l''}\,=\,\CF_{k,l}\,,
\end{equation}
of equations for the coefficients $\CD_{k;l,m}$ of $\CD_k(y,z_1)$.
This is an infinite system of equations for a finite number of
unknowns, so uniqueness of the solutions seems clear, while
existence may not be obvious. The existence of solutions 
is here assured by the representation
theoretic construction of the conformal blocks, as discussed in the above.

Inserting the relation \rf{GkDG0} between $\CF_k(y;z_1)$ and $\CF_0(y;z_1)$ into 
\rf{Fkreconstr} gives us the relation 
\begin{equation}
F^{}_k\,=\,y^{-b\al}\cdot\CD_{k}(y;0)\cdot y^{b\al}
=\sum_{m=0}^k \CD_{k;0,m}\left(b\al\right)^m\,,\qquad \al=\al_0+b/2\,.
\end{equation}
As the coefficients $\CD_{k;l,m}$ can be calculated from the expansion coefficients
$\CF_{k,l}$ by solving \rf{CDklm}, we thereby get a procedure to calculate the
coefficients $F_k$ from the differential equation satisfied by 
the modified conformal blocks $\CF(y;z_1,z_2)$.

\subsection{The case of an irregular singularity or rank 2}

We now want to consider the insertion of a degenerate vertex operator in the 
conformal block with a regular singularity at infinity, and a rank $2$ irregular singularity at the origin. 
The main idea is to use the degenerate field as a probe of
the internal structure of the irregular singularity. 

In order to
make this idea more precise we will show that
there exists a unique solution to the null vector 
decoupling equations that can be identified with the 
conformal blocks denoted as 
\begin{align}\label{rk2-confbl-yto0}
&\CF^{(2)}(y;c_1,c_2):=\langle\,\al_0\,|\,
\Psi^{r,1}_{\al_0,\be'-b/2}(c_1)
V^{(1)}_+(y)\,\Psi^{1,2}_{\be',\al''}(\bc)\,|\,I_2(\al'')\,\rangle\,,
\end{align}
using the notations of Section \ref{sec:vertex}. This should a priori
not be confused with 
\begin{align}\label{rk2-confbl-ytoinfty}
&\CE^{(2)}_\pm(y;c_1,c_2):=\langle\,\al_0\,|\,V^{}_\pm(y)\,
\Psi^{r,1}_{\al_0\pm b/2,\be'}(c_1)\Psi^{1,2}_{\be',\al''}(\bc)\,|
\,I_2(\al'')\,\rangle\,.
\end{align}
The conformal blocks defined in \rf{rk2-confbl-yto0}
and \rf{rk2-confbl-ytoinfty} turn out to be closely related, however.
$\CF^{(2)}(y;c_1,c_2)$, initially being characterized near $y\ra 0$
by an asymptotic
double series in powers of $y/c_1$ and $c_2/yc_1$, can be 
analytically continued into the region where 
$y\ra\infty$. As we will show later in this 
subsection, one may represent the result of this analytic 
continuation as a linear combination of the two 
conformal blocks $\CE^{(2)}_\pm(y;c_1,c_2)$ 
defined in \rf{rk2-confbl-ytoinfty}.

To begin with, let us use the expansion \rf{eq:rk2ansatz} to 
represent the conformal blocks $\CE^{(2)}_\pm(y;c_1,c_2)$ as
series in powers of $c_2$. The resulting expansion takes the form
\begin{align}
&\CE^{(2)}_\pm(y;c_1,c_2)\,=\,c_2^{\nu_2}c_1^{\nu_1}
e^{(\al''-\beta') \frac{c_1^2}{c_2}}\,
\sum_{k=0}^{\infty}c_2^k\,\CE^{(2)}_{k;\pm}(y;c_1)\\
&{\rm where} \quad\CE^{(2)}_{k;\pm}(y;c_1):=
\langle\,\al_0\,|\,V^{}_\pm(y)\,|\,I^{(1)}_{2k}(c_1,\beta')\,\rangle\,.
\end{align}
Assuming that the vectors $|\,I^{(1)}_{2k}(c_1,\beta')\,\rangle$ 
can be represented as generalized descendants of 
$|\,I^{(1)}(c_1,\beta')\,\rangle$, as we had proposed in 
Section \ref{sec:vertex}, we may move the Virasoro generators
in $|\,I^{(1)}_{2k}(c_1,\beta')\,\rangle$ to the left
and get recursive relations of the form 
\begin{equation}\label{rk2-recrel}
\CE^{(2)}_{k;\pm}(y;c_1)
\,=\,\CD_k(y,c_1)\,\CE^{(2)}_{0;\pm}(y;c_1)\,.
\end{equation}
Below we will show that $\CE^{(2)}_{0;\pm}(y;c_1)$ can be expressed
in terms of the confluent hypergeometric functions.
Taking into account \rf{rk2-recrel} allows us to 
conclude that the coefficients $\CE^{(2)}_{k;\pm}(y;c_1)$
are analytic multivalued functions in $y$ for all values of $k$. 

We can then find a linear combination of 
$\CE^{(2)}_{0;+}(y;c_1)$ and $\CE^{(2)}_{0;-}(y;c_1)$,
\begin{equation}
\CF^{(2)}_{0}(y;c_1)\,=\,K_1^{(2)}\CE^{(2)}_{0;+}(y;c_1)+
K_2^{(2)}\CE^{(2)}_{0;-}(y;c_1)\,,
\end{equation}
that has the correct leading asymptotics for $y\ra 0$
to be identified with the leading term of the expansion in powers of $c_2$
of the conformal blocks $\CF^{(2)}(y;c_1,c_2)$ we are interested 
in, 
\begin{align}
&\CF^{(2)}(y;c_1,c_2)\,=\,c_2^{\nu_2}c_1^{\nu_1}
e^{(\al''-\beta') \frac{c_1^2}{c_2}}\,
\sum_{k=0}^{\infty}c_2^k\,\CF^{(2)}_{k}(y;c_1)\,.
\end{align}
The leading asymptotics of $\CF^{(2)}_{0}(y;c_1)$ should be 
proportional to $e^{-b\frac{c_2}{2y^2}-b\frac{c_1}{y}}
y^{b\be'}$. Up to a normalization factor, there is going to be 
a unique solution
which has this property. It furthermore follows from \rf{rk2-recrel}
that we have 
\begin{equation}
\CF^{(2)}_{k}(y;c_1)\,=\,K_1^{(2)}\CE^{(2)}_{k;+}(y;c_1)+
K_2^{(2)}\CE^{(2)}_{k;-}(y;c_1)\,,
\end{equation}
for all values of $k$. As in the discussion of the regular case before,
we can calulate the differential operators $\CD_k(y,c_1)$ in 
\rf{rk2-recrel} with the help of the differential equations, and
ultimately recover the conformal block 
$\langle\,\al\,|\,
\Psi^{r,1}_{\al_0,\be'}(c_1)
\Psi^{1,2}_{\be',\al''}(c_2)\,|\,I_2(\al'')\,\rangle$
by sending $y\ra\infty$ in the end.

The lesson we want to extract from these observations is that 
alternative ways to define bases for the spaces of irregular 
conformal blocks can be found by probing the 
internal structure of irregular vectors
with the help of degenerate fields. The parameters 
labeling the elements of such bases are identified with 
the parameters describing the
asymptotic behavior of the degenerate fields for $y\ra 0$, 
here in particular by the parameter $\be'$.
The details are worked out for the 
three main examples at hand in Appendix \ref{app:PDE}.

\section{Physical correlation functions} \label{sec:measure}

\setcounter{equation}{0}

In the present section we are going to formulate our main 
conjectures concerning physical Liouville correlation 
functions with irregular singularities.

\subsection{Existence of collision limits}

We conjecture that the collision limits exist on the level of 
physical correlation functions after dividing by the corresponding 
free field correlator.
In the case of a four-point function
we conjecture in particular existence of the limit
\begin{equation}
\lim_{(0)\ra(2)}\,\frac{\bra\!\bra \,\al_0\,|\,
V_{\al_1}(z_1,\bz_1)V_{\al_2}(z_2,\bz_2)\,|\,\al_3\,\ket\!\ket_{\rm \mu}}
{\bra\!\bra \,\al\,|\,
V_{\al_1}(z_1,\bz_1)V_{\al_2}(z_2,\bz_2)\,|\,\al_3\,\ket\!\ket_{\rm 0}}\,,
\end{equation}
where the correlator in the denominator is evaluated in the
free boson theory obtained from Liouville theory by setting $\mu=0$
and $\al=\al_1+\al_2+\al_3$. 
The result represents an overlap of the form
 \begin{equation}
\Phi^{(2)}(c_1,c_2;\al_0,\al)\,=\,
\bra\!\bra \,\al_0\,|\,I^{(2)}(c_2,c_1;\al)\,.
\ket\!\ket\,,\end{equation}
The vector $|\,I^{(2)}(c_2,c_1;\al)\,
\ket\!\ket$ represents the insertion of an irregular singularity 
of order $2$ into physical correlation functions.

The evidence we may offer in favor of this proposal 
is obtained from two different sources: 

In Appendix \ref{nullcoll}
we demonstrate in two different ways that the series expansions for 
the conformal blocks can be rearranged in such a way that we 
have well-defined
collsion limits order by order in the series expansions. 
The first argument is essentially based on the observation that the 
Ward identities which define the series expansions for the
conformal blocks have a well-defined limit after extracting
the divergent free-field parts.  The second argument discussed
in Appendix \ref{nullcoll} uses the null vector equations. 
After factoring out the free field part, one obtains differential
equations that have a well-defined limit. The details of these 
arguments turn out to be delicate, however, as one needs to 
consider conformal blocks constructed from intermediate representations
whose highest weights diverge. We refer to Appendix \ref{nullcoll}
for further details.

A rather different approach to the existence of collision limits may
be based on the free field representation described in Section 
\ref{sec:screen}. Whenever this representation can be used, 
it will make our claim nearly obvious. The collision 
limit is defined in such a way that the numbers of screening charges
stay constant. A reordering as done explicitly in \rf{reorder} above
will therefore identify the operator product expansion of normal
ordered exponentials as the origin of all divergencies.
What needs further discussion, though, is the treatment of the
cases where noninteger screening powers appear. 
In the regular case one may use the observation \cite{TL} 
that $\SQ_{\ga}$ is a positive self-adjoint operator for 
$\ga$ being any interval on the unit circle. It follows
that $(\SQ_\ga)^s$ is well-defined even for non-integer values of $s$.
It is not completely clear to us how to generalize this approach to
the irregular case at the moment as the positivity may be lost.

\subsection{Expansion into conformal blocks}

We conjecture that the ``irregular correlation function''
$\Phi^{(2)}(c_1,c_2;\al_0,\al)$ can be expanded into irregular 
conformal blocks 
as follows:
\begin{equation}
\Phi^{(2)}(c_1,c_2;\al_0,\al)=\!\int\limits_{{Q}+i\BR}\!d\be\;
C^{(2)}(\al_0,\al,\be)\,
F^{(2)}_\be(c_1,c_2;\al_0,\al)
F^{(2)}_\be(\bar{c}_1,\bar{c}_2;\al_0,\al)\,,
\end{equation}
where 
\begin{itemize}
\item
$F^{(2)}_\be(c_1,c_2;\al_0,\al)$ are conformal blocks which
have an asymptotic expansion in powers of $c_2/c_1^2$ of the form
\begin{align}\label{F(2)asym}
F^{(2)}_\be & (c_1,c_2;\al_0,\al)\,=\,\\
& \quad=c_1^{\De_0-\De_{\al}}
\left(\frac{c_2}{c_1^2}\right)^{\nu_2}\!
e^{-(\be-\al)\frac{c_1^2}{c_2}}\,\left(1+\sum_{k=1}^{\infty}\left(\frac{c_2}{c_1^2}\right)^k F^{(2)}_\be(k;\al_0,\al)\right)\,,
\notag\end{align}
where $2\nu_2=(\be-\al)(3Q-3\be-\al)$, and the higher orders in the expansion 
are determined by the procedures described in Sections 
\ref{sec:vertex} or \ref{sec:degenerate}, respectively,
\item the conformal blocks 
$F^{(2)}_\be(c_2;\al_0,\al):=
F^{(2)}_\be(1,c_2;\al_0,\al)$ are
multivalued analytic functions of $c_2$ 
on $\BP^1\setminus\{0,\infty\}$ which can be  characterized  uniquely by
having the asymptotic expansion  \rf{F(2)asym} for $c_2\ra 0$ 
in a Stokes sector of width $\pi/2$,
\item the structure constants $C^{(2)}(\al_0,\al,\be)$ are explicitly
given by the following expression:
\begin{align}\label{irrDOZZ}
C^{(2)}(\al_0,\al,\be)&=\La^{\frac{1}{b}(Q-\al_0-\al)}
\frac{\up^2_0\up(2\al_0)}
{\up(\al_0+\be-Q)\up(\be-\al_0)\up(\al-\be)
}2^{2 \De_0-2 \De_{\be}-\De_{\be-\al}}\end{align}
where $\Lambda:=
\pi \mu\ga(b^2)b^{2-2b^2}$ 
\item 
The ``irregular correlation function''
$\Phi^{(2)}(c_2;\al_0,\al)\equiv
\Phi^{(2)}(1,c_2;\al_0,\al)$ is real analytic as function of $c_2$ on 
$\BP^1\setminus\{0,\infty\}$, and the expansion of the 
correlation function into conformal blocks is independent
of the choice of the Stokes sector used to characterize the 
conformal blocks by their asymptotic expansion \rf{F(2)asym}.
\end{itemize}
The last property is an analog of the crossing symmetry or modular
invariance of the physical Liouville correlation 
functions in the presence of an irregular singularity.

It is important to note that the precise form of the structure
functions $C^{(2)}(\al_0,\al,\be)$ is linked to the precise 
definition of the conformal blocks $F^{(2)}_\be(c_1,c_2;\al_0,\al)$.
By means of analytic reparameterizations of the variables 
$c_1$ and $c_2$ one could change the form of $C^{(2)}(\al_0,\al,\be)$,
but the series expansion of $F^{(2)}_\be(c_1,c_2;\al_0,\al)$
would also be changed. We had fixed the precise definition 
of the series $F^{(2)}_\be(c_1,c_2;\al_0,\al)$ in 
Section \ref{sec:irregular} or, equivalently in 
Section \ref{sec:degenerate}.

This conjecture would follow from the existence of the collision 
limits, and the fact 
that the collision limits will preserve the single-valuedness
of the physical correlation functions.
The precise form of the structure 
functions $C^{(2)}(\al_0,\al,\be)$ 
proposed in 
\rf{irrDOZZ} above was determined by carefully analyzing the
collision limits. The details of this analysis 
are described in Appendix 
\ref{nullcoll}.

\section{Gauge theory perspective} \label{sec:gauge}

\setcounter{equation}{0}

\subsection{Overview}

The purpose of this section is to discuss a possible application of our results on 2d CFT to the study 
of certain four-dimensional ${\cal N}=2$ gauge theories. It can be seen as a natural generalization of the
relations between expectation values of supersymmetric observables in a certain class of $SU(2)$ gauge 
theories and correlation functions in Liouville theory 
discovered in \cite{Alday:2009aq}. 
The gauge theories in question are associated to Riemann surfaces $C$, possibly with punctures \cite{Gaiotto:2009we,Gaiotto:2009hg}.
In a certain limit the above-mentioned relations reduce to relations between the 
Seiberg-Witten geometry describing the IR physics of the relevant gauge theories 
and the Teichm\"uller theory of the surfaces $C$ \cite{Witten:1997sc,Gaiotto:2009we}.

We propose that similar relations exist between correlation functions in Liouville theory with
irregular singularities and gauge theories of Argyres-Douglas type.
Even if the lack of a Lagrangian formulation of the Argyres-Douglas theories 
makes it 
difficult to directly generalize the calculations supporting the correspondence between 
gauge theories and Liouville theory, we may still describe the IR physics of the Argyres-Douglas
theories with the help of a variant of Seiberg-Witten theory. Our proposed 
relation between Argyres-Douglas theories and Liouville theory with irregular singularities will be supported by showing 
that it implies a relation between the IR physics 
of the Argyres-Douglas theories and the
Teichm\"uller theory for Riemann surfaces 
with irregular singularities that naturally generalizes the 
previously found relations.

The link between Seiberg-Witten- and Liouville theory can be described
a bit more concretely as follows. 
In the relevant limit the conformal blocks turn into the 
prepotential of the gauge theory, schematically
\begin{equation}
\CZ \;\sim\; e^{\cal F}\,.
\end{equation}
The expectation value of the energy-momentum tensor
of Liouville theory furthermore becomes the quadratic differential $\phi$ on $C$ which defines the 
Seiberg-Witten curve $\la^2=\phi$. The fact that insertions of the energy-momentum 
tensor generate derivatives $\pa_{\tau_a}$ of the conformal blocks with respect to the 
complex structure moduli of $C$ turns into the statement that the quadratic differential $\phi$
describes the behavior of the prepotential under variations of the gauge couplings, 
\begin{equation}\label{genMatone}
u_a\,=\,-\pa_{\tau_a}\CF\,,
\end{equation}
where $u_a$ can be computed from $\phi$ and the Beltrami differentials $\mu_a$ 
representing $\pa_{\tau_a}$
by 
\begin{equation}
u_a = \int_{C}\mu_a\phi. 
\end{equation}

The relation \rf{genMatone} generalizes well-known relations in 
Seiberg-Witten theory going back to \cite{Bonelli:1996qc}. 
We will review the derivation of this result from Seiberg-Witten theory
and extend it to the case of Argyres-Douglas theories. It will coincide with 
the appropriate limit of the conformal 
Ward identities \rf{irrOPE} in the presence of irregular singularities.

In this section we will also go through the gauge theory version of 
our collision limits and decoupling limits, to give a four-dimensional interpretation of
the results of the previous sections. 

Our main example will be the behavior of the $SU(2)$ $N_f=4$ gauge theory,
which corresponds to the four-punctured sphere, under the collision limits which 
reduce it to a famous Argyres-Douglas ($N_f=3$ in \cite{Argyres:1995xn}) theory, which corresponds to the correlation function with one irregular puncture of rank $2$ and one regular puncture. 
We will also give a physical interpretation to the ansatz for our bases of solutions of irregular Ward identities. 

Thus, we leverage the 2d CFT description in order to both probe the behavior of
protected correlation functions of asymptotically free four-dimensional 
${\cal N}=2$ gauge theories at strong coupling,
and compute protected correlation functions for Argyres-Douglas theories. 

\subsection{The six-dimensional perspective}

The class of four-dimensional gauge theories with ${\cal N}=2$ supersymmetry we are considering
(``class ${\cal S}$'') of four dimensional gauge theories  arises from the 
twisted compactification  on a Riemann surface of six-dimensional field theories 
with  $(2,0)$ superconformal symmetry. The six-dimensional origin  
represents an important source of inspiration for the study of the theories
in class $\CS$. 

The rules of the twisted compactification allow one to insert codimension two half-BPS defects at points on the Riemann surface.
The six-dimensional theory is labeled by a choice of simply-laced Lie algebra $\fg$. Thus theories in class ${\cal S}$ are labelled by the choice of Lie algebra, of Riemann surface $C$
with punctures, and of the type of punctures. 

Many conventional four-dimensional ${\cal N}=2$ gauge theories admit an alternative description as
theories in the class ${\cal S}$. The main advantage of the six-dimensional description of the theory is that several protected quantities in the four-dimensional theory have an hidden 
geometric description in six dimensions. In particular, there is an exact correspondence between certain 
correlation functions of the four-dimensional theory and correlation functions or conformal blocks on $C$ of two-dimensional non-rational CFTs. 
In some cases, both sides of the correspondence are fully computable, and match. In many cases, the two-dimensional CFT 
allows us to compute answers which are much harder to get at in the four-dimensional gauge theory. Sometimes, 
the four-dimensional gauge theory interpretation can help uncover hidden truths about two-dimensional CFTs, or forces us to ask new questions. 

Much of the flexibility in the construction arises from the possibility to choose which codimension two half-BPS defects are placed at points ("punctures") in $C$.
Each of the six-dimensional theories comes with a standard array of ``regular'' defects whose existence can be gleaned from the 
basic properties of the six-dimensional theory. Theories of class ${\cal S}$ with regular defects typically have four-dimensional 
superconformal symmetry in the IR. Most regular defects carry flavor symmetry currents localized at the defect, in some subalgebra $\fh$ of $\fg$
which coincide with $g$ itself for a ``full'' regular puncture. In four-dimensional ${\cal N}=2$ supersymmetry, every flavor symmetry current is associated with a mass deformation parameter.
The regular punctures typically give rise to standard highest weight vertex operators in the dual two-dimensional CFTs. The 
mass deformation parameters map to quantities such as the conformal dimension of the vertex operators. 

Regular punctures hardly exhaust the set of possible codimension two defects in the six-dimensional theory. 
Indeed, the mere existence of regular punctures with a non-Abelian flavor symmetry, say $\fg$ for simplicity, on their world-volume 
allows one to define many more defects: add four-dimensional degrees of freedom at the defect, with flavor symmetry $\fg$, and 
add $\fg$ four-dimensional gauge fields coupled to both the defect flavor symmetry and the 4d degrees of freedom. As long as the $\beta$ function of the 
$\fg$ gauge theory is negative or zero, this is a UV-complete definition of a new type of defect. Notice that the flavor currents for a full regular puncture 
cancel half of the beta function from the 4d gauge fields, so there is some scope for adding extra degrees of freedom at the puncture.

This general class of defects should map to some local defect in the two-dimensional CFT side of the duality, which is not a standard highest weight operator. 
Following the dictionary of the duality, it is natural to expect that the procedure of weakly ``gauging in'' extra degrees of freedom into a regular puncture 
should correspond to a ``sewing in'' description of the new local defect, akin to an OPE: one can cut a small circle around the defect, 
and insert a complete set of states for the 2d CFT on the circle. The power series expansion in the sewing parameter should match the instanton expansion of 
the gauge theory partition function. The coefficients in the expansion depend on the choice of four-dimensional 
degrees of freedom which are gauged in, which affect the instanton measure. 

The duality becomes useful if the new defect can be given an independent definition directly in the 2d CFT. 
Then CFT methods allow us to probe the properties of the system away from weak coupling, and possibly to compute the 
correlation functions of the four-dimensional degrees of freedom which were gauged in.  

There is a useful class of non-regular defects which have a simple, if unfamiliar, independent description in the 2d CFT:
they correspond to local operators at which the Ward identities for the energy-momentum tensor and the other currents 
of the CFT have poles of unusually high degree. We will denote such defects as ``irregular defects''. Irregular defects 
appear naturally whenever one considers scaling limits in the four-dimensional gauge theory, 
where some UV parameters such as masses or other scales in the UV description are sent to infinity, 
but UV gauge couplings are tuned so that the IR effective gauge couplings are kept finite. 

Simple scaling limits can be used to 
define asymptotically free theories as a limit of superconformal field theories in the class ${\cal S}$. 
One simply sends the mass of some hypermultiplet flavors to infinity while keeping the renormalized gauge couplings finite. 
More refined scaling limits give rise to situations where the extra degrees of freedom at the defect 
define a four-dimensional theories of the Argyres-Douglas type, 
which is a rather mysterious non-trivial superconformal, strongly interacting fixed point with no exactly marginal couplings.
Currently, not much is known about AD theories, besides their Seiberg-Witten geometry. 

We will focus on the six dimensional theory associated to the  $sl(2)$ algebra, and the Liouville two-dimensional CFT,
based on the Virasoro algebra. 
The $A_1$ six-dimensional theory admits a single basic codimension two half-BPS  defect, 
the full regular defect with $su(2)$ flavor symmetry, which maps to the standard highest weight 
vertex operator in Liouville theory. Regular $A_1$ theories 
admit simple four-dimensional descriptions as $SU(2)$ gauge theories with exactly marginal gauge couplings. 
The space of gauge couplings can be identified with the space of complex structure deformations of the Riemann surface $C$. 

At first, we can add asymptotically free $SU(2)$ gauge groups to the construction. This can be done by decoupling some fundamental matter 
in regular $A_1$ theories. If one follows the manipulation of parameters in detail, the result is that on the Liouville theory side of the story 
two standard vertex operators will collide to give a new operator in Liouville theory at which the stress tensor Ward identity has poles of degree $3$ or $4$,
that is a rank $1/2$ or $1$ irregular vector. 
The behavior of the correlation functions when regular singularities approach irregular singularities, which we have learned to describe through new bases of irregular conformal blocks, 
corresponds to a strong coupling region for the asymptotically free gauge groups. 
This is precisely the regime which is relevant for further decoupling limits which produce AD theories. 

There is a tower $AD_n$ of Argyres-Dougles theories with $SU(2)$ flavor symmetry which can be ``gauged in'' 
at a regular puncture to define ``irregular'' $A_1$ theories. They contribute to the beta function of the new $SU(2)$ gauge group 
slightly less than the amount required for conformal symmetry. Thus the ``gluing'' $SU(2)$ gauge group is still asymptotically free. 
They are expected to correspond to irregular vertex operators in Liouville theory of rank higher than $1$. 

\subsection{Relations between 4d gauge theory and 2d CFT - a dictionary}
There is a well developed dictionary between geometric objects associated to a
Riemann surface $C$ with genus $g$ and $n$ regular punctures, and protected quantities in the corresponding 
regular $A_1$ theory $T_{g,n}$. We will denote it as ``the 2d dictionary'' in this section. 

\subsubsection{Lagrangian formulation}

The class of (mass deformed) ${\cal N}=2$ superconformal gauge theories which we denote as 
regular $A_1$ theories admits Lagrangian descriptions based on $SU(2)^{3g-3+g}$ gauge groups 
coupled to fundamental, bifundamental, trifundamental and adjoint matter hypermultiplets.
It is useful to group the matter hypermultiplets in $2g-2+n$ blocks of eight complex fields, 
where each block carries three independent $SU(2)$ doublet indices. 

Each of the $SU(2)$ gauge groups should have zero beta function. This is accomplished by 
requiring each $SU(2)$ gauge group to gauge the diagonal combination of two of the $SU(2)$ flavor symmetries of 
the matter hypermultiplets. If the two $SU(2)$ flavor symmetries belong to two distinct 
blocks, the two blocks  behave as four fundamental flavors, and cancel the beta function. 
If the two $SU(2)$ flavor symmetries belong to the same block, the block behaves as the sum of an adjoint and a singlet, and again it cancels the 
beta function.  As the beta function is zero, the complexified gauge coupling $\tau$ of the $SU(2)$ gauge theory is exactly marginal. 
It is useful to define the corresponding instanton factor $q_a = e^{\pi i \tau_a}$.

Clearly, the structure of the Lagrangian is captured by an unrooted binary tree, where each block of hypermultiplets 
is a trivalent vertex, each gauge group an internal edge, and each residual flavor group is an outer edge. 
Different topologies of the tree correspond to different Lagrangian descriptions of the same theory $T_{g,n}$,
related by S-dualities. The parameter space of exactly marginal deformations of $T_{g,n}$ is identified with
the Teichmuller space of complex structure deformations of $C$. Different Lagrangians correspond 
to different ways to sew the Riemann surface from a pair of pants decomposition, labelled by the corresponding unrooted binary tree. 
The instanton factors are mapped to the sewing parameters, so that the Lagrangian is weakly coupled and useful when the Riemann surface 
is almost degenerate to a collection of three-punctured spheres.

\subsubsection{Seiberg-Witten theory}

The Lagrangian description makes it clear that $T_{g,n}$ should have $3g-3+n$ dimension $2$ Coulomb branch order parameters 
$u_r = {\mathrm Tr} \Phi_r^2$, and $n$ dimension $2$ Casimirs $d_i = {\mathrm Tr} M_i^2 $ for the mass parameters $M_i$ of the ungauged $SU(2)$ 
flavor symmetries. 
The six-dimensional description of the theory indicates that it is useful 
to package the $u_a$ and $d_i$ together in a single quadratic differential 
$\phi$ on the Riemann surface, which will allow us to describe the Coulomb 
branch and Seiberg-Witten geometry of $T_{g,n}$ in an S-duality covariant way.
The quadratic differential has double poles at the puncture of 
$C$, with coefficient equal to the corresponding mass Casimir $d_i$.
In a local sewing coordinate $z_r$, $\phi$ should have a local Laurent 
expansion 
\begin{equation}
\phi \sim \cdots + u_r \frac{dz_r^2}{z_r^2} + \cdots
\end{equation}  

The basic ingredients of Seiberg-Witten theory are the 
central charge functions $a_r$, and $a_r^D$ on the Coulomb branch
together with the prepotential $\CF(a)$ relating them as
$a_r^D=\pa_{a_r}\CF(a)$. In order
to define these objects we may first 
define the Seiberg-Witten curve $\Sigma$ by the equation
\begin{equation}
\lambda^2 = \phi\,.
\end{equation}
This equation defines a Riemann surface $\Sigma$ in $T^* C$, 
together with a canonical one-form $\lambda$ on $\Sigma$.
The periods of $\la$ along a canonical set of homology cycles
on $\Sigma$ give the central charges $(a, a_D)$ of the IR theory.

We want to show that the Coulomb branch parameters $u_r$ represent 
the effect of variations of the UV gauge couplings on the prepotential,
as expressed by \rf{genMatone}. To this aim it is natural to 
consider an enlarged parameter space $\CM$ which is the 
fibration of the Coulomb branch over the space $\CT$ of 
UV couplings of the theory. 
In our case we may observe that the space $\CM$ is naturally identified
with the cotangent bundle $T^*\CT(C)$ 
over the Teichm\"uller space of $C$.
Indeed, let us recall that there is 
a natural dual pairing between quadratic differentials $\phi$ and the
Beltrami-differentials $\mu$ that describe variations
of the complex structure of $C$, given by 
$\langle \phi,\mu\rangle=\int_C \phi\mu$. This identifies the
spaces of quadratic differentials
on $C$ with the fibers of $T^*\CT(C)$,
and it may be used introduce coordinates $u_r$ on $T^*\CT(C)$ 
which are conjugate to a given set of coordinates $\tau_s$ on the
Teichm\"uller space in the sense that 
$\langle \phi,\mu\rangle=u_r\tau_r$. 

Considering the prepotential $\CF$ as a function on $\CM$, 
i.e. a function of both $a$ and $\tau$, we may first observe that
the periods $a_k^D=\pa_{a_k}\CF=a_k^D(a,\tau)$ can be varied 
for fixed $a$ by varying $\tau$. At least locally, we may therefore consider
the collection of $(a_k,a_k^D)$ as coordinates on $\CM$. 
The first step in our proof of the 
relations \rf{genMatone} will be to observe that these
relations are equivalent to the statement that the 
canonical symplectic form on $\CM=T^*\CT(C)$ can be 
rewritten in terms of the coordinates $(a_k,a_k^D)$ as
\begin{equation}\label{aa-utau}
\sum_{r} \de u_r \wedge  \de\tau_r =\sum_{k} \de a_k \wedge \de a_k^D \,.
\end{equation} 
Indeed, assuming that the change 
of variables from $(a_k,a_k^D)$ to $(u_r,\tau_r)$ is such that 
\rf{aa-utau} holds, we may locally consider the 
difference $a_k^D \de a_k^{}-u_r^{}\de\tau_r^{}$  of one-forms on $\CM$
which is closed due to \rf{aa-utau}, therefore locally on $\CM$ 
representable as $\de\CF(a,\tau)$ with 
$a_k^D=\pa_{a_k}\CF$ and
$u_r=-\pa_{\tau_r}\CF$. The converse is proven by
a straightforward calculation.  

In order to prove relation \rf{aa-utau}, let us first note
that the right
hand side of this equation 
can be written in terms of the Seiberg-Witten differential
$\la$ as $\frac{1}{2} \int_{\Sigma} \delta \lambda \wedge \delta \lambda$.
This follows easily from the Riemann bilinear identity. 
In the expression 
$\frac{1}{2} \int_{\Sigma} \delta \lambda \wedge \delta \lambda$
we consider $\lambda$ as a family of closed one-forms on 
the same smooth manifold $\Sigma$, and define the variations 
$\delta \lambda$ in the integral 
that way. 
The $\wedge$ in this expression indicates the 
wedge operation both on $\Sigma$ and $\CM$. 
Note that variations along the Coulomb branch give normalizable 
deformations of the SW curve, i.e. 
$\delta_{u_r} \lambda$ define holomorphic differentials on $\Sigma$. Thus the integral $\frac{1}{2} \int_{\Sigma} \delta \lambda \wedge \delta \lambda$ 
is zero when the two variations are along the Coulomb branch, as it should be.
On the other hand, variations of the couplings 
$\tau_r$ give $\delta \lambda$ with crucial $(0,1)$ components. 

It follows that the equation \rf{genMatone} we want to prove becomes
equivalent to 
\begin{equation} \label{eq:lala}
\sum_{r}\de u_r\wedge \de \tau_r \,=\,
\frac{1}{2} \int_{\Sigma} \delta \lambda \wedge \delta \lambda\,.
\end{equation}
It remains to observe that
this equation \rf{eq:lala} has a strikingly simple proof: 
The $(0,1)$ part of $\delta \lambda$ only receives a 
contribution from the change of complex structure of $C$ 
under the variation $\delta \tau$. If we parameterize a 
complex structure deformation of $C$ by a Beltrami differential 
$\mu \delta \tau$, 
the holomorphic differential $dz$ gets deformed into 
$dz + \mu_{\bar z}^{z} d \bar z \delta \tau$.
The $(0,1)$ part of $\delta \lambda$ 
 is therefore just $\mu \lambda\, \delta \tau $, allowing us to calculate 
\begin{equation} \label{eq:lala2}
\frac{1}{2}  \int_{\Sigma} \delta \lambda \wedge \delta \lambda\, =\int_{\Sigma}  \delta \lambda \wedge (\mu_r \lambda \delta \tau_r ) =  \frac{1}{2}
\int_{C} \delta \phi \wedge \mu_r \delta \tau_r =
\frac{1}{2}\delta u_r \wedge \delta \tau_r\,.
\end{equation}

A weak-coupling limit in gauge theory will correspond to 
a component of boundary of the Teichm\"uller space 
represented by surfaces $C$ which look like collections 
of three-punctured spheres glued together by identifying annuli 
around their punctures. 
One may naturally use the gluing 
parameters introduced in this construction
as coordinates for the Teichm\"uller space near
such a boundary component. The gluing parameters can be labelled by
a collection of 
closed curves $\ga_r$, $r=1,\dots,3g-3+n$
embedded into the annuli used in the gluing 
construction, $q_r\equiv q_{\ga_r}$. 
We may then assume that
the Beltrami differential $\mu_r$ describing a variation of $q_r$
is distributionally supported on the closed curve $\ga_r$ 
where it defines a local vector field 
$v_r$. The coordinate $u_r$ on the Coulomb branch conjugate to $q_r$
is then given as
\begin{equation}
u_r = \int_C\mu_r\phi=\int_{\ga} v_r \phi\,.
\end{equation}
Recalling that each curve $\ga_r$ also parameterizes an $SU(2)$ factor 
$SU(2)_r$ of the 
gauge group in the Lagrangian formulation associated to the given
boundary component of $\CT(C)$, we may finally 
identiy the geometrically defined
Coulomb branch parameter $u_r$ with the order parameter
\begin{equation}\label{orderpar}
u_r\,=\,{\rm Tr}(\Phi_r^2)\,,
\end{equation}
where $\Phi_r$ is the value of the 
scalar field from the vector multiplet associated to
the $SU(2)$ factor $SU(2)_r$ at infinity. We thereby arrive
at the relations
\begin{equation}\label{genMatone2}
u_r\,=\,-\pa_{\tau_r}\CF\,,
\end{equation}
relating the order parameter $u_r$ defined in \rf{orderpar} 
to the derivative of the prepotential with respect to the
gauge coupling $q_r=e^{2\pi i \tau_r}$, $\tau_r=\frac{4\pi i}{g_r^2}+
\frac{\theta_r}{2\pi}$.

Notice that the proof we have given for the relations
\rf{genMatone2} generalizes easily to higher rank gauge theories.
It also remains valid for the Argyres-Douglas
theories we are interested in, and the reformulation given in 
\rf{eq:lala2}
will be useful for the explicit comparison between our 
CFT results with the Seiberg-Witten theory of these theories.

\subsubsection{Supersymmetric partition functions and expectation values}

There is a variety of protected correlation functions in ${\cal N}=2$ four-dimensional gauge theories 
which can be computed by localization techniques. They typically involve a careful definition of the theory 
on some four-manifold, which preserves a supercharge which squares to an isometry of the manifold. 
The partition function is reduced to an integral over some zeromodes of one-loop 
determinants and contributions localized at the fixed points of the isometry. 
On four-manifolds with boundary, the answer will be a function of the choice of boundary conditions.

Although the modifications are generically implemented in a concrete Lagrangian description of the theory,
it is believed that they are sufficiently canonical not to actually depend on the choice of Lagrangian description. 
Thus one can hope that the partition function on compact manifolds will be S-duality invariant. 
The partition function on manifolds with boundary is more subtle, because different sets of boundary conditions will be
 natural and computable in different  S-duality frames. The existence of Janus domain walls and duality walls, 
 codimension one interfaces which can be used to compare boundary conditions at different values of the couplings
 and different S-duality frames,  can be used to argue that the partition function on 
a manifold with boundary will live in some linear space which has a flat connection on the space of 
couplings, and different natural bases in different weakly coupled regions of parameter space. 

The original example of partition function on a manifold with boundary is Nekrasov's partition function 
on $\Omega$-deformed flat space. The $\Omega$ deformation has two deformation parameters $\epsilon_1 = b \epsilon$ and
$\epsilon_2 = b^{-1} \epsilon$. The parameter $\epsilon$ is simply a scale. The partition function is traditionally defined 
with Dirichlet boundary conditions for the gauge fields, so it is a function of the vev of vectormultiplet scalars at infinity. 
For $b=1$, the same partition function is expected to arise for compactification on the upper hemisphere of a round $S^4$ of radius $\epsilon^{-1}$, 
with Dirichlet boundary conditions at the equator. It is conjectured that some deformation $S^4_b$ of the round sphere 
exists, which is related to the $\Omega$-deformed flat space at general $b$. 
The partition function on the full round $S^4$, or the conjectural $S^4_b$, are computed by Pestun's localization 
as an integral over the vectormultiplet zeromodes of the square modulus of the instanton partition function. 
The holomorphic and anti-holomorphic halves arise from localization at the North and South pole of the sphere. 
 
For theories of class ${\cal S}$, Nekrasov's partition function conjecturally 
coincides with conformal blocks 
for W-algebras. \footnote{This was proven for linear quiver gauge theories.}
The $S^4_b$ partition function combines holomorphic and anti-holomorphic conformal blocks into 
modular-invariant correlation functions for the Toda theory on $C$. In particular, for $A_1$ theories,
one obtains Virasoro conformal blocks and Liouville theory correlation functions at central charge $1+6 Q^2$, $Q=b+b^{-1}$.  
The Dirichlet boundary conditions for a given Lagrangian definition of the theory correspond to conformal blocks 
built by sewing along the corresponding pair of pants decomposition. The instanton partition function is computed by 
a sum over instanton sectors which coincide with the power series over sewing parameters for the conformal blocks. 
The tree-level gauge theory action produces the leading power of the sewing parameters, and the one-loop measure 
for the matter fields reproduces the specific integration measure which defines Liouville theory correlation functions 
out of conformal blocks. 

The identification of the parameters between Liouville theory and gauge theory is straightforward. 
Most subtleties can be attributed to the relation between conventional flat space and $\Omega$-deformed flat space,
or four-sphere. Basically, the quadratic differential $\phi$ is identified with $\epsilon^2 T(z)$, where $T(z)$ is the energy-momentum tensor. 
In particular, the mass Casimirs $d_i$ control the conformal dimensions of primary fields $d_i = \epsilon^2 \Delta_i$. 
In terms of the eigenvalues $m_i$ of the $SU(2)$ flavor mass matrix, we can write $m_i = \epsilon \mu_i$, 
in terms of the Liouville momentum $\mu_i$ defined by $\Delta_i = \mu_i(Q-\mu_i)$.
The insertion of a Coulomb branch order parameter $u_r$ at the origin (North pole of the sphere)
can be traded for a derivative with respect to the corresponding gauge coupling $\tau_r$: $u_r \to \epsilon^2 \partial_{\tau_r}$. This is 
the reason Matone-like relations \cite{M} 
in $A_1$ theories are promoted to Virasoro Ward identities. 

On the other hand, the choice of conformal dimensions $\Delta_k$, or better Liouville momenta $\alpha_k$ such that $\Delta_k = \alpha_k(Q-\alpha_k)$,
in the intermediate sewing channels of the conformal blocks 
is controlled by the vevs $a_i = \epsilon \alpha_i$  of vectormultiplets selected at infinity. 
In supersymmetric configurations, the vevs are not ``read'' by the expectation value of local operators. 
Rather, they can be related to the vev of fundamental Wilson line defects on a large circle. There are two possible locations for the circle, 
which give $\cos 2 \pi b \alpha_i$ and $\cos 2 \pi b^{-1} \alpha_i$. Supersymmetric line defects in gauge theory are 
associated to the vevs of Verlinde line operators in the two-dimensional CFT, defined by the insertion and transport of 
degenerate fields in conformal blocks. Boundary conditions in gauge theory which fix the vev of a line defect 
can thus be matched with conformal blocks which are eigenvectors of the corresponding Verlinde line defect.

\subsection{Collision limits in $A_1$ theories}
We are now ready to go beyond the regular $A_1$ theories, to irregular $A_1$ theories. 
\subsubsection{Irregular punctures of rank $1$}
Consider a Lagrangian description associated to a pants decomposition where two punctures share a pair of pants. 
This means that the corresponding block of eight complex scalar fields is coupled to the gauge theory 
as two fundamental hypermultiplets to the $SU(2)$ gauge group
represented by the tube which attach to the three-punctured sphere. 
The two fundamental hypermultiplets have an $SO(4)$ flavor symmetry, which is just the combination of the two $SU(2)_\pm$ flavor symmetries 
associated to the two punctures. It is useful to rewrite the mass parameters $m_\pm$ at the punctures in terms of 
the masses of each of the two doublets of hypermultiplets: $m_\pm = m_1 \pm m_2$. 

The simplest way to get an asymptotically free theory is to remove one of the two doublets, by sending the corresponding mass parameter $m_2$ to infinity.
A little bit of care is needed in order to keep a finite renormalized gauge coupling below the $m_2$ energy scale. By dimensional transmutation, 
the gauge coupling of the resulting asymptotically free theory is described by a scale $\Lambda = q m_2$ which we want to keep fixed in the limit 
$m_2 \to \infty$. Here $q$ is the usual instanton factor. In the 2d description, we can have the two punctures at positions $z=0$ and $z=q$ in a 
local coordinate system on the Riemann surface. We will relabel $m_1 \to m/2$.

The quadratic differential locally behaves as 
\begin{equation}
\phi = \frac{m_-^2}{z^2} + \frac{c_-}{z} + \frac{m_+^2}{(z-q)^2} + \frac{c_+}{z-q}+ \cdots
\end{equation}
It is interesting to see the dimensional transmutation happen in the SW curve. Suppose we want to send $q \to 0$, but keep the IR physics non-singular. 
That means that we do not want to make the SW curve singular. The geometry of the SW curve is controlled by the zeroes of $\phi$, which are the branch points of the 
double cover $\Sigma \to C_{g,n}$. The simplest thing to do is to keep the zeroes of $\phi$ away from the region where we are colliding the two singularities. 
In particular, $\phi$ should have a well-defined square root in the region around the two singularities. 

If we write 
\begin{equation}
\phi = \left( \frac{m_-}{z} + \frac{m_+}{z-q}\right)^2 + \frac{u}{z(z-q)} + \frac{c}{z}+ \cdots = \left( \frac{q m_2 + m z}{z(z-q)} \right)^2 + \frac{u}{z(z-q)} + \frac{c}{z}+ \cdots
\end{equation}
then we can readily take the limit $m_2 \to \infty$, $q m_2= \Lambda$ finite to 
\begin{equation}
\phi =  \left( \frac{\Lambda}{z^2}+ \frac{ m }{z} \right)^2 + \frac{u}{z^2} + \frac{c}{z}+ \cdots = \frac{\Lambda^2}{z^4} + \frac{2 m \Lambda}{z^3} + \frac{m^2 + u}{z^2} + \frac{\tilde u}{z} + \cdots
\end{equation}
The first form is useful to read off the behavior of the SW differential
\begin{equation}
\lambda = \pm \left( \frac{\Lambda}{z^2}+ \frac{ m }{z} + \cdots \right) dz
\end{equation}
This verifies that $m$ is still a mass parameter, and $\Lambda$ a non-normalizable deformation. 

Of course, we recognize the basic collision limit of two regular punctures in a Virasoro conformal block. Repeating it in the gauge theory context we have learned the physical meaning of the various ingredients. 
Notice that in the limit we kept $u$, $\tilde u $ finite. This specified the region of the Coulomb branch of the original theory which has a good limit 
as $q \to 0$. We can sharpen our understanding of the Coulomb branch operators if we expand $\lambda$ further: 
\begin{equation}
\lambda = \pm \left( \frac{\Lambda}{z^2}+ \frac{ m }{z} + v + \cdots \right) dz
\end{equation}
A straightforward calculation of $da \wedge d a_D$ shows that the normalizable parameter $v$ is dual to the gauge coupling $\Lambda$, 
i.e. it can be added to the prepotential to shift $\Lambda$. Thus, expanding out $\phi = \lambda^2$, we find that 
\begin{equation}
u\,=\,2\Lambda v\,,
\end{equation}
which shows that $u$ can be added to the prepotential to rescale $\Lambda$ (i.e. shift the bare UV gauge coupling, as it should),
i.e. 
\begin{equation} u \,\sim \,\Lambda \partial_\Lambda {\cal F}\,. \end{equation}
On  the other hands, $\tilde u$ can be added to the prepotential to shift the location of the puncture. 
In an $\Omega$ background, $\Lambda$ is mapped to $c_1$, $m$ to $\al'$, and the relation between $u$, $\tilde u$ and derivatives of the prepotential 
is the semiclassical limit of the Virasoro Ward identities we derived for a rank $1$ irregular vector!

Let's summarize our conclusions about the construction of ``rank $1$'' irregular punctures where $\phi$ has a pole of degree $4$. 
We started from a Riemann surface $C_{g,n}$ with two close regular punctures $p^\pm$, 
and represented them as a trinion glued to a single puncture $p$ a Riemann surface $C_{g,n-1}$ by a long thin tube of sewing parameter $q$, i.e. an $SU(2)$ gauge group weakly 
coupled to a block of hypermultiplets and to the $SU(2)$ flavor symmetry at $p$. The decoupling limit removes half of the hypers, and leaves us with the gauge theory description of the irregular puncture: 
an asymptotically free $SU(2)$ gauge theory coupled to the $SU(2)$ flavor symmetry of the regular puncture at $p$ 
and to a single doublet of mass parameter $m$. This can be taken to be the {\it definition} of the irregular puncture in the six-dimensional $A_1$ theory.
The mass parameter $m$ is associated to the $SO(2)$ flavor symmetry acting on the lone doublet. 

It is interesting to look more closely to the parameter space of the theory. Locally, it is parameterized by the remaining sewing parameters $q_r$ 
together with $\Lambda$. But the definition of $\Lambda$ depends on the choice of local coordinate at 
the irregular puncture. A change in the local coordinate definition will rescale $\Lambda$. If we pick a different 
pair of pants decomposition of $C$, i.e. we do some S-duality at the gauge groups which still have 
exactly marginal couplings, we will naturally change the choice of local coordinate at the irregular puncture, and 
thus rescale $\Lambda$ by some function of the couplings. This simply corresponds to a relative finite renormalization of the 
bare UV gauge coupling in the two Lagrangian descriptions of the theory. The main consequence for us is that 
$\Lambda$ lives in a line bundle over the complex structure moduli space of $C$, and the parameter space of the theory is the total 
space of that bundle. This should be identified with the moduli space of complex structure parameters of an irregular conformal block with a rank $1$ puncture. 

We should ask what are the boundaries of this new parameter space. In the regular case, all the boundaries of parameter space 
corresponded to weakly coupled UV-complete Lagrangian descriptions of the theory. The story is more intricate in the irregular case. 
A simple boundary of parameter space is $\Lambda \to 0$. This corresponds to making the theory very weakly coupled,
especially if we are looking at the Nekrasov partition function, or sphere partition function, which have a natural IR cutoff. 
This boundary is akin to the usual $q_i \to 0$ boundaries of parameter space. 

But we can also ask what happens to the theory if we send $\Lambda \to \infty$. Let's pick some Lagrangian description of the theory.
We have the asymptotically free $SU(2)$ gauge group, coupled to the lone doublet of mass $m$ and to another block of hypermultiplets,
which carries two more, possibly gauged, flavor symmetries $SU(2)_1$ and $SU(2)_2$. This $SU(2)$ $N_f=3$ theory has an $SO(6)$ flavor symmetry, inside which the
$SO(2) \times SU(2)_1 \times SU(2)_2$ sit as a block diagonal $SO(2) \times SO(4)$. At strong coupling $\Lambda$, the theory looks like an Abelian gauge theory, 
with BPS particles formed as bound states of a dyonic particle with no flavor symmetry, and a quartet of monopoles in a spinor representation of $SO(6)$. 
Under the $SO(2) \times SU(2)_1 \times SU(2)_2$ subgroup of flavor symmetry, the monopoles split into a doublet of $SU(2)_1$, of positive $SO(2)$ charge, 
and a doublet of $SU(2)_2$ of negative $SO(2)$ charge. In most of the Coulomb branch, all BPS particles are massive, but there is a region where the 
monopoles are light, and all other particles have masses which scale with $\Lambda$. In an electric-magnetic duality frame where the monopoles are electrically charged, 
the only light degrees of freedom are the two doublets of monopoles, and a photon under which the monopoles have charge $1$. 

If the $SU(2)_1$ and/or $SU(2)_2$ groups are gauged, at scales smaller than $\Lambda$ they will be coupled not to the original block of hypermultiplets, but only to the single light monopole doublets. 
Thus we should renormalize their gauge couplings to new finite scales $\Lambda_1 = \Lambda q_1$ and $\Lambda_2 = \Lambda q_2$. Thus the description of the theory, in the 
limit of large $\Lambda$ with finite $\Lambda_1$ and $\Lambda_2$, is that of two separate $A_1$ theories, each with a rank $1$ irregular singularity, coupled by a $U(1)$ 
gauge fields which gauges the diagonal combination of the two $SO(2)$ flavor symmetries at the irregular singularities. This is not a fully UV complete description of the theory,
but it is a possible tool to study the large $\Lambda$ behavior. In particular, we may hope to be able to express the partition function of the theory in a form adapted to this limit, possibly 
as an expansion in inverse powers of $\Lambda$, representing a tower of non-renormalizable effective corrections to the Abelian gauge theory Lagrangian. 

The basis of conformal blocks \rf{eq:regrk1not} which we have built, where a regular puncture (possibly part of a larger conformal block) together with a rank $1$ irregular puncture 
are realized in a rank $1$ irregular module is precisely adapted to this limit. The effective $U(1)$  gauge field which is weakly coupled in this description should be identified with the momentum 
$\beta'$ in the intermediate rank $1$ irregular channel. The asymptotic power series expansion in $z/c_1$ for \rf{eq:regrk1not}  should be matched to the expected asymptotic expansion in 
inverse powers of $\Lambda$. 

The existence of such conformal blocks is presumably related to the existence of boundary conditions and line defects which have a simple behavior in the $\Lambda \to \infty$ limit,
and tend to Dirichlet boundary conditions and Wilson loops in the effective Abelian description we gave. 
Basic line defects in regular $A_1$ theories are labelled by closed, non-self-intersecting paths on $C$. Under the collision limit which gives rise to a rank $1$ irregular singularity, 
line defects labelled by curves which are not pinched between the colliding punctures have a finite limit. They have no monopole charge under the asymptotically free gauge group.  
Line defects labelled by curves which are pinched between the 
colliding punctures require some renormalization in the limit, and give rise to line defects with 't Hooft charge under the asymptotically free gauge group. 
They are labelled by ``laminations'', collections of curves which can end in specific ways at the irregular puncture. 
The basic 't Hooft monopole operator for the asymptotically free gauge group, in a given Lagrangian description of the theory, 
is labelled by a curve which starts at the irregular puncture, goes around the trinion and comes back to the irregular puncture. 
It arises from the collision limit of the basic 't Hooft loop for the same gauge group. 

It is easy to argue, say by looking at the vev of line defects on circle compactifications of the theory, 
that the basic 't Hooft loop has a finite vev in the above-defined $\Lambda \to \infty$ limit, and becomes a Wilson loop for the 
Abelian gauge field. We can use this intuition to sketch how one could produce a boundary condition for the theory with a rank $1$ irregular singularity, which would go in the 
$\Lambda \to \infty$ to a Dirichlet boundary condition for the Abelian gauge field. We can start from a boundary condition in the regular $A_1$ theory 
which fixes the vev of the 't Hooft loop: this is just the S-duality image of a standard Dirichlet boundary condition. 
Then we can carry the boundary condition through the collision limit, and tentatively define a boundary condition 
which fixes the vev of the basic  't Hooft loop in the asymptotically free theory. 

Notice that although $\Lambda \to e^{2 \pi i} \Lambda$ is a symmetry of the theory, 
it is not a symmetry of the 't Hooft loop, which by Witten's effect is mapped to a 't Hooft-Wilson loop 
of electric charge $n$ by $\Lambda \to e^{2 \pi i n} \Lambda$. Thus, it is not a symmetry of the 
boundary condition adapted to the $\Lambda \to \infty$ limit. This is the first piece of evidence of the fact that 
some sort of Stokes phenomenon happens in the $\Lambda \to \infty$ limit: the analytic continuation in $\Lambda$ and the 
$\Lambda \to \infty$ limit do not commute. Later, this will suggest that the expansion in inverse powers of $\Lambda$ of correlation functions 
is really an asymptotic series, which approximates well the correct answer in a given Stokes sector around $\Lambda \to \infty$. 

\subsubsection{Irregular puncture of rank $1/2$}
We can readily decouple the remaining doublet, if we send $m \to \infty$ but leave the new gauge coupling scale 
$\tilde \Lambda^2 = 2 m \Lambda$ finite in the limit. We get a behavior  
\begin{equation}
\phi =  \frac{\tilde \Lambda^2}{z^3} + \frac{u}{z^2} + \frac{c}{z} + \cdots
\end{equation}
and
\begin{equation}
\lambda = \pm \left( \frac{\tilde \Lambda}{z^{\frac{3}{2}}}+ \cdots \right) dz
\end{equation}
Hence the irregular puncture where $\phi$ has a pole of degree $3$ can be defined in the $A_1$ theory by coupling an asymptotically free four-dimensional  $SU(2)$  
gauge group to the flavor symmetry of a regular puncture. We denote such a puncture as a ``rank $1/2$'' irregular puncture. 
We have not studied this puncture in the conformal field theory analysis. 

\subsubsection{Irregular puncture of rank $2$} 
Now, we are ready for more interesting limits. For example, suppose we want to collide one more regular puncture with the irregular 
puncture where $\phi$ has a degree four pole.
We can write again 
\begin{equation}
\phi =  \left( \frac{\Lambda}{z^2}+ \frac{ m }{z} + \frac{m'}{z-q} \right)^2 + \frac{u'}{z^2(z-q)} + \frac{u}{z^2}+ \frac{\tilde u}{z}+ \cdots \end{equation}
in order to keep the zeroes of $\phi$ away from the collision region. 
Then we can renormalize the three parameters $\Lambda$,$m$, $m'$ as $q \to 0$ 
\begin{equation}
m + m' = \tilde m \qquad \Lambda + m' q = \Lambda_1 \qquad m' q^2 = \Lambda_2
\end{equation}
in order to get a finite limit 
\begin{equation}
\phi =  \left( \frac{\Lambda_2}{z^3}+ \frac{\Lambda_1}{z^2}+ \frac{ \tilde m }{z} \right)^2 + \frac{\tilde u}{z^3} + \frac{u}{z^2}+ \frac{c}{z}+ \cdots \end{equation}
Again, we picked a form of the answer which makes manifest the behavior of the SW differential
\begin{equation}
\lambda =  \pm \left( \frac{\Lambda_2}{z^3}+ \frac{\Lambda_1}{z^2}+ \frac{ \tilde m }{z}  \cdots \right) dz \end{equation}
Clearly, this is the field theory version of the collision limit to a rank $2$ irregular singularity. 

The physical meaning of this limiting procedure is actually rather transparent: we sent some mass parameters to infinity, 
removing degrees of freedom, but we kept the periods of $\lambda$, which encode the low energy Lagrangian, all finite. 
In particular, the dimension of the Coulomb branch did not change.  In order to explore the physics further, 
we can cut the tube which connects this new ``irregular puncture'' to the rest of the Riemann surface, i.e. 
turn off the gauge couplings of the $SU(2)$ gauge group whose order parameter is $u$, and whose gauge coupling controls
the sewing parameter which glues the pair of punctures to the rest of the surface. 

Notice that a shift of that gauge coupling corresponds to a rescaling of the $z$ coordinate. It is easy to see that 
$\Lambda_2^{1/2}$ is playing the tole of the renormalized gauge coupling of the $SU(2)$ gauge group, which is now asymptotically free, 
with a peculiar beta function, which would naively correspond to a coupling to three and a half hypermultiplet doublets. 
If we turn off the gauge coupling, we are left with an irreducible theory, described by the quadratic differential 
\begin{equation}
\phi =  \left( \frac{1}{z^3}+ \frac{\Lambda_1}{z^2}+ \frac{ \tilde m }{z} \right)^2 + \frac{\tilde u}{z^3} + \frac{m^2-\tilde m^2}{z^2} \end{equation}
We replaced the $u$ order parameter by the mass parameter $m^2$ of the now ungauged $SU(2)$ flavor symmetry, which controls the residue of $\lambda$ at infinity . 
The limiting procedure we followed is well-known for this irreducible theory: we went from $SU(2)$ $N_f=3$ 
to an Argyres-Douglas theory with an $SU(3)$ flavor symmetry, by giving all three doublet flavors the same large mass, but adjusting 
the Coulomb branch parameters and gauge coupling to keep them light, and have simultaneously a light  monopole.
In the current context, we will only consider an $SU(2)\times U(1)$ subgroup of that $SU(3)$ flavor symmetry. 
The full theory corresponding to the Riemann surface with such an irregular singularity can be described by gauging the diagonal combination of that 
$SU(2)$ flavor group and the flavor group of a regular singularity. The AD theory appears to contribute to the beta function of the $SU(2)$ 
theory as one and a half doublet of free fields.

The Argyres Dougles theory has a Coulomb branch parameter, and a coupling. Gauging the $SU(2)$ flavor symmetry adds a new Coulomb branch parameter, and a new coupling.
If we expand $\lambda$ further, 
\begin{equation}
\lambda =  \pm \left( \frac{\Lambda_2}{z^3}+ \frac{\Lambda_1}{z^2}+ \frac{ \tilde m }{z} + v_1 + v_2 z +  \cdots \right) dz \end{equation}
it is easy to see that $\Lambda_2$ can be shifted by adding $v_2$ to the prepotential, and the same is true for $\Lambda_1$ and $v_1$. 
In this parameterization, $\phi$ becomes 
\begin{equation}
\phi =  \frac{\Lambda^2_2}{z^6}+ \frac{2 \Lambda_2 \Lambda_1}{z^5}+ \frac{2 \Lambda_2 \tilde m + \Lambda_1^2}{z^4} + \frac{2 \Lambda_2 v_1 + 2 \Lambda_1 m}{z^3} + \frac{2 \Lambda_2 v_2 + 2 \Lambda_1 v_1 + m^2}{z^2}+ \frac{\tilde u}{z}+ \cdots \end{equation}
and we see that adding $u'$ to the prepotential shifts $\Lambda_1$ by a multiple of $\Lambda_2$, while adding $u$ rescales both $\Lambda_2$ and $\Lambda_1$. Finally, $\tilde u$ shifts the location of $z$. 
These are the semiclassical limits of the Virasoro Ward identities. 

Now we have an interesting parameter space which is parameterized locally by $(\Lambda_2, \Lambda_1, q_r)$, and it is a bundle over the space of 
complex structure deformations of the Riemann surface. We should ask about possible limits in this parameter space. We can surely consider a degeneration limit where the gauge coupling which couples the AD theory to the rest of the theory 
is turned off. But this limit does not probe the AD theory at all, it simply decouples it from the rest of the $A_1$ theory. 
There is a more subtle limit: $\Lambda_2 \to 0$ for finite $\Lambda_1$. This limit replaces the rank $2$ puncture in $\phi$ with a rank $1$ puncture. What is the physics of this limit? 
At low energies, for $\Lambda_1 \gg \Lambda_2$, the AD theory is roughly the theory of an $SU(3)$  triplet of monopoles and of a singlet dyonic particle. In a generic region of the Coulomb branch, 
both sets of particles are massive, but we can look at the region where the monopoles are light. Thus the $\Lambda_2 \to 0$ physics is somewhat familiar: 
a light $SU(2)$ doublet of monopoles of unit flavor $U(1)$ charge, another monopole  of charge $-2$ under the $U(1)$ flavor symmetry, all charged under an Abelian gauge field. 
If the AD theory is part of a larger $A_1$ theory, the $SU(2)$ gauge theory coupled to the doublet of monopoles will have a renormalized coupling $\Lambda =q \Lambda_2$,
and represent the residual rank $1$ irregular singularity. 

Of course, the conformal blocks which are adapted to this limit are exactly the ones where the rank $2$ puncture is built our of descendants of a rank $1$ puncture, whose Liouville momentum 
corresponds to the Coulomb branch parameter of the $U(1)$ gauge field. A key observation is that 
this Abelian description at small $\Lambda_2$ arises from the collision limit of a regular puncture and rank $1$ puncture in the 
Abelian description we gave for the rank $1$ irregular singularity. Conjecturally, the line defects of the irregular $A_1$ theories 
are still labelled by appropriate laminations. It is not difficult to identify laminations which become Abelian Wilson loops in the Abelian  
$\Lambda_2 \to 0$ limit. We will come back to this point in future work

\subsubsection{Irregular puncture of other rank } 

We can also reduce the degree of the pole of $\phi$ from $6$ to $5$ (rank $3/2$ irregular puncture) if we turn $\Lambda_2$ off, but keep $\Lambda_2 \Lambda_1$  fixed,
to get \begin{equation}
\phi =  \left( \frac{ \Lambda_{\frac{3}{2}}}{z^{\frac{5}{2}}}+ \frac{\Lambda_{\frac{1}{2}}}{z^{\frac{3}{2}}} \right)^2 + \frac{\tilde u}{z^3} + \frac{u}{z^2}+ \frac{c}{z}+ \cdots \end{equation}
This can be understood as the AD theory which arises from $SU(2)$ $N_f=2$, and has a $SU(2)$ flavor symmetry.

It is clear that we can repeat this exercise further, and derive theories associated to Riemann surfaces with punctures 
where $\phi$ has poles of even degree $2d+2$
\begin{equation}
\phi =  \left( \frac{\Lambda_d}{z^{d+1}}+ \cdots  + \frac{m }{z} \right)^2 + \frac{u_{d-1}}{z^{d+1}}  + \cdots + \frac{c\tilde u}{z} +\cdots  \end{equation}
or odd degree $2d+1$
\begin{equation}
\phi =  \left( \frac{ \Lambda_{d-\frac{1}{2}}}{z^{d+\frac{1}{2}}}+\cdots + \frac{\Lambda_{\frac{1}{2}}}{z^{\frac{3}{2}}} \right)^2 + \frac{u_{d-1}}{z^{d+1}}  + \cdots + \frac{c}{z} +\cdots  \end{equation}

If we look at the generalized AD theories which describe the physics when the UV gauge couplings are turned off,
say for even degree
\begin{equation}
\phi =  \left( \frac{1}{z^{d+1}}+  \frac{\Lambda_{d-1}}{z^{d}}+\cdots  + \frac{m }{z} \right)^2 + \frac{u_{d-1}}{z^{d+1}}  + \cdots + \frac{\tilde m^2- m^2}{z^2}  \end{equation}
we can give a straightforward interpretations of the various parameters. The scaling dimension of $\phi dz^2$ is $[\phi]=2$,
and hence $[z] = -\frac{1}{d}$. The Coulomb branch parameters $u_i$ have dimension $[u_i] = 2-\frac{i}{d}$, and are vevs of 
operators $\hat u_i$ with the same dimension. The non-normalizable parameters $\Lambda_i$ have scaling dimension 
$[\Lambda_i] = 1-\frac{i}{d}$. Roughly, we should be able to do a change $\delta \Lambda_i$ by adding a prepotential 
deformation $\delta \Lambda_i u_{d-i}$, but a more precise statement should map each $u_i$ to a vector field acting on the space of $\Lambda_j$. 
The 2d dictionary suggests immediately the details of the map: we can extract $u_i$ from $\phi$ by contracting with the 
vector field $v_i = z^{i+1} \frac{d}{dz}$ and integrating on a small loop  around the irregular puncture. 
This Beltrami differential correspond to a specific redefinition of the local coordinate 
$z \to z+ \epsilon_i z^{i+1}$, which gives $\delta \Lambda_{j-i} = - (j+1) \epsilon_i \Lambda_j$. 
Hence we expect that the Coulomb branch order parameter $\hat u_i$
changes the couplings $\Lambda_i$ according to the vector field 
\begin{equation}
\ell_i = - \sum (j-i) \Lambda_j  \frac{\partial}{\partial \Lambda_{j-i} } 
\end{equation}
This expression is also valid in the general case where the irregular singularity sits in a full Riemann surface. 
In that case we also have the operator $\hat u_0$, which is the Coulomb branch order parameter for the $SU(2)$ 
gauge group which is coupled to the AD theory, and maps to a simple rescaling of the local coordinate $z$, and hence of the $\Lambda_i$.
The Coulomb branch parameter $\tilde u$ maps to a translation of the local coordinate, and maps to a Beltrami differential which moves the puncture. 
Finally, the parameter $m$ is simply the mass parameter of a $U(1)$ flavor symmetry. 

The correspondence between the parameters $u_i$ and the variations of the couplings 
can be expressed in a suggestive fashion if we expand further 
\begin{equation}
\lambda =  \frac{\Lambda_d}{z^{d+1}}+ \cdots  + \frac{m }{z} + v_{1} + v_{2} z + \cdots + v_{d} z^d + \cdots
\end{equation}
Then \begin{equation} \label{eq:end}
da \wedge da_D = \frac{2}{\pi i}\sum_k \frac{1}{k} d v_{k} \wedge d \Lambda_k
\end{equation}
and hence $v_k$ can be added to the prepotential to shift $\Lambda_k$. 

Though this formula follows from our general analysis of the relation between Coulomb branch order parameters and 
Beltrami differentials, it is entertaining to re-derive it in a slightly different manner. 
If we want to compute the variation of the periods as we vary the $\Lambda_k$ parameters in $\lambda$, in terms of a normalizable $\de \lambda$,
we need to correct the naive $\de \lambda$ in a region near the origin. 

Near the origin, we can find a primitive $w(z)$ such that $\lambda = d w$. We can regularize $\lambda$ to 
$\tilde \lambda = d (f w)$, where $f$ is a smooth function which goes to zero exponentially fast at the origin, and
goes to $1$ away from the origin. Then the $(0,1)$ part of $\de \lambda$ is $\delta w \bar \partial f$. Notice that $\de w$ is single-valued, as $m$ 
is not varied!

Thus the canonical symplectic form becomes 
\begin{equation}
da \wedge da_D = \int_\Sigma \de \lambda \wedge \delta w \bar \partial f = 2\oint_{|z|=\epsilon} \de \lambda \de w
\end{equation}
which coincides with \ref{eq:end} 
We thus recovered the semiclassical limit of the Virasoro Ward identities for a general irregular vector. 

For odd degree, everything works in the same way, except that there is no $U(1)$ flavor symmetry, and the $\Lambda$ parameters 
have half-integral grading. 

\subsection{Gauge theory conclusions}
Now we are ready to compare the physical properties of the $AD_2$ theory with 
the final form of the irregular correlation function $\Phi^{(2)}(c_1,c_2;\alpha_0, \alpha)$.
In the $\Lambda_2 \to 0$ limit, we expect to see an Abelian gauge theory coupled to a triplet of hypermultiplets

Indeed, we recognize three $\Upsilon$ functions in the denominator of the 
structure constant. They correspond to the one-loop determinants on $S^4$ 
of three hypers of gauge charge $1$. 
We see that two hypers sit in a doublet of $SU(2)$ flavor symmetry with mass $\alpha_0$, 
and the third hyper has charge under the $U(1)$ flavor symmetry of mass $\alpha$,
as expected from the gauge theory analysis. 

The conformal blocks have a ``tree level'' prefactor which should be identified with the 
exponent of the prepotential of the Abelian gauge theory. The magnetic period dual to $\beta$ 
can be computed by taking the first $\beta$ derivative of the prepotential, and is controlled by 
the constant $\frac{c_1^2}{c_2}$ term. This controls the mass of the heavy magnetic particles in the $AD_2$ theory.
Thus we interpret the power series in $\frac{c_2}{c_1^2}$ as an expansion in inverse powers of the mass of the heavy particles, 
which have apparently been integrated out. 

We expect that the physical interpretation of correlation functions written in a general basis of irregular conformal blocks will follow the same general  pattern.
Standard regular internal legs of the conformal block will correspond to weakly coupled UV non-Abelian gauge groups. Internal legs corresponding to 
irregular intermediate channels will map to Abelian gauge groups, emerging in an effective weakly coupled description valid in the appropriate corner of the space of couplings of the theory. 
The structure constants will keep track of the contribution of light matter hypermultiplets. The expansion of conformal blocks in power series will keep track systematically of 
instanton contributions for the non-Abelian gauge groups, and of the effect of integrating out heavy magnetically charged particles for the effective Abelian gauge groups. 
It is reasonable to expect that the effective expansion in inverse powers of the masses of magnetic particles should be only asymptotic, while the instant on expansion should  have a finite radius of convergence. 
This expectation seems to be supported by the conformal field theory calculation. 

\section{Discussion and future directions}

\setcounter{equation}{0}

In this paper we have initiated the study of Virasoro conformal blocks and Liouville theory correlation functions 
in the presence of irregular singularities. Regular BPZ conformal blocks are usually defined through the sewing construction, 
 which provides a convergent power series expansion around the corners of the complex structure 
 moduli space where the Riemann surface degenerates. Irregular conformal blocks are functions 
 of an enlarged complex structure moduli space, which cannot be parameterized fully by the usual sewing construction,
 and has a more intricate boundary structure.  
 
 Ultimately, we would like to find a straightforward, fully computable, definition of several bases of irregular conformal blocks, 
 each adapted to a different degeneration limit in the enlarged complex structure moduli space, and equipped with explicit generalization of
 braiding and fusion integral kernels relating these bases. Each basis should also be equipped with an integration measure to package the holomorphic and anti-holomorphic 
 conformal blocks into a Liouville theory correlation function, invariant under generalized fusion and braiding transformations. 
 
In this  paper we completed some basic steps towards that goal. First, we 
extended the familiar notion of OPE expansion, which replaces some punctures on the Riemann surface with a 
sum over descendants of a regular puncture, by considering formal sums over descendants of an irregular puncture. 
In the regular case, the OPE is an equivalent reformulation of the sewing procedure, and gives convergent power series expansions.
In the irregular case, the sums over descendants of an irregular puncture are sufficiently versatile to 
cover all interesting corners of the extended complex structure moduli space, but appear to be formal power series only,
possibly asymptotic. A crucial feature of such expansions in descendants of irregular vectors is that they commute with fusion and braiding transformations 
done on the rest of the Riemann surface. 

Experimentally, this type of expansion appears to exits and be unique at all orders which we could test. 
We also devise a collision limit of regular conformal blocks 
which can give a solution to our ansatz at all orders of the expansion, order by order in the formal power series. 
Based on our analysis of the free field 
representation in Section \ref{sec:screen}, 
we conjectured that irregular conformal blocks 
have Stokes phenomena in the extended complex structure moduli space, and that 
the true bases of irregular conformal blocks could be characterized uniquely by their asymptotic expansion in appropriate Stokes sectors. 

We expect that Verlinde-like line defects will be important in characterizing the properties of irregular conformal blocks. 
As a preparation to define them, we provide an alternative characterization of our bases of conformal blocks:
they are uniquely specified by requiring the existence of certain series expansions for conformal blocks with a degenerate puncture. 
The series expansion is tailored to be well-behaved under transport of degenerate fields across the irregular conformal block,
and seems to arise from the careful collision limit of regular conformal blocks with degenerate insertions. Indeed, in this formalism we can actually prove that such 
collision limits make sense, and produce well-defined formal power series. 

The last step we take in this paper is to use the collision limit 
to predict the integration measure which gives Liouville correlation functions from conformal blocks with irregular singularities. 
In a future publication, we plan to use such collision limits to derive the generalized fusion and braiding transformations, 
 transport of degenerate insertions and Verlinde line defects for irregular conformal blocks.
 We expect the resulting integral kernels to provide a more intrinsic definition of our bases of irregular conformal blocks:  a general strategy is to define the conformal blocks through a Riemann-Hilbert problem in the extended complex structure moduli space,
specifying the fusion and braiding transformations which relate bases of conformal blocks adapted to the possible degeneration limits of the Riemann surface. 
\appendix
\section{Conventions}

\setcounter{equation}{0}

We will write the mode expansion of the 
chiral free field on the cylinder as
\begin{equation}
\tilde{\phi}(x)=q+px+\sum_{k\neq 0}\frac{i}{n}a_ne^{-inx}\,,
\end{equation}
where
\begin{equation}
[q,p]\,=\,\frac{i}{2}\,\qquad
[a_n,a_m]\,=\,\frac{n}{2}\de_{n,-m}\,.
\end{equation}
The corresponding expansion of the 
chiral free field on the plane 
is obtained via \[
\phi(z)=\tilde{\phi}(w(z))-\frac{Q}{2}
\log\frac{\pa z}{\pa w}\,,\]
and it takes the form
\begin{equation}
\phi(z)=q-\al \log z+\sum_{k\neq 0}\frac{i}{n}a_nz^{-n}\,,\qquad
\al:=ip+\frac{Q}{2}\,.
\end{equation}
If $a_n|c\rangle=-ic_n|c\rangle$, this implies
\begin{equation}
\pa\phi(z)\,\sim\,-\sum_{k=1}^{n}\frac{c_k}{z^{k+1}}-\frac{\al}{z}+\dots
\end{equation}
in the vicinity of an irregular singularity of order $n$.

\section{Irregular chiral vertex operators}\label{app:CVO}

\setcounter{equation}{0}

This appendix describes the evidence that is available for the 
existence of the intertwining operators between irregular
modules from Section \ref{sec:vertex} from the purely 
algebraic point of view.

\subsection{The standard constructions revisited}

In this section we will review some standard facts, 
recast in a way which is suitable to 
generalization.
 
 \subsubsection{The chiral vertex operator}\label{sec:rev}

It is useful some review the properties of the chiral vertex operator,
in a way which highlights the parallelism with the irregular case, and 
sets up the problem for collision limits. 
Consider first the image under $\Psi^{\De_z}_{\De_f,\De_i}(z)$ of 
the highest weight vector in the module $\CV_{\De_i}$
\begin{equation}\label{eq:R1not}
|\,R^{(1)}(z)\,\rangle = \Psi^{\De_z}_{\De_f,\De_i}(z)|\,\De_i\,\rangle\,.
 \end{equation}
This is a vector in $\CV_{\De_f}$ defined by the action of raising Virasoro operators:
\begin{align}
L_k |\,R^{(1)}(z)\,\rangle &=\, z^k
\left(z\paz+\De_z(k+1)\right)|\,R^{(1)}(z)\,\rangle \quad
k> 0 \, ,\cr 
L_0 |\,R^{(1)}(z)\,\rangle &=\, 
\left(z\paz+\De_z + \De_i\right)|\,R^{(1)}(z)\,\rangle \, . \end{align}
 
The vector  $|\,R^{(1)}(z)\,\rangle$ can be built as a unique power series in $z$ 
in terms of descendants of the highest weight vector $|\,\De_f \,\rangle$:
\begin{equation}\label{eq:R1ansatz}
|\,R^{(1)}(z)\,\rangle=\,z^{\De_f-\De_i-\De_z}
\sum_{k=0}^{\infty}z^k\,|\,\De_f;k\,\rangle\,.
\end{equation}

The Ward identities take a recursive form on the coefficients of the expansion
\begin{align}\label{R1constr-rec}
L_0 |\,\De_f;k\,\rangle\,&=\, \left(\De_f+ k\right) |\,\De_f;k\,\rangle \cr
L_n  |\,\De_f;k\,\rangle \,&=\, 
\left(\De_f-\De_i+n \De_z +k-n\right) |\,\De_f;k-n\,\rangle  \quad
n> 0\,,
\end{align}
so that one can set $|\,\De_f;0\,\rangle = |\,\De_f \,\rangle$ and in principle 
solve the recursion order-by-order in $k$. 

Of course, for the standard chiral vertex operator, we can solve the Ward identities directly.   
We write 
\begin{equation}
|\,\De_f;k\,\rangle\,=\,\sum_{{I};|{I}|=k}
C_{{I}}\,\BL_{-{I}}
|\,\De\,\rangle\,
\end{equation}
with $\BL_{-{I}}$ being a monomial in Virasoro generators,
$|{I}|$ being the $L_0$-weight of $L_{-{I}}$.
The coefficients $C_{{I}}$ can be computed right away 
\begin{equation}
C_{I}\,=\,\sum_{{I}'}M^{-1}_{{I}{I}'}(\De_f)\,\langle\,\De_f\,|\,
\BL_{{I}'}|\,R^{(1)}(z)\,\rangle\,,
\end{equation}
where $M^{-1}_{{I}{I}'}(\De_f)$ are defined by
\[
M_{{I}{I}'}(\De_f):=\langle\De_f|L_{{I}}L_{-{I}'}|\De_f\rangle\,,\qquad
{\rm and}\qquad
\sum_{{I}'}M_{{I}{I}'}^{}(\De_f)M_{{I}'{I}''}^{-1}(\De_f)=\de_{{I}{I}''}\,.
\]
and $\langle\,\De_f\,|\,\BL_{{I}'}|\,R^{(1)}(z)\,\rangle$ is computed from 
\begin{equation}
\langle\,\De_f\,|\,R^{(1)}(z)\,\rangle = z^{\De_f-\De_i-\De_z}
\end{equation}
simply by applying the Ward identities.

Starting from $|\,R^{(1)}(z)\,\rangle$, we can define the action of the chiral vertex operator $\Psi^{\De}_{\De_f,\De_i}(z)$ over descendants of 
 the highest weight vector $|\,\De_i\,\rangle$ recursively by 
\begin{equation}\label{eq:chiralrec}
\Psi^{\De_z}_{\De_f,\De_i}(z) L_k |\,v\,\rangle= \left( L_k -z^k
\left(z\paz+\De_z(k+1)\right) \right) \Psi^{\De_z}_{\De_f,\De_i}(z) |\,v\,\rangle \quad
k<0\,,
\end{equation}
For the standard chiral vertex operator we can solve this recursion directly, 
by computing $\langle\,\De_f\,|\,\BL_{{I}'} \Psi^{\De}_{\De_f,\De_i}(z) |\,\De_i \,\rangle$ via Ward identities
and acting with $M^{-1}_{{I}{I}'}(\De_f)$.

\subsubsection{The rank $1$ irregular vector}\label{sec:rk1}
The rank $1$ irregular vector is a vector in $\CV_{\De_f}$ defined by the action of raising Virasoro operators:
\begin{align} \label{eq:rank1}
L_0 |\,I^{(1)}(c_1)\,\rangle &= \left(\Delta_{\al'} + c_1 \partial_{c_1}  \right) |\,I^{(1)}(c_1)\,\rangle \cr 
L_1 |\,I^{(1)}(c_1)\,\rangle&=  - 2 c_1 (\al' - Q) |\,I^{(1)}(c_1)\,\rangle\cr 
L_2|\,I^{(1)}(c_1)\,\rangle&= -c_1^2|\,I^{(1)}(c_1)\,\rangle \cr
 L_n |\,I^{(1)}(c_1)\,\rangle &= 0 \qquad n>2
\end{align} 

The vector  $|\,I^{(1)}(c_1)\,\rangle$ can be built as a unique power series in $c_1$ 
in terms of descendants of the highest weight vector $|\,\De_f \,\rangle$:
\begin{equation}\label{eq:rank1ansatz}
|\,I^{(1)}(z)\,\rangle=\,c_1^{\De_f-\De_{\al'}}
\sum_{k=0}^{\infty}c_1^k\,|\,\De_f;k\,\rangle\,,
\end{equation}
We can find the $|\,\De_f;k\,\rangle$ in three ways: solving a recursion relation, by direct calculation, or by collision limit on $|\,R^{(1)}(z)\,\rangle$. 

Plugging the series ansatz into the Ward identities, we get the recursive definition 
\begin{align} \label{eq:rank1-rec}
L_0|\,\De_f;k\,\rangle &= \left(\Delta_f + k \right) |\,\De_f;k\,\rangle\cr 
L_1|\,\De_f;k\,\rangle&=  - 2 (\al' - Q) |\,\De_f;k-1\,\rangle\cr 
L_2|\,\De_f;k\,\rangle&= - |\,\De_f;k-2\,\rangle \cr
 L_n|\,\De_f;k\,\rangle &= 0 \qquad n>2
\end{align} 
which can be solved order-by-order starting from $|\,\De_f;0\,\rangle = |\,\De_f\,\rangle$

The solution can also be derived by computing directly 
$\langle\,\De_f\,|\,\BL_{{I}'}|\,I^{(1)}(c_1)\,\rangle$
via the Ward identities and acting with $M^{-1}_{{I}{I}'}(\De_f)$.

From this point of view, the collision limit from 
$|\,R^{(1)}(z)\,\rangle$ to $|\,I^{(1)}(c_1)\,\rangle$ with constant $c_1 = \al_z z$ and $\al' = \al_i + \al_z$ is obvious: 
the Ward identities for $|\,R^{(1)}(c_1)\,\rangle$ go to the Ward identities for 
$|\,I^{(1)}(c_1)\,\rangle$ in the collision limit, so obviously
\begin{equation} \label{eq:rk1lim1}
\langle\,\De_f\,|\,\BL_{{I}'}|\,R^{(1)}(z)\,\rangle \to \langle\,\De_f\,|\,\BL_{{I}'}|\,I^{(1)}(c_1)\,\rangle
\end{equation}
as long as 
\begin{equation} \label{eq:rk1lim2}
\langle\,\De_f\,|\,R^{(1)}(z)\,\rangle \to \langle\,\De_f\,|\,I^{(1)}(c_1)\,\rangle
\end{equation}

In the simple (BPZ) normalization, a rescaling is needed for this to be true. 
We already know that we should strip off a divergent power of $z$. 
But we also need to multiply $|\,R^{(1)}(z)\,\rangle$ by $\al_z^{\De_f-\De_{\al'}}$
to convert $z^{\De_f-\De_{\al'}}\to c_1^{\De_f-\De_{\al'}}$.
Thus we need to take the limit of 
\begin{equation}\label{eq:rk1lim}
z^{2 \al_z \al_i} \al_z^{\De_f-\De_{\al'}} |\,R^{(1)}(z)\,\rangle
\end{equation}
with constant $c_1 = \al_z z$ and $\al' = \al_i + \al_z$. 

Finally, it is useful to understand the collision limit from the point of view of the 
recursion relation. The coefficient of $|\,\De_f;k-1\,\rangle$ in the recursion
\ref{R1constr-rec} grows linearly with $\al_z$, all the others grow quadratically. 
This means that $|\,\De_f;k\,\rangle$ grows at most as $\al_z^k$, and the coefficient 
of $\al_z^k$ satisfies the recursion relations \ref{eq:rank1-rec}. 

In the following, we will expand various objects as sums of descendants of $|\,I^{(1)}(c_1)\,\rangle$, defined as the vectors of the form 
\begin{equation} \label{eq:rk1-des}
\BL_{-{I}} \partial_{c_1}^k |\,I^{(1)}(c_1)\,\rangle
\end{equation}
Crucially for us, action of a lowering Virasoro generator on a descendant of $|\,I^{(1)}(c_1)\,\rangle$ can be rewritten as a sum over descendants in a straightforward way: one can commute the Virasoro generator through $\BL_{-{I}}$ 
in the usual way, act on $|\,I^{(1)}(c_1)\,\rangle$ via the Ward identities, 
and then bring to the left any factors of $c_1$, passing them through  
$\partial_{c_1}^k$ in the obvious way. 

If we define the ``weight'' of the descendant \ref{eq:rk1-des} as
$|I| + k$, the following is true: acting with $L_n$ on a weight $k$ descendant 
gives a sum over descendants of weight $t$ in the range $k-n \leq t \leq k-n+2$,
multiplied by a power $c_1^{t-k+n}$, with coefficients which are polynomials in $\al'$. 
Thus the ``rank 1 irregular module $\CI_1$'' defined as the space of descendants 
of $|\,I^{(1)}(c_1)\,\rangle$ carries an interesting action of the Virasoro algebra. 

We can give simple examples Ward identities for of weight $1$ descendants.  
\begin{align} \label{eq:rank1des1}
\left(L_0 - \Delta_{\al'} - c_1 \partial_{c_1} \right) \partial_{c_1} |\,I_1\,\rangle &= \partial_{c_1} |\,I_1\,\rangle \cr 
\left( L_1 + 2 c_1 (\al' - Q)\right) \partial_{c_1}  |\,I_1\,\rangle&=  - 2  (\al' - Q) |\,I_1\,\rangle\cr 
\left(L_2 + c_1^2 \right) \partial_{c_1}|\,I_1\,\rangle&=  - 2 c_1 |\,I_1\,\rangle \cr
 L_n \partial_{c_1} |\,I_1\,\rangle &= 0 \qquad n>2
\end{align} 
and

\begin{align} \label{eq:rank1des2}
\left(L_0 - \Delta_{\al'} - c_1 \partial_{c_1} \right) L_{-1} |\,I_1\,\rangle &= L_{-1} |\,I_1\,\rangle \cr 
\left( L_1 + 2 c_1 (\al' - Q) \right)  L_{-1} |\,I_1\,\rangle&=  2  \left(\Delta_{\al'} + c_1 \partial_{c_1}  \right) |\,I_1\,\rangle \cr 
\left(L_2 + c_1^2 \right)L_{-1} |\,I_1\,\rangle&=- 6 c_1 (\al' - Q) |\,I_1\,\rangle \cr
L_3 L_{-1} |\,I_1\,\rangle&= -4 c_1^2 |\,I_1\,\rangle \cr
 L_n L_{-1} |\,I_1\,\rangle &= 0 \qquad n>3
\end{align}

By definition, every descendant of $|\,I^{(1)}(c_1)\,\rangle$ can be evaluated as a vector 
in the standard Verma module $\CV_{\De_f}$. But it is important that the Virasoro action on 
$\CI_1$ does not make any reference to $\De_f$. We will denote this evaluation map $\CI_1 \to \CV_{\De_f}$ as $\Psi^{r,1}_{\De_f,\al'}(c_1)$. 

\subsubsection{Direct construction of series expansions for regular vectors}

It may be useful to observe that the expansion \rf{eq:Regexp} for the
regular vectors $|R^{(2)}(z,\be)\rangle$ can be constructed 
in a way that is closely analogous to the procedure we will 
use for the irregular vectors later in this section.
To this aim we will think about the 
$|\,R^{(1)}_k(w)\,\rangle$ as solutions of a recursion relation
inside the space $\CR_1$ of descendants of $|\,R^{(1)}(w)\,\rangle$.
Indeed, much as it happened for the action of Virasoro generators on descendants of $|\,I^{(1)}(c_1)\,\rangle$, the result of acting with Virasoro generators on a descendant of 
$|\,R^{(1)}(w)\,\rangle$ can be rewritten in terms of descendants of $|\,R^{(1)}(w)\,\rangle$,
with coefficients polynomial in $w$, with no reference on the ambient module 
$\CV_{\De_0}$. 

For example, at weight $1$ (defining the weight of descendants in $\CR_1$ as we did for the descendants in $\CI_1$),
\begin{align}\label{R1ex1}
\left( L_0 -w\paw-\De_1 -\De_{\beta} \right) \partial_w |R^{(1)} \rangle &=\,  \partial_w |R^{(1)} \rangle \cr
\left( L_n -w^{n+1}\paw-\De_1 (n+1)w^n \right) 
\partial_w |R^{(1)} \rangle &=\, 
(n+1)\left(w^{n}\pa_w+n \De_1 w^{n-1}\right) |R^{(1)}\rangle \,
\end{align}
and 
\begin{align}\label{R1ex2}
\left( L_0 -w\paw-\De_1 -\De_{\beta} \right) L_{-1}|R^{(1)} \rangle &=\, L_{-1} |R^{(1)} \rangle \cr
\left( L_1 -w^2\paw-2 \De_1 w \right)L_{-1} |R^{(1)}\rangle &=\, 
2\left(w\pa_w+ \De_1 + \De_{\beta} \right) |R^{(1)} \rangle \\
\left( L_n -w^{n+1}\paw-\De_1 (n+1)w^n \right)L_{-1} |R^{(1)}\rangle &=\, 
(n+1)\left(w^{n}\pa_w+n \De_1 w^{n-1}\right) |R^{(1)} \rangle \,
\notag\end{align}
The recursion relations which follow from 
\begin{align}\label{R2}
L_0 |\,R^{(2)}(w,z)\,\rangle &=\, \left(z\paz+\De_2 +w\paw+\De_1 + \De_3\right)|\,R^{(2)}(w,z)\,\rangle \\
L_n |\,R^{(2)}(w,z)\,\rangle &=\, 
\left(z^{n+1}\paz+\De_2 (n+1)z^n+w^{n+1}\paw+\De_1 (n+1)w^n\right)|\,R^{(2)}(w,z)\,\notag\rangle
\end{align}
take the form 
\begin{align}\label{R2k}
\left( L_0 -w\paw-\De_1 -\De_{\beta} \right) |R^{(1)}_{k} \rangle &=\, k |R^{(1)}_{k} \rangle \cr
\left( L_n -w^{n+1}\paw-\De_1 (n+1)w^n \right) |R^{(1)}_{k} \rangle &=\, 
(\De_{\beta}  + n\De_2 -\De_3+ k-n) |R^{(1)}_{k-n} \rangle \,.
\end{align}
As an example, consider $k=1$. Then the recursion relation has a source only for $n=1$. 
We can easily solve 
\begin{equation}
 |R^{(1)}_{1} \rangle = \frac{\De_{\beta} + \De_2 - \De_3}{2\De_{\beta}} \left( L_{-1} - \partial_w \right) |R^{(1)} \rangle\,.
\end{equation}
It is straightforward to generalize this procedure to higher 
orders in the expansion.

\subsection{Maps to rank 1} \label{sec:rk1rk2}
Now we are ready to describe how solutions of Ward identities can be expanded 
recursively in descendants of a rank $1$ irregular vector. 

\subsubsection{Rank $2$ to rank $1$}
In order to define the image $|\,I^{(2)}(c,\al'')\,\rangle$ of a rank $2$ irregular vector 
of parameters $c_2, c_1, \al''$ under $\Psi^{1,2}(c_2)$, we 
can start from the formal series ansatz 
\begin{equation}\label{irrdeg}
|\,I^{(2)}(c,\al'')\,\rangle\,=\,c_2^{\nu_2}c_1^{\nu_1}\,
e^{(\al''-\beta') \frac{c_1^2}{c_2}}\,
\sum_{k=0}^{\infty}c_2^k\,|\,I^{(1)}_{2k}(c_1,\beta')\,\rangle\,,
\end{equation}
where the vectors $|\,I^{(1)}_{2k}(c_1,\beta')\,\rangle$  for $k>0$
can be represented as generalized
descendants of the rank 1 irregular vector $|\,I^{(1)}_0(c_1,\beta')\,\rangle$
of parameters $c_1, \beta'$.  Assigning weight $k$
to the Virasoro generator $L_{-k}$ and weight 1 to both $c_1^{-1}$
and $\pa_{c_1}$, the vector
$|\,I^{(1)}_{2k}(c_1,p)\,\rangle$ must be a descendant
of total weight $2k$.

Let's test this ansatz. We want 
\begin{align} \label{eq:rank2}
L_0 |\,I^{(2)}(c,\al'')\,\rangle &= \left(\De_{\al''} + c_1 \partial_{c_1}  + 2 c_2 \partial_{c_2} \right)  |\,I^{(2)}(c,\al'')\,\rangle \cr 
L_1 |\,I^{(2)}(c,\al'')\,\rangle &= \left( c_2 \partial_1 - 2 c_1 (\al'' - Q) \right)|\,I^{(2)}(c,\al'')\,\rangle \cr 
L_2|\,I^{(2)}(c,\al'')\,\rangle &= -\left(c_1^2+ c_2 (2\al'' - 3 Q) \right) |\,I^{(2)}(c,\al'')\,\rangle   \cr
 L_3 |\,I^{(2)}(c,\al'')\,\rangle &= - 2 c_2 c_1|\,I^{(2)}(c,\al'')\,\rangle \cr 
 L_4 |\,I^{(2)}(c,\al'')\,\rangle &= - c_2^2|\,I^{(2)}(c,\al'')\,\rangle   \cr 
 L_n |\,I^{(2)}(c,\al'')\,\rangle &= 0 \qquad n>4
\end{align} 
If we insert the ansatz into the Ward identities, the prefactor  
$e^{(\al''-\beta') \frac{c_1^2}{c_2}}$ shifts $\al'' \to \beta'$ in the action of $L_1$, and 
we should set $\nu_1 + 2 \nu_2 = \De_{\beta'} - \De_{\al''}$ in order to shift $\al'' \to \beta'$ in the $L_0$ equation. 

Then the ansatz is consistent: the equations are satisfied by $|\,I_1\,\rangle= |\,I^{(1)}(c_1,\beta')\,\rangle$
at the leading order, and at higher orders we get 
\begin{align} \label{eq:rank2order}
L_0 |\,I^{(1)}_{2k}(c_1,\beta')\,\rangle &= \left(\De_{\beta'} + 2 k + c_1 \partial_{c_1} \right)  |\,I^{(1)}_{2k}(c_1,\beta')\,\rangle \cr 
\left(L_1 + 2 c_1 (\beta' - Q) \right)   |\,I^{(1)}_{2k}(c_1,\beta')\,\rangle &= \left(\partial_1 + \nu_1 c_1^{-1} \right) |\,I^{(1)}_{2k-2}(c_1,\beta')\,\rangle \cr 
\left( L_2 + c_1^2 \right) |\,I^{(1)}_{2k}(c_1,\beta')\,\rangle &= -(2\al'' - 3 Q) |\,I^{(1)}_{2k-2}(c_1,\beta')\,\rangle  \cr
 L_3 |\,I^{(1)}_{2k}(c_1,\beta')\,\rangle &= - 2 c_1|\,I^{(1)}_{2k-2}(c_1,\beta')\,\rangle \cr 
 L_4 |\,I^{(1)}_{2k}(c_1,\beta')\,\rangle &= - |\,I^{(1)}_{2k-4}(c_1,\beta')\,\rangle   \cr 
 L_n |\,I^{(1)}_{2k}(c_1,\beta')\,\rangle &= 0 \qquad n>4
\end{align} 

Extensive experiments indicate that the solution to these recursion
equations exists for generic values of the parameters $c_1$, $c_2$, $\al''$, $\beta'$, and is always
unique. At the first stages of the calculations, $\nu_1$ and $\nu_2$ are also fixed uniquely.
The solution for $|\,I^{(1)}_{2k}(c_1,p)\,\rangle$ turns out to be a a sum over descendants at level $s$, $0 \leq s \leq k$, multiplied by a power $c_1^{s-2k}$ . 
The linear equations for the coefficients of the various descendants have a triangular form: no non-trivial matrix inversion appears to be needed. Indeed, the coefficients 
of the expansion are polynomials in $\al''$, $\beta'$. This contrasts with the usual chiral vertex operator, which gives an expansion in rational functions of the Liouville momenta, with poles due to the existence of null vectors for special values of $\De_f$. 

As an example, set $k=1$. We can reproduce the source in the $L_3$ equation 
by $(2 c_1)^{-1} L_{-1}|\,I_1\,\rangle$. Then the source in the $L_2$ equation 
by adding $(2 c_1)^{-1} ( 2 \al'' - 3 \beta') \partial_{c_1} |\,I_1\,\rangle$ to that.
Then the $L_1$ equation is satisfied if we set 
\begin{align}\label{eq:rk1exp}
\nu_1 &= 2 ( \al''-\beta' )(Q - \beta') \cr
\nu_2 &= (\beta'- \al'')\left(\frac{3}{2} Q - \frac{3}{2} \beta' - \frac{1}{2} \al'' \right) \,. 
\end{align} 
Thus we will set 
\begin{equation}
|\,I^{(1)}_{2}(c_1,\beta')\,\rangle = (2 c_1)^{-1} L_{-1}|\,I_1\,\rangle + (2 c_1)^{-1}  ( 2 \al'' - 3 \beta') \partial_{c_1} |\,I_1\,\rangle + \nu_3 c_1^{-2} |\,I_1\,\rangle
\end{equation}

The constant $\nu_3$ is undetermined at this order. At the next order of the recursion $\nu_3$ will be fixed, and a new undetermined multiple of $c_1^{-4} |\,I_1\,\rangle$ will appear. Etcetera.

It is useful to elaborate on why the solution, if it exists, is unique. The difference of two solutions will satisfy at each order 
of the expansion the same Ward identities as the irregular vector $|\,I^{(1)}(c_1,\beta')\,\rangle$, except for a shift of $L_0$. 
At any order we checked, we could not find any non-trivial ``null irregular descendant'' in $\CI_1$ which 
satisfies the same Ward identities of the irregular vector $|\,I_1\,\rangle$ defining $\CI_1$. Hence the only ambiguity at order $2k$ 
is by $\nu_{k+2} c_1^{-2k} |\,I_1\,\rangle$. The ambiguity is always fixed at the next order of the expansion.

It should not be hard to prove uniqueness at all order in the expansion, 
assuming a solution of this general form, by making the triangular form of the Ward identities constraints 
more manifest. More precisely, it should not be hard to prove that at any order, the only homogeneous solution 
for the Ward identities is $v_{2k} = c_1^{-2k} |\,I_1\,\rangle$, and that the equations at the next order cannot be solved with the source
induced by $v_{2k}$.

Thus, at least at the level of this formal power series in $c_2$, we 
have a definition of the basis 
\begin{equation}
\Psi^{r,1}_{\De_0,\beta'}(c_1) \Psi^{1,2}_{\beta',\al''}(c_2)\,
|\,I_2\,\rangle\,
\end{equation}
labeled by the intermediate Liouville momentum $\beta'$ of the rank 1 irregular module,
at least as a power series in positive powers of $\frac{c_2}{c_1}^2$ and $c_1$. 
In order to make statements which go beyond this formal power series analysis, 
we will need more refined tools. But observe that the exponential 
prefactor for the power series in $c_2$ is suggestive of an asymptotic series, 
rather than a convergent power series. One may 
imagine that the crucial conformal block
\begin{equation}
\langle \De_0|\Psi^{r,1}_{\De_0,\beta'}(c_1) \Psi^{1,2}_{\beta',\al''}(c_2)\,
|\,I_2\,\rangle\,,
\end{equation} if well defined as a function, 
may be uniquely determined by a choice of Stokes sector at $c_2 \to 0$, where the asymptotic expansion is valid.

\subsubsection{Rank $1$ plus regular to rank $1$} 
In a similar fashion, we can look for a formal power series of descendants of a rank $1$ vector  of parameters $c_1$ and $\beta'$
which represents a regular vector at $z$ and an irregular vector of rank $1$ at the origin, of parameters $c_1$, $\al'$.
We will denote the solution of the problem as 
\begin{equation}\label{irrdeg2}
|\,IR_1\,\rangle= \Psi^{\De_2}_{\beta',\al'}(z)|\,I_1\,\rangle =\,z^{\mu_z}c_1^{\mu_1}\,
e^{(\beta'-\al') \frac{2c_1}{z}}\,
\sum_{k=0}^{\infty}z^k\,|\,I^{(1)}_k(c_1,\al_i)\,\rangle\,,
\end{equation}
and plugging in the Ward identities
\begin{align} 
L_0 |\,IR_1\,\rangle &= \left(\De_{\al'} + c_1 \partial_{c_1} + z {\pa_z} +  \De_2 \right)  |\,IR_1\,\rangle  \cr 
L_1 |\,IR_1\,\rangle  &= \left( - 2 c_1 (\al' - Q)+ z^2 {\pa_z} + 2 \De_2 z \right) |\,IR_1\,\rangle  \cr 
L_2 |\,IR_1\,\rangle &= \left(-c_1^2 + z^3 {\pa_z} + 3 \De_2 z^2\right)  |\,IR_1\,\rangle    \cr
 L_n |\,IR_1\,\rangle  &= z^n
\left(z{\pa_z}+\De_2(n+1)\right)  |\,IR_1\,\rangle \qquad n>2
\end{align} 
we can again compute the coefficient recursively (after setting $\mu_z+ \mu_1= \De_{\beta'}- \De_2-\De_{\al'}$) from 
\begin{align} \label{eq:rank1rr}
\left( L_0-\Delta_{\beta'}- c_1 \partial_{c_1} \right)|\,I^{(1)}_k(c_1,\al')\,\rangle &= k |\,I^{(1)}_k(c_1,\al')\,\rangle\cr 
\left( L_1+2 c_1 (\beta' - Q) \right)|\,I^{(1)}_k(c_1,\al')\,\rangle  &=  (\mu_z+ 2 \De_2+k-1) |\,I^{(1)}_{k-1}(c_1,\al')\,\rangle  \cr 
\left( L_2+c_1^2 \right)  |\,I^{(1)}_k(c_1,\al')\,\rangle &=  (\mu_z+3 \De_2+k-2) |\,I^{(1)}_{k-2}(c_1,\al')\,\rangle\\
&\quad - 2 c_1 (\beta'-\al')  |\,I^{(1)}_{k-1}(c_1,\al')\,\rangle   \cr
 L_n |\,I^{(1)}_k(c_1,\al')\,\rangle  &= (\mu_z+\De_2(n+1)+k-n)|\,I^{(1)}_{k-n}(c_1,\al')\,\rangle \cr
&\quad- 2 c_1 (\beta'-\al')  |\,I^{(1)}_{k-n+1}(c_1,\al')\,\rangle \qquad n>2
\notag\end{align} 

The solution appears to exist and be unique for 
generic values of the parameters $c_1$, $z$, $\al'$, $\beta'$, $\De_2$. 
At the first stages of the calculations, $\mu_1$ and $\mu_z$ are also fixed.

Let's look at the first non-trivial level, $k=1$. We only have non-trivial sources for the $L_2$ equation and $L_1$ equation. 
We can satisfy both equations with $(\beta'-\al') \partial_{c_1} |\,I_1\,\rangle$, if 
\begin{align}
\mu_z &= - 2 \De_2- 2 (\beta' - Q)(\beta' - \al') \cr
\mu_1 &=\De_2+(\al'- \beta' + Q)(\al' - \beta')  \,.
\end{align}
Again, we can add to that a $\mu_3 |\,I^{(1)}(c_1,\al')\,\rangle$, with the expectation that $\mu_3$ will be fixed at the next order, etc. 

We expect that it should not be hard to prove uniqueness at all order in the expansion, 
assuming a solution of this general form, by making the triangular form of the Ward identities constraints 
more manifest. More precisely, it should not be hard to prove that at any order, the only homogeneous solution 
for the Ward identities is $v_{k} = c_1^{-k} |\,I_1\,\rangle$, and that the equations at the next order cannot be solved with the source
induced by $v_{k}$.

We take it to define, as a formal power series, the basis 
\begin{equation}
\Psi^{r,1}_{\De_0,\beta'}(c_1) \Psi^{(1)\De_2}_{\beta',\al'}(z)|\,I_1\,\rangle
\end{equation}
Again, this expansion has the form of an asymptotic expansion in $z$, and the conformal blocks 
$\langle \De_0|\Psi^{r,1}_{\De_0,\beta'}(c_1) \Psi^{(1)\De_2}_{\beta',\al'}(z)|\,I_1\,\rangle$, if they exist as actual functions, may be labeled by the extra data of a Stokes sector 
as $z \to 0$. 

\subsubsection{More general maps}
It should be clear how one can pursue this strategy further, given sufficient amount of patience. 
At the next stage, we can define an irregular module $\CI_2$ of descendants of an
irregular vector of rank $2$ $|I_2 \rangle$.  
For later convenience, we define $\CI_2$ as the span (with coefficients which are function of $c_1, c_2$) of vectors of the form
\begin{equation}
\tilde \BL_{-{I}} \partial_{c_1}^{k_1} \partial_{c_2}^{k_2} |\,I_2\,\rangle
\end{equation}
where the symbols $\partial_{c_1}$ and $\partial_{c_2}$ are taken to commute with the Virasoro generators,
and $\tilde L_{-n} = L_{-n}$ for $n >1$, but we find convenient in our explicit calculations to define 
\begin{equation}
\tilde L_{-1} =  L_{-1} - 2 c_1 \partial_{c_2}.
\end{equation}

We say that such a vector is a descendant of weight $|I|+k_1 + 2 k_2$. 
We define an action of the Virasoro algebra on $\CI_2$
in the obvious way: commute raising operators through to hit $ |\,I_2\,\rangle$,
apply the Ward identities for the rank $2$ irregular vector and then act with the derivatives. 

The irregular module depends on the choice of Liouville momentum $\alpha''$ in the Ward identities, not of the specific choice of ambient Verma module or of the specific realization of the rank $2$ irregular vector 

It is straightforward, if tedious, to seek formal power series solutions for various useful maps. 
The most important for our purposes is of course the image $\Psi^{2,3}_{\beta'',\al'''}(c_3)\,
|\,I_3\,\rangle$ of a rank $3$ irregular vector in $\CI_2$. This can be built as a power series in $c_3$,
\begin{equation}\label{irrdeg3}
|\,I^{(3)}(c,\al^{(3)})\,\rangle\,=\,c_3^{\rho_3}c_2^{\rho_2}\,
e^{(\al^{(3)}-\al'')S_3(c)}\,
\sum_{k=0}^{\infty}c_3^k\,|\,I^{(2)}_{3k}(c_2,c_1,\al'')\,\rangle\,,
\end{equation}

It takes some work to find the correct prefactor exponent
\begin{equation}
S_3(c) = \frac{2 c_1 c_2}{c_3} - \frac{c_2^3}{3 c_3^2} - \frac{c_1^2}{ c_2}
\end{equation}
so that the ansatz works at the leading order. 
The vectors $|\,I^{(2)}_{3k}(c_2,c_1,\al')\,\rangle$ are expanded as a sum of level $s$ descendants, $s \leq k$, 
multiplied by positive powers of $c_1^t$, $0 \leq t \leq 3(k-s)$, and by $c_2^{-(3k-s+t)/2}$. 
The Liouville momenta appear polynomially. 

This allows a definition of a basis of the form
\begin{equation}
\Psi^{r,1}_{\De_0,\beta'}(c_1) \Psi^{1,2}_{\beta',\beta''}(c_2) \Psi^{2,3}_{\beta'',\al'''}(c_3)\,
|\,I_3\,\rangle\,,
\end{equation}
at least as a formal power series. 
The series involves positive powers of 
$\frac{c_3 c_1^3}{c_2^3}$, $\frac{c_2}{c_1^2}$ and $c_1$. 

\section{Bases of conformal blocks of null vector equations}\label{app:PDE}

\setcounter{equation}{0}

We here describe in detail how the null vector equations
can be used to define certain bases for the space of conformal 
blocks, and to construct the series expansions of their elements.

\subsection{Solutions to the null vector equations - the regular case}

We'll now describe how to realize this program in detail before
we apply the same method to calculate series expansions
for conformal blocks containing irregular singularities.

\subsubsection{Power series solutions}

For future work with the differential equation \rf{PDEreg0}
it will be useful to factor out 
the corresponding conformal block of the Gaussian free field,
\begin{equation}
\CF^{}(y):=e^{-b\phi_s(y)}\CG^{}(y)\,,
\end{equation}
where
\begin{align}
e^{-b\phi_s(y)}=&\,y^{b\al_3}(z_1-y)^{b\al_1}(y-z_2)^{b\al_2}
z_1^{-2\al_1\al_3}z_2^{-2\al_2\al_3}(z_1-z_2)^{-2\al_1\al_2}\,.\notag
\end{align}
We find that the functions $\CG^{}(y;z_1,z_2)$, satisfy the
equations
\begin{subequations}\label{PDEG0}
\begin{align}\label{PDEG0a}
&0\,=\,\bigg[\,\frac{1}{b^2}\frac{\pa^2}{\pa y^2}+\frac{1}{y-z_1}\frac{z_1}{y}\frac{\pa}{\pa z_1}
+\frac{1}{y-z_2}\frac{z_2}{y}\frac{\pa}{\pa z_2}+\\
&\notag \hspace{3.5cm}+\frac{2}{b}\bigg(
\frac{\al_1}{y-z_1}+\frac{\al_2}{y-z_2}+
\frac{\al_3}{y}\bigg)\frac{\pa}{\pa y}-\frac{1}{y}
\frac{\pa}{\pa y}\,\bigg]\CG^{}(y)\,,\\
&0\,=\,\bigg[\,y\frac{\pa}{\pa y}+{z_1}\frac{\pa}{\pa z_1}
+{z_2}\frac{\pa}{\pa z_2}-\la\,\bigg]\CG^{}(y)\,,
\label{PDEG0b}\end{align}
\end{subequations}
where $\la$ is defined as
\begin{equation}\label{la'def}
\la:=\De_{0}-\De_{\al_1+\al_2+\al_3-b/2}\,.
\end{equation}

It will furthermore be useful to factorize the relevant solutions
$\CG(y;z_1,z_2)$ to \rf{PDEG0} 
into a part that is singular when $z_2\ra 0$, $y\ra 0$, 
and a regular part,
\begin{equation}\label{Gfactor}
\CG^{}(y;z_1,z_2)\,=\,\CG_s(y;z_1,z_2)
\CG_r({y}/{z_1},
{z_2}/{y})\,,
\end{equation}
where 
\begin{itemize}
\item
$\CG_s(y;z_2,z_1)$ is the function defined as 
\begin{equation}
\CG_s\,:=\,G_0\;z_1^{\la}
\left(\frac{z_2}{z_1}\right)^{\mu}
\left(\frac{y}{z_1}\right)^{\nu}\,,
\end{equation}
where $G_0$
is constant with respect to $y$, $z_1$ and $z_2$, but
may depend on $\al_1$, $\al_2$, $\al_3$, $\al_0$ and $\be$,
the constant $\la$ is defined in \rf{la'def}, and $\mu$ and $\nu$ 
are defined as 
\begin{equation}\label{munu'def}
\mu\,=\,\De_{\be}-\De_{\al_2+\al_3}\,,\qquad
\nu\,=\,b(\be-\al_2-\al_3)\,.
\end{equation}
\item $\CG_r(u,v)$ is a power series of the form
\begin{equation}
\CG_r(u,v)\,=\,\sum_{k=0}^\infty v^{k}
\sum_{l=0}^\infty u^{l}
\CG^{}_{k,l}\,,\qquad \CG^{}_{0,0}\,=\,1\,.
\end{equation}
\end{itemize}
The equation \rf{PDEG0b} is automatically satisfied by 
the ansatz \rf{Gfactor}, while differential 
equation \rf{PDEG0a} is equivalent to 
\begin{equation}\label{PDE0'}
{\CD_{\CG_s}^{}}\cdot \CG_r\left({y}/{z_1},
{z_2}/{z_1}\right)\,=\,0\,,\qquad
{\CD_{\CG_s}^{}}:=\CG_s^{-1}\cdot\left(\CD_0-\frac{z_2}{y}\CD_1\right)
\cdot\CG_s^{}\,,
\end{equation}
where  
\begin{subequations}
\begin{align}
\CD_0:=
& \;\frac{1}{b^2}
y^2\frac{\pa^2}{\pa y^2}
+z_1\frac{\pa}{\pa z_1}+z_2\frac{\pa}{\pa z_2}
+\frac{2}{b}\left(\al_1+\al_2+\al_3-\frac{b}{2}\right)y\frac{\pa}{\pa y}
\notag \\
& \qquad-\frac{z_1}{y}\left[
\frac{1}{b^2}
y^2\frac{\pa^2}{\pa y^2}
+\frac{2}{b}\left(\al_2+\al_3-\frac{b}{2}\right)y
\frac{\pa}{\pa y}+z_2\frac{\pa}{\pa z_2}\right]\,, \label{CD0def}  \\
\CD_1:= &
\frac{1}{b^2}
y^2\frac{\pa^2}{\pa y^2}
+\frac{2}{b}\left(\al_1+\al_3-\frac{b}{2}\right)
y\frac{\pa}{\pa y}+z_1\frac{\pa}{\pa z_1}\notag\\
& \qquad-\frac{z_1}{y}
\left[\frac{1}{b^2}
y^2\frac{\pa^2}{\pa y^2}
+\frac{2}{b}\left(\al_3-\frac{b}{2}\right)
y\frac{\pa}{\pa y}\right] \,.
\label{CD1def}\end{align}
\end{subequations}
We may construct solutions to \rf{PDE0'} in the form of a
double series expansion simlar to \rf{F0series}. 
The recursion relations resulting from \rf{PDE0'} take the
form
\begin{equation} \label{recrel}
A_l^{}\,\CG^{}_{k,l}-B_{k,l}^{}\,\CG^{}_{k,l+1}\,=\,
C_{k,l}^{}\,\CG^{}_{k-1,l}-D_l^{}\,\CG^{}_{k-1,l+1}\,.
\end{equation}
We are looking for a solution with $\CG^{}_{k,l}=0$ for $k<0$
and $\CG^{}_{k,l}=0$ for $l<0$. A  solution with
$\CG^{}_{0,0}\neq 0$ will exist only if
$\mu$ and $\nu$ satisfy the relation $B_{0,-1}=0$, which is equivalent to 
the equation $\CD_{s}\CG_s=0$, where $\CD_s$ is the differential
operator proportional to $z_1$ in \rf{CD0def}.
It is easy to check that $B_{0,-1}=0$ follows from our definitions 
\rf{munu'def}.
The recursion relations \rf{recrel} will then
determine $\CG^{}_{k,l}$ uniquely in terms of
the first term $\CG^{}_{0,0}$.

\subsubsection{Lowest order term}

The term of lowest order in the expansion of $\CG(y;z_1,z_2)$
in powers of $z_2$, defined by
\begin{equation}
\CG(y;z_1,z_2)\,=\,z_1^{\la}\,\sum_{k=0}^{\infty}
\left(\frac{z_2}{y}\right)^{\mu+k}
\CG_k(y;z_1)\,,
\end{equation}
can be calculated 
explicitly. Indeed, it is clear that
$\CG^{}_0(y;z_1)$ must satisfy 
\begin{equation}
\left(\CD_0+\la-\mu\frac{z_1}{y}\right)\CG^{}_0(y;z_1)=0\,.
\end{equation}
Writing
$\CG^{}_0(y;z_1)=y^{b(\al_0+b/2-\al_1-\al_2-\al_3)}\CH^{}_0(y;z_1)$, 
we find that
$\CH^{}_0(y;z_1)$ must satisfy 
\begin{equation}\label{hypgeom}
\bigg[\,(y-{1})y^2\frac{\pa^2}{\pa y^2}-
[(C-2)y-{A+B-1}]\frac{\pa}{\pa y}-{AB}
\bigg]
\CH^{}_0(y;z_1)=0\,,
\end{equation}
where $A,B,C$ are given as
\begin{equation}
\begin{aligned}
&A=b(\beta+\al_1-\al_0-b/2)\,,\\
&B=1-b(\beta+\al_0-\al_1-b/2)\,,
\end{aligned}\qquad
C=2-b(2\al_0-b)\,.
\end{equation}
The equation \rf{hypgeom}
is satisfied by the hypergeometric function $F(A,B;C;1/y)$.
Picking the solution 
which has the required behavior for $y\ra 0$ gives us 
\begin{equation}\label{hypsol0}
\CG_0(y;z_1)\,=\,G_0^{(0)}\,w^{b(\be-\al_2-\al_3)} 
F(A,1-C+A;1-B+A;w)\,.\qquad w:=\frac{y}{z_1}\,.
\end{equation}
The constant prefactor $G_0^{(0)}$ is dependent on the normalization
of the chiral vertex operators $\Psi_{\al_1\al_3}^{\al_2}(y)$ and
will be specified when it becomes relevant.



The behavior of the lowest order term $\CF^{}_0(y;z_1)$
of the $z_2$-expansion for $y\ra \infty$ 
follows from the well-known formula
\begin{align}
F(A,B,C;z)\,=\,& K_1\,(-z)^{-A}F(A,1-C+A;1-B+A;1/z)\\ \notag &+
K_2\,(-z)^{-B}K_1F(B,1-C+B;1-A+B;1/z)\,,
\end{align}
which implies the braid relation \rf{degbraid}. The coefficients are
explicitly given as
\[
K_1\,=\,\frac{\Ga(C)\Ga(A-B)}{\Ga(B)\Ga(C-A)}\,,\qquad
K_2\,=\,\frac{\Ga(C)\Ga(B-A)}{\Ga(A)\Ga(C-B)}\,.
\]
For $\Re(2\al_0-Q)>0$ we will therefore get a relation of the form
\rf{Fkreconstr} which allows us to calculate the coefficients $F_k$ from $\CF_k(y;z_1)$ as described in the main text.

\subsection{Case $n=2$}\label{sec:n2}

\subsubsection{Differential equations}

The function $\CF^{(2)}(y;c_1,c_2)$ defined in \rf{rk2-confbl-yto0}
satisfies the differential equations
\begin{subequations}\label{PDEreg}\begin{align}
& \bigg[\,\frac{1}{b^2}\frac{\pa^2}{\pa y^2}+\CT^{(2)}\,\bigg]
\CF^{(2)}(y;z_1,z_2)=0\,,\\
&\bigg[\,y\frac{\pa}{\pa y}+{c_1}\frac{\pa}{\pa c_1}
+2{c_2}\frac{\pa}{\pa c_2}+\De_{\al''}+\de_b-
\De_{\al_0}\,\bigg]\CF^{(2)}(y;c_1,c_2)=0\,,
\label{grading}\end{align}
\end{subequations}
where $\De_\al=\al(Q-\al)$,
$\de_b=-\frac{1}{2}-\frac{3}{4}b^2$,
\begin{align*}
\CT^{(2)}:=&-\frac{1}{y}
\frac{\pa}{\pa y}+\frac{\La_4}{y^6}+\frac{\La_3}{y^5}+\frac{\La_2}{y^4}
+\frac{1}{y^3}\bigg(\La_1+c_2\frac{\pa}{\pa c_1}\bigg)
+\frac{1}{y^2}
\bigg(2c_2\frac{\pa}{\pa c_1}+
c_1\frac{\pa}{\pa c_1}+\De_{\al''}\bigg)\,.
\end{align*}
Stripping off the free field part as above,
\begin{equation}
\CF^{(2)}(y;c_1,c_2):=e^{-b\phi_s(y)}\CG^{(2)}(y;c_1,c_2)\,,
\end{equation}
where
\begin{align}
\phi_s(y)=\,\frac{c_2}{2y^2}+\frac{c_1}{y}-\al''\log y
\,.
\end{align}
the functions $\CG^{(2)}(y;c_1,c_2)$ satisfy the
equations
\begin{subequations}\label{PDE(2)}
\begin{align}
&0\,=\,\bigg[\,\frac{1}{b^2}\frac{\pa^2}{\pa y^2}+\frac{1}{y^3}c_2\frac{\pa}{\pa c_1}
+\frac{1}{y^2}\bigg(2c_2\frac{\pa}{\pa c_2}+{c_1}\frac{\pa}{\pa c_1}\bigg)
+\\
&\notag \hspace{5cm}+\frac{2}{b}\bigg(
\frac{c_2}{y^3}+\frac{c_1}{y^2}+\frac{\al''}{y}\bigg)\frac{\pa}{\pa y}-\frac{1}{y}
\frac{\pa}{\pa y}\,\bigg]\CG(y)\,,\\
&0\,=\,\bigg[\,y\frac{\pa}{\pa y}+{c_1}\frac{\pa}{\pa c_1}
+2c_2\frac{\pa}{\pa c_2}-\la\,\bigg]\CG(y)\,,
\end{align}
\end{subequations}
where $\la=\De_{\al_0}-\De_{\al''-b/2}$.
The first of these equations can be written as 
$\CD^{(2)}\CG^{(2)}(y;c_1,c_2)=0$
\begin{equation}
\begin{aligned}
\CD^{(2)}:=\, \,y^2\frac{\pa^2}{\pa y^2}+
&\bigg(\frac{2bc_1}{y}+2b\al''-b^{2}\bigg)
y\frac{\pa}{\pa y}+b^2\left(2c_2\frac{\pa}{\pa c_2}+c_1\frac{\pa}{\pa c_1}\right) \\
& \qquad\quad +\frac{c_2}{c_1 y}\,\bigg(
b^2 c_1\frac{\pa}{\pa c_1}
+\frac{2bc_1}{y}\,
y\frac{\pa}{\pa y}\bigg)\,.
\end{aligned}
\end{equation}

\subsubsection{Series solutions}

We will look for a  
solution to \rf{PDE(2)} in the form
\begin{equation}
\label{Gfactor(2)}
\CG^{(2)}(y;c_1,c_2)\,=\,\CG_s^{(2)}(y;c_1,c_2)
\CG_r^{(2)}\left({y}/{2b c_1},
{c_2}/{c_1y}\right)\,,
\end{equation}
where 
\begin{itemize}
\item
$\CG_s^{(2)}(y;c_1,c_2)$ is the function defined as 
\begin{equation}\label{CG0(2)}
\CG_s^{(2)}(y;c_1,c_2)\,:=\,G_0^{(2)}\;c_1^{\mu_0+\nu_1}
\,c_2^{\nu_2}\,
\left(\frac{y}{2bc_1}\right)^{\nu}
\,e^{-\si\frac{c_1^2}{c_2}}
\,,
\end{equation}
where $G_0^{(2)}$
is constant with respect to $c$, $c_1$ and $c_2$. 
We will set
\begin{equation}
\si_2\,=\,\be'-\al''\,,\qquad\nu\,=\,b\si_2\,,\qquad \mu_0+\nu_1+\nu_2=\la\,,
\end{equation}
to get the asymptotic behavior for $c_2\ra 0$ characteristic for 
an irregular singularity with $n=2$ as discussed in the previous sections. 
The exponents $\nu$, $\nu_2$ will 
be determined with the help of the differential equation below.
\item $\CG_r^{(2)}(u,v)$ is a power series of the form
\begin{equation}
\CG_r^{(2)}(u,v)\,=\,\sum_{k=0}^\infty v^{k}
\sum_{l=k}^\infty u^{l}
\CG^{(2)}_{k,l}\,,\qquad \CG^{(2)}_{0,0}\,=\,1\,.
\end{equation}
\end{itemize}

The recursion relations resulting from \rf{PDE(2)} take the
form 
\begin{equation}\label{recrel(2)}
a_{k,l}\CG^{(2)}_{k,l}+b_{k,l}\CG^{(2)}_{k,l+1}+c_{k,l}\CG^{(2)}_{k-1,l}+d_{k,l}\CG^{(2)}_{k-1,l+1}=0\,,
\end{equation}
where 
\begin{align}\label{akldef}
& a_{k,l}=(\nu+l-k)(\nu+l-k+b(2\al''-Q)+b^2(\la-\nu+k-l)\,,\quad b_{k,l}=\nu-\si+l+1-k\,,\notag\\
& c_{k,l}=b^2(\la-\nu_2-\nu-(k-1)-l)\,,
\qquad\qquad d_{k,l}=\nu-(k-1)+l+1\,.
\end{align}
We are again looking for a solution with $\CG^{(2)}_{k,l}=0$ for $k<0$
and $\CG^{(2)}_{k,l}=0$ for $l<0$. This 
requires the relation $B_{0,-1}=0$, again.
It is then easily found that 
\begin{subequations}\label{CGlow}\begin{align}
&\CG^{(2)}_{0,1}\,=\,-A_{0,0}\CG^{(2)}_{0,0}\,,\\
&\CG^{(2)}_{1,0}\,=\,\nu\CG^{(2)}_{0,0}\,. 
\end{align}
\end{subequations}
For $k=1$, $l=0$ we observe that $B_{1,0}=0$. Instead of determining $\CG^{(2)}_{1,1}$ we
therefore get a constraint on the exponents $\la$, $\nu_2$, $\nu$:
\begin{equation}
A_{1,0}\CG^{(2)}_{1,0}+(\la-\nu_2-\nu)\CG^{(2)}_{0,0}+
(\nu+1)\CG^{(2)}_{0,1}\,=\,0\,.
\end{equation}
Inserting \rf{CGlow} and \rf{akldef} yields the equation
\begin{equation}
\nu_2\,=\,\frac{\nu}{2b^2}(3bQ-3\nu-4\al'')\,.
\end{equation}
Then the subsequent equations fix the other
$\CG^{(2)}_{1,l}$ in terms of $\CG^{(2)}_{1,1}$.
At $k=2$, similarly, $\CG^{(2)}_{2,0}$ and $\CG^{(2)}_{2,1}$ 
are solved for in terms of
$\CG^{(2)}_{1,1}$. 
In the next equation, $\CG^{(2)}_{2,2}$ does not appear, and instead
the equation fixes $\CG^{(2)}_{1,1}$. 
Then the subsequent equations fix $\CG^{(2)}_{2,l}$
in terms of $\CG^{(2)}_{2,2}$
The equations for $k=3$ will fix $\CG^{(2)}_{2,2}$ but leave 
$\CG^{(2)}_{3,3}$ undetermined. 
This pattern continues when we go to higher values of $k$. 
The recursion relations \rf{recrel} will therefore
determine $\CG^{(2)}_{k,l}$ uniquely in terms of
the first term $\CG^{(2)}_{0,0}$.

\subsubsection{Lowest order terms}

We are looking for solutions of \rf{PDE(2)} which take the form
of a power series expansion of the form 
\[
\CG^{(2)}(y;c_1,c_2)\,=\,G_0^{(2)}\;c_1^{\la}
\left(\frac{c_2}{c_1^2}\right)^{\nu_2}
\,e^{-\si\frac{c_1^2}{c_2}}\,\big(\,\CG^{(2)}_0(y;c_1)\,+
\CO(c_2/c_1y)\,\big)\,.
\]
The behavior for
$y\ra \infty $ is of the form
$\CG^{(2)}_0(y;c_1)=Ny^{b(\al_0+b/2-\al'')}(1+\CO(y))$, where 
$N$ does not depend on $y$. Writing
$\CG^{(2)}_0(y;c_1)=y^{b(\al_0+b/2-\al'')}\CH^{}_0(y;c_1)$, 
we find that
$\CH^{(2)}_0(y;c_1)$ must satisfy the differential equation
\begin{equation}\label{confhypg}
\bigg[\,y^3\frac{\pa^2}{\pa y^2}+\big[1-(C-2)y\big]y\frac{\pa}{\pa y}-A
\bigg]
\CH^{(2)}_0(y;c_1)=0\,,
\end{equation}
where $A$ and $C$ are now given as
\begin{equation}\label{AC}
A=b(\be'-\al_0-b/2)\,,\qquad
C=2- b(2\al_0-b)\,.
\end{equation}
Equation \rf{confhypg} is the equation satisfied by the confluent 
hypergeometric function $F(A;C;1/y)$.

The function $\CG^{(2)}_0(y;c_1)$ can be represented as
\begin{equation}
\CG^{(2)}_0(y;c_1)\,=\,w^{b\si_2}\Psi(A;C;1/w)\,,\qquad
w:=\frac{y}{2bc_1}\,,
\end{equation}
where $\Psi(A;C;z)$ is the function defined by the integral representation
\begin{equation}\label{PsiEuler}
\Psi(A;C;z)=\frac{1}{\Ga(A)}
\int_0^{\infty}d\tau\;\tau^{A-1}(1+\tau)^{C-A-1}
e^{-z\tau}\,.
\end{equation}
This function is the unique solution to the confluent hypergeometric
equation which behaves at infinity as $\Psi(A;C;z)=z^{-A}(1+\CO(z^{-1}))$.

\subsubsection{Reconstructing the conformal blocks}

Assuming that $|\,I_2(c_2,c_1,\al'')\,\rangle$ has an expansion 
of the form \rf{irrdeg} leads us to the claim that 
$\CF^{}(y)$ must have an expansion of the form
\begin{equation}\label{z2exp(2)}
\CF^{(2)}(y;c_1,c_2)\,=\,e^{\frac{c_1^2}{c_2}(\al''-\be')}
\sum_{k=0}^{\infty}\left(\frac{c_2}{yc_1}\right)^{\nu_2+k}
\CF^{(2)}_k(y;c_1)\,,
\end{equation}
where the higher order terms $\CG^{}_k(y;c_1)$, $k>0$ are obtained
from $\CG^{}_0(y;c_1)$ by acting with differential operators $\CD_k(y,c_1)$,
\begin{equation}\label{GkDG2}
\CF^{(2)}_k(y;c_1):=\,\CD_k^{(2)}(y,c_1)\CF^{(2)}_0(y;c_1)\,.
\end{equation}
The differential operators
$\CD_k^{(2)}(y,c_1)$
are again of the form
\begin{equation}
\CD_k^{(2)}(y,c_1)\,=\,\sum_{l=-k}^k\left(\frac{c_1}{y}\right)^{l}\,
\sum_{m=0}^k \CD_{k;l,m}^{(2)}\left(y\frac{\pa}{\pa y}\right)^m\,,
\end{equation}
and can be calculated from \rf{z2exp(2)} as soon as the
power series expansions of $\CF^{}_k(y;c_1)$ have been calculated from 
the differential equation.

The rest of the analysis
proceeds as in the case $n=0$ above. We thereby get another
useful algorithm for computing the power series 
expansion of conformal blocks with irregular singularities
with the help of the null vector equations.

\subsection{Case $n=1$}\label{n=1PDE}

Let us now consider conformal blocks with a degenerate insertion, two regular punctures and a rank $1$ irregular puncture. 
We will look at a solution which could be written as  
\begin{align}
&\CF^{(1)}(y;c_1',z_2):=\langle\,\al_0\,|\,\Psi^{r,1}_{\al_0, \beta'-b/2}(c_1)V^{(1)}_+(y)\,\Psi_{\beta', \al'}^{(1)\al_2}(z_2)
|\,I_1(\al')\,\rangle
\end{align}
in the notations of \ref{sec:degvec}, which will turn out to be unique. 

\subsubsection{Differential equations}

This function satisfies the differential equations
\begin{align}
& \bigg[\,\frac{1}{b^2}\frac{\pa^2}{\pa y^2}+\CT^{(1)}\bigg]
\CF^{(1)}(y;z_1,z_2)=0\,,\\
&\bigg[\,y\frac{\pa}{\pa y}+{c_1'}\frac{\pa}{\pa c_1'}
+{z_2}\frac{\pa}{\pa z_2}+\De_{\al_2}+\De_{\al'}-\De_{\al_0}\,\bigg]\CF^{(0)}(y)=0\,,
\end{align}
where
\begin{align*}
\CT^{(1)}:=\frac{\La_2'}{y^4}+\frac{\La_1'}{y^3}+\frac{1}{y^2}\bigg(
c_1'\frac{\pa}{\pa c_1'}+\De_{\al'}\bigg)
+\frac{\De_{\al_2}}{(y-z_2)^2}
+\frac{1}{y(y-z_2)}{z_2}\frac{\pa}{\pa z_2}-\frac{1}{y}\frac{\pa}{\pa y}\,.
\end{align*}
It will again be useful to factor out the free field part:
\begin{equation}
\CF^{(1)}(y):=e^{-b\phi_s(y)}\CG^{(1)}(y)\,,
\end{equation}
where
\begin{align}
e^{-b\phi_s(y)}=&\,e^{-b\frac{c_1'}{y}}(y-z_2)^{b\al_2}z_2^{-2\al_2\al'}
e^{2\al_2\frac{c_1'}{z_2}}\,.
\end{align}
We find that the functions $\CG^{(1)}(y;c_1',z_2)$, $n=0,1,2$ satisfy the
equations
\begin{subequations}\label{PDE-G-n=1}
\begin{align}\label{PDE(1)a}
&0\,=\,\bigg[\,\frac{1}{b^2}\frac{\pa^2}{\pa y^2}+\frac{1}{y^2}{c_1'}\frac{\pa}{\pa c_1'}
+\frac{1}{y-z_2}\frac{z_2}{y}\frac{\pa}{\pa z_2}+\\
&\notag \hspace{5cm}+\frac{2}{b}\bigg(
\frac{\al_2}{y-z_2}+\frac{c_1'}{y^2}+\frac{\al'}{y}\bigg)\frac{\pa}{\pa y}-\frac{1}{y}
\frac{\pa}{\pa y}\,\bigg]\CG^{(1)}(y)\,,\\
&0\,=\,\bigg[\,y\frac{\pa}{\pa y}+{c_1'}\frac{\pa}{\pa c_1'}
+{z_2}\frac{\pa}{\pa z_2}-\la\,\bigg]\CG^{(1)}(y)\,,
\label{scaling1}\end{align}
\end{subequations}
where $\la$ is defined in \rf{la'def}.

\subsubsection{Series solutions}

We will look for a  
solution to \rf{PDE-G-n=1} in the form
\begin{equation}
\label{Gfactor(1)}
\CG^{}(y;c_1',z_2)\,=
\,\CG_s^{}(y;c_1',z_2)
\,\CG_r({y}/{2b c_1'},{z_2}/y)\,,
\end{equation}
where 
\begin{itemize}
\item
$\CG_s^{}(y;c_1',z_2)$ is the function defined as 
\begin{equation}\label{CG0(1)}
\CG_s(y;c_1',z_2)\,:=\,G_0^{}\;(c_1')^{\mu_0+\mu_1}
(c_2)^{\mu_2'}
\left(\frac{y}{2bc_1'}\right)^{\nu}
\,e^{2(\be'-\al'-\al_2)\frac{c_1'}{z_2}}
\,,
\end{equation}
where $G_0$
is constant with respect to $c$, $c_1$ and $c_2$,
and
\begin{equation}\label{mu''def}
\nu=b(\be'-\al'-\al_2)\equiv b\si_2\,,\qquad \mu_0+\mu_1+\mu_2'=\la\,.
\end{equation}
The coefficient $\mu'$ will be determined by the differential equation.
\item $\CG_r(u,v)$ is a power series of the form
\begin{equation}
\CG_r(u,v)\,=\,\sum_{k=0}^\infty v^{k}
\sum_{l=k}^\infty u^{l}
H^{}_{k,l}\,,\qquad H^{}_{0,0}\,=\,1\,.
\end{equation}
\end{itemize}
In order to calculate the series expansion it is useful to
rewrite
equation \rf{PDE(1)a} as 
\begin{equation}\label{PDE(1)a'}
\left(\CD_0^{(1)}-
\frac{z_2}{y}\CD_1^{(1)}\right)
\CG^{(1)}(y;c_1',z_2)\,=\,0\,,
\end{equation}
where 
\begin{align}
\CD_0^{(1)}:=&\,\frac{1}{b^2}y^2
\frac{\pa^2}{\pa y^2}+
\frac{2}{b}\left(\al'+\al_2-\frac{b}{2}\right)
y\frac{\pa}{\pa y}
+c_1'\frac{\pa}{\pa c_1'}+z_2\frac{\pa}{\pa z_2} 
+\frac{1}{b^2}\frac{2bc_1'}{y}\,y\frac{\pa}{\pa y}\,,\\
\CD_1^{(1)}:=&
\,\frac{1}{b^2}y^2\frac{\pa^2}{\pa y^2}+
\frac{2}{b}\left(\al'-\frac{b}{2}\right)\frac{\pa}{\pa y}
+c_1'\frac{\pa}{\pa c_1'}+\frac{1}{b^2}\frac{2bc_1'}{y}\frac{\pa}{\pa y}
\,.\notag
\end{align}
The recursion relations will take the same form \rf{recrel(2)} 
as in the case $n=2$ above, with coefficients now given as
\begin{align}\label{akldef(1)}
& a_{k,l}=(\nu+l-k)(\nu+l-k+b(2\al'+2\al_2-Q)+b^2(\la-\nu+k-l)\,,\notag \\ 
& b_{k,l}=\nu-\si+l+1-k\,,\notag\\
& c_{k,l}=(\nu+l-k+1)(\nu+l-k+1-bQ+2b\al')+b^2(\la-\nu_2-\nu-(k-1)-l)\,,
\notag\\
& d_{k,l}=\nu-(k-1)+l+1\,.
\end{align}
The following discussion is very similar to the discussion 
we gave in the case $n=2$ after equation \rf{akldef}. It is 
found, in particular, that the exponent $\mu'_2$ equals
\begin{equation}
\mu'_2=\nu_2+\frac{1}{b^2}\nu(\nu-bQ+2b\al')=
\frac{2}{b^2}\nu(bQ-\nu-b\al'-2b\al_2)\,,
\end{equation}
which is exactly as required by our representation-theoretic 
analysis in the previous section.

It is worth noting that the lowest order term is given by the same
confluent hypergeometric function we found in the case $n=2$.

\section{Existence of collision limits}\label{nullcoll}

\setcounter{equation}{0}

In this appendix we'll describe two approaches to prove the existence
of the collision limits in the sense of formal series expansions.
To be more precise, we'll show that the series expansions
for the conformal blocks can be rearranged in such a 
way that the collision limits exist order by order in the
series expansion.

These results provide further evidence for
our conjectures on the existence of irregular vertex operators.
They are also used in order to support our conjectures
on physical correlation function in Section \ref{sec:measure}. 

\subsection{Warmup} \label{sec:warm}
It is useful to look first at the most basic collision limit, of two regular vectors into a rank $1$ irregular vector. 
Remember that the collision limit at the level of conformal blocks, from $|\,R^{(1)}(z)\,\rangle$ to $|\,I^{(1)}(c_1)\,\rangle$ with constant $c_1 = \al_z z$ and $\al' = \al_i + \al_z$,
requires a rescaling both by an uninteresting power of $z$, but also a much more important $\al_z^{\De_f-\De_{\al'}}$. Inside a correlation function, $\al_f$ will be integrated over,
and thus such a divergent factor would be troublesome. 
On the other hand, in correlation functions, 
$|\,R^{(1)}(z)\,\rangle$ will be accompanied by a normalization factor
\begin{align}
G_0\,=\,\sqrt{C_0}\,,\qquad C_0:=C(\al_f,\al_z,\al_i)
\end{align}
where $C(\al_1,\al_2,\al_3)$ is the function proposed in \cite{DO,ZZ}
\begin{align}
C(\al_1,\al_2,\al_3) & =
(\mu_0)^{\frac{1}{b}(Q-\al_1-\al_2-\al_3)}\times \\
&\times\frac{\up_0\up(2\al_1)\up(2\al_2)\up(2\al_3)}
{\up(\al_1+\al_2+\al_3-Q)\up(\al_1+\al_3-\al_2)
\up(\al_1+\al_2-\al_3)\up(\al_2+\al_3-\al_1)
}
\,,
\notag\end{align}
which was later shown \cite{T01,TL} to represent the three point function 
of Liouville theory. Here $\mu_0 = \pi \mu\ga(b^2)b^{2-2b^2}$. 

The Barnes double Gamma function $\Ga_b(x)$
is known
to have the following asymptotic behavior \cite{Sp}:
\begin{align}
\log
\Ga_b(x) = \frac{1}{2} x(Q-x)\log x-\frac{1}{12}(1+Q^2)
\log x+\frac{3}{4}x^2-\frac{Q}{2}x+\CO(x^0)\,.
\end{align}
This implies the following asymptotic behavior of the 
function $\up(x)$
\begin{align}\label{Upasym}
\log
\up(x) = -\frac{1}{2}\De_x\log \De_x+\frac{1}{12}(1+Q^2)
\log\De_x+\frac{3}{2}\De_x+\CO(x^0)\,,
\end{align}
using the notation $\De_x=x(Q-x)$.
This expansion is valid for large imaginary $x$. 

If we take the limit of large $\al_z$ with fixed $\al'$, we have a neat asymptotic behavior
\begin{align}
C(\al_f,\al_z,\al'-\al_z) & =
(\mu_0)^{\frac{1}{b}(Q-\al_f-\al')}
\times\frac{\up_0\up(2\al_f)}
{\up(\al_f+\al'-Q)\up(\al'-\al_f)
} (2\al_z)^{2 \De_f-2 \De_{\al'}}
\,,
\notag\end{align}

Thus the normalization factor produces exactly the divergent rescaling required for the proper limit $\al_z^{\De_f-\De_{\al'}}|\,R^{(1)}(z)\,\rangle \to |\,I^{(1)}(c_1)\,\rangle$.\footnote{For clarity, we
omitted a sign: $(2\al_z)^{2 \De_f-2 \De_{\al'}}$ should have been $(2\al_z)^{\De_f- \De_{\al'}}(-2\al_z)^{ \De_f- \De_{\al'}}$. The two factors produce the rescalings 
for holomorphic and antiholomorphic conformal blocks, as $\al_z$ is imaginary in the collision limit! We omit similar signs in the following}
We are left with the proper normalization factor for the $\Psi^{r,1}_{\al_f, \al'}$ map: 
\begin{align}
G_{r,1}\,=\,\sqrt{C_{r,1}}\,,\qquad C_{r,1}:=(\mu_0)^{\frac{1}{b}(Q-\al_f-\al')}
&\times\frac{\up_0\up(2\al_f)}
{\up(\al_f+\al'-Q)\up(\al'-\al_f)
}2^{2 \De_f-2 \De_{\al'}}
\end{align}
This result is consistent with the AGT relation.

\subsection{Direct approach}\label{sec:existence}

The existence of a solution to the Virasoro constraints in the form
of a formal series like \rf{irrdeg} is far from obvious. The corresponding
statement \rf{eq:Regexp} about the
vectors $|\,R^{(2)}(z)\,\rangle$
was obtained from well-known results
in Subsection \ref{sec:vertintro}. We are now going to argue that
there exists a limit of $|\,R^{(2)}(z)\,\rangle$
in which the formal series \rf{irrdeg} can be obtained from
a rearrangement of the series \rf{eq:Regexp}.

\subsubsection{First collision limit} \label{sec:first}
The recursion relations \rf{R2k} do not have a good collision limit as $w \to 0$, $\beta + \al_1 = \beta'$, $\al_1 w =c_1$. Although the differential operator in $w$ does, 
the $\De_{\beta}$ term blows up. As we aim to derive the expansion for $|\,RI_1\,\rangle$,
representing an irregular vector at the origin and a regular vector at $z$, it is actually natural to keep $\al_3 + \al_1= \al'$ fixed in the collision limit. This means that $\De_{\beta} - \De_3 = (Q- \beta - \al_3) (\beta' - \al')$ only blows up linearly in the limit, but still blows up. 

There is a useful reorganization of this sum which does have a good limit. 
Let's modify the ansatz to 
\begin{equation}
|\,R^{(2)}(z)\,\rangle = z^{\De_{\beta} - \De_2- \De_3} 
\bigg(1-\frac{z}{w}\bigg)^A \sum_{k=0}^{\infty} z^k |R^{(1)}_{k} \rangle
\end{equation}
for some constant $A$. This amounts to a redefinition of 
$|R^{(1)}_{k}\rangle$ by multiples of $w^{-s} |R^{(1)}_{k-s}\rangle$, 
with $s>0$. 
The action of $z^{k+1}\frac{\pa}{\pa z}+w^{k+1}\frac{\pa}{\pa w}$ on the new prefactor is easy to compute, and we get 
a modified recursion relation
\begin{align}
\left( L_0-w\paw-\De_1- \De_{\beta} \right) |R^{(1)}_{k} \rangle &=\, k|R^{(1)}_{k} \rangle \\
\left( L_n -w^{n+1}\paw-\De_1(n+1)w^k\right) |R^{(1)}_{k} \rangle &=\, 
(\De_{\beta} + n\De_2-\De_3+ k-n) |R^{(1)}_{k-n} \rangle + \cr &\quad
+ A \sum_{s=1}^n  w^{n-s} |R^{(1)}_{k-s} \rangle \quad
n> 0\,,\notag
\end{align}

Now we are in a good shape to take a collision limit. At order $s=n$, we have now the combination $\De_{\beta} - \De_3 + A$, which can be finite if
$A$ grows as $2 \al_1 (\al' - \beta')$. Then the term $s=n-1$ is also finite, controlled by $A w$, and all other terms in the sum drop out.
The recursion relations take exactly the form of \rf{eq:rank1rr} as long as $A w \to 2 c_1 (\al' - \beta')$, and
$A=\mu_z + \De_2 - \De_{\beta} + \De_3 = 2 \al_1 (\al' - \beta') - \De_z- (Q-\beta' + \al')(\al' - \beta')$. 

Thus we have the following situation. We have certain descendants 
$|R^{(1)}_{k} \rangle$ which satisfy the recursion relations, are built our of 
$|R^{(1)} \rangle$ by acting with Virsoro generators, and powers of $\partial_w$ and $w^{-1}$, 
with  coefficients which are rational functions of the Liouville momenta. 
The denominators of the rational functions are the usual Kac determinants for $\De_{\beta}$. 

In the collision limit, $|R^{(1)} \rangle \sim w^{-2 \al_1 \beta}|I^{(1)} \rangle$.
We can just write $|R^{(1)} \rangle = w^{-2 \al_1 \beta}|\tilde R^{(1)} \rangle$,
and for any descendant $|R^{(1)}_k \rangle= w^{-2 \al_1 \beta}|\tilde R^{(1)}_k \rangle$ 
for some descendant $|\tilde R^{(1)}_k \rangle$ of $|\tilde R^{(1)} \rangle$,
and replace $w = c_1/\al_1$. 
The coefficients will be rational functions of the Liouville momenta, and we can ask if these rational functions have a finite collision limit. 

If so, the collision limit of $|\tilde R^{(1)}_k \rangle$ will give us some $|I^{(1)}_k \rangle$
which automatically satisfy the recursion relation for the coefficients of the expansion of $|RI^{(1)} \rangle$!
For example 
\begin{equation}
 |\tilde R^{(1)}_{1} \rangle = \left(  \frac{\De_{\beta} + \De_2 - \De_3}{2\De_{\beta}} L_{-1} -  \frac{\De_{\beta} + \De_2 - \De_3}{2\De_{\beta}} \al_1 \partial_{c_1} -  \frac{\De_{\beta} + \De_2 - \De_3}{2\De_{\beta}} \frac{2 \al_1 \beta}{w} + \frac{A}{w} \right) |\tilde R^{(1)} \rangle
\end{equation}
has a finite limit 
\begin{equation}
 |I^{(1)}_{1} \rangle = \left( (\al'-\beta') \partial_{c_1} + \frac{(\beta'-Q)(\De_2 + (\al'- \beta')(Q+\al'-\beta')}{c_1} \right) |\tilde R^{(1)} \rangle\,.
\end{equation}

Crucially, the action of the left hand side of the Ward identities on descendants of $|\tilde R^{(1)} \rangle$ goes smoothly to the action on the corresponding descendants of $|I^{(1)} \rangle$. 
The coefficients on the right hand sides of the Ward identities are finite.
So the only way the coefficients of the descendants in $|\tilde R^{(1)}_k \rangle$
could start blowing in the collision limit up at some order $t$ of the recursion, rather than having a finite limit, is if the left hand side of the Ward identities
annihilates the part of the answer which is blowing up. 

If we assume the uniqueness property for $|I^{(1)}_k \rangle$,  then the divergent piece at order $t$ 
must be a multiple of $v_{k} = c_1^{-k} |\,I_1\,\rangle$, but then the equations at the next order cannot be solved because the source at the leading order in $\al_1$ is the one 
induced by $v_{k}$. Thus the collision limit $|R^{(2)} \rangle \sim w^{2 \al_1 \al_3}|IR^{(1)} \rangle$ will give us a solution 
for $|IR^{(1)} \rangle$, as long as the uniqueness assumption is true. 
 
Notice that the prefactor also behaves reasonably well in the limit. We should remember that in order to 
have a good collision limit for the Ward identities, $|\,R^{(2)}(z)\,\rangle$ is multiplied by $(-w)^{2 \al_1 \al_3}$, and use  $|R^{(1)} \rangle = (-w)^{-2 \al_1 \beta}|\tilde R^{(1)} \rangle$ (we inserted the minus signs for convenience), 
so that the overall prefactor of the expansion of $(-w)^{2 \al_1 \al_3}|\,R^{(2)}(z)\,\rangle$ in $|\tilde R^{(1)}_k \rangle$ is 
\begin{equation}\begin{aligned}
 w^{-2 \al_1 \beta+2 \al_1 \al_3} z^{\De_{\beta} - \De_2- \De_3} \bigg(1-\frac{z}{w}\bigg)^A & = (-w)^{-2 \al_1(\beta'-\al')-A} z^{\nu}  \bigg(1-\frac{w}{z}\bigg)^A \\ & \qquad\to (-w)^{\De_z- \De_{\beta'-\al'}} z^{\nu} 
e^{\frac{2 c_1}{z} (\beta' - \al')}\,,
\end{aligned}\end{equation}
so that we have almost recovered the prefactor in $|IR^{(1)} \rangle$. We dropped some phase which cancels against the anti-holomorphic conformal blocks. 

This is not enough to convert $(-w)^{\De_z- \De_{\beta'-\al'}} $ to $c_1^{\De_z- \De_{\beta'-\al'}} $ and reproduce $|IR^{(1)} \rangle$.
For that, we need an extra power $\al_1^{\De_z- \De_{\beta'-\al'}}$. Furthermore, we need a rescaling factor to convert $|\tilde R^{(1)} \rangle$ to $|I^{(1)} \rangle$, as in section \ref{sec:warm}. We will find both momentarily in the collision limit of the normalization factors. 

\subsubsection{Normalization}
The original vector $|\,R^{(2)}(z)\,\rangle$
is normalized by the square root of $C(\al_0,\al_1,\be)C(Q-\be,\al_2,\al_3)$. 
In the collision limit, $C(\al_0,\al_1,\be)$ behaves as in section \ref{sec:warm}
and goes to $C_{r,1}(\al_0,\be') \al_1^{2 \De_0-2 \De_{\be'}}$. That divergence cancels out as in section \ref{sec:warm}
and leaves the correct normalization for $\Psi^{r,1}_{\al_0, \beta'}$.

On the other hand, 
\begin{align}
& C(Q-\be'+\al_1,\al_2,\al'-\al_1)  \sim \\
& \sim(\mu_0)^{\frac{1}{b}(\be'-\al_2-\al')}\times\frac{\up_0\up(2\al_2)}
{\up(\al'-\be'+\al_2)\up(\al'-\be'+Q-\al_2)
}(2\al_1)^{2\De_z- 2\De_{\beta'-\al'}}
\,,
\notag\end{align}
and thus produces exactly the other required normalization factor. 

We are left with the proper normalization factor for $\Psi^{(1) \De_2}_{\beta',\al'}$: the square root of 
\begin{align}
C^{(1) \De_2}_{\beta',\al'}= 
(\mu_0)^{\frac{1}{b}(\be'-\al_2-\al')}\times \frac{\up_0\up(2\al_2)}
{\up(\al'-\be'+\al_2)\up(\al'-\be'+Q-\al_2)
}2^{2\De_z- 2\De_{\beta'-\al'}}
\,,
\notag\end{align}
\subsubsection{Second collision limit}
Next, we would like to take a limit from $|IR^{(1)} \rangle$ to the vector 
$|\,I^{(2)}(c,\al)\,\rangle$ representing an irregular vector of rank $2$ in the irregular module of rank $1$. 
We know the appropriate collision limit: $z \to 0$ with finite $\al_2 z^2 = c_2$, $\al_2 z + c_1$, $\al_2 + \al' = \al''$. 
All that we need to find out is how to reorganize the ansatz for $|IR^{(1)} \rangle$ so that it has a finite limit. 

Notice that the resulting rank $2$ irregular vector has parameters $c_2, \al_2 z + c_1$, but we are working in a rank 1 module of parameter $c_1$. 
In our original ansatz for $|\,I^{(2)}(c,\al)\,\rangle$ we kept for simplicity the same $c_1$ both in the rank 2 irregular vector, and in the rank 1 irregular module.

It is much more natural to take a different starting point:  rather than  $|IR^{(1)} \rangle$ we can start from a 
 $|\tilde IR^{(1)} \rangle$ which represents inside a rank 1 module of parameter $c_1$ a regular puncture at $z$ and an irregular 
vector of parameters $c_1' = c_1 - \al_2 z$ and $\al'$ at the origin. Then the collision limit will give back a rank $2$ vector of parameters $c_2$,$c_1$. 

The vector $|\tilde IR^{(1)} \rangle$  is simply produced by acting on $|IR^{(1)} \rangle$ by 
$ \exp \left( -\al_2 z \partial_{c_1} \right)$. This produces an expression of the form 
\begin{equation}\label{irrdeg2a}
|\,\tilde IR_1\,\rangle= \,z^{\mu_z}(c_1 - \al_2 z)^{\mu_1} \,
e^{(\beta'-\al') \frac{2c_1}{z}}\,
\sum_{k=0}^{\infty}z^k\,|\,\tilde I^{(1)}_k(c_1)\,\rangle\,,
\end{equation}
up to an overall factor $e^{-2 \al_2(\beta'-\al')}$ which we will deal with momentarily.

In order to have a good, term-by-term collision limit we need to further rearrange the sum, by pulling out an 
overall factor of $(1-\frac{\al_2 z}{c_1})^{-\nu_1}$. Thus we re-define the $|\,\tilde I^{(1)}_k(c_1)\,\rangle$ 
appropriately, and write 
\begin{equation}
|\,\tilde IR_1\,\rangle= \,z^{\mu_z}c_1^{\nu_1}(c_1 - \al_2z)^{\mu_1-\nu_1} \,
e^{(\beta'-\al') \frac{2(c_1 - \al_2z)}{z}}\,
\sum_{k=0}^{\infty}z^k\,|\,\tilde I^{(1)}_k(c_1)\,\rangle\,,
\end{equation}

The Ward identities satisfied by  $|\,\tilde IR_1\,\rangle$ are
\begin{align} 
L_0 |\, \tilde IR_1\,\rangle &= \left(\De_{\al'} + c_1  \partial_{c_1} + z \paz +  \De_2\right)  |\, \tilde IR_1\,\rangle  \cr 
L_1 |\,\tilde IR_1\,\rangle  &= \left( - 2 (c_1- \al_2 z) (\al' - Q)+ z^2 \left( \paz+\al_2\pa_{c_1}\right) + 2 \De_2 z \right) |\,\tilde IR_1\,\rangle  \cr 
L_2 |\, \tilde IR_1\,\rangle &= \left(-(c_1- \al_2 z)^2 + z^3 \left( \paz+\al_2 \pa_{c_1}\right) + 3 \De_2 z^2\right)  |\,\tilde IR_1\,\rangle    \cr
 L_n |\,\tilde IR_1\,\rangle  &= z^n
\left(z\left( \paz+\al_2 {\pa}_{c_1}\right)+\De_z(n+1)\right)  |\, \tilde IR_1\,\rangle \qquad n>2
\end{align} 

This gives a recursion for the $|\,\tilde I^{(1)}_k(c_1)\,\rangle$:
\begin{align} 
L_0 |\,\tilde I^{(1)}_k(c_1)\,\rangle &= \left(\De_{\beta'} + c_1 \partial_{c_1} \right)  |\,\tilde I^{(1)}_k(c_1)\,\rangle\\ 
L_1|\,\tilde I^{(1)}_k(c_1)\,\rangle  &= - 2 c_1 (\beta' - Q) |\,\tilde I^{(1)}_k(c_1)\,\rangle+ (\mu_z+ 2 \De_2+k-1+ 2 \al (\beta' - Q)) |\,\tilde I^{(1)}_{k-1}(c_1)\,\rangle \cr &\quad+\al_2(\partial_{c_1}+ \nu_1/c_1)|\,\tilde I^{(1)}_{k-2}(c_1)\,\rangle  \cr 
L_2 |\,\tilde I^{(1)}_k(c_1)\,\rangle &= -c_1^2 |\,\tilde I^{(1)}_k(c_1)\,\rangle  + (\mu_z+3 \De_2+k-2- \al_2^2+ 2 \al_2(\beta' - \al')) |\,\tilde I^{(1)}_{k-2}(c_1)\,\rangle \cr &\quad- 2 c_1 (\beta'-\al'- \al_z)  |\,\tilde I^{(1)}_{k-1}(c_1)\,\rangle +\al_2(\partial_{c_1}+ \nu_1/c_1)|\,\tilde I^{(1)}_{k-3}(c_1)\,\rangle  \cr
 L_n |\,\tilde I^{(1)}_k(c_1)\,\rangle  &= (\mu_z+\De_2(n+1)+k-n+2 \al_2(\beta'-\al'))|\,\tilde I^{(1)}_{k-n}(c_1)\,\rangle  \cr &\quad- 2 c_1 (\beta'-\al')  |\,\tilde I^{(1)}_{k-n+1}(c_1)\,\rangle+\al_2 (\partial_{c_1}+ \nu_1/c_1)|\,\tilde I^{(1)}_{k-n-1}(c_1)\,\rangle \qquad n>2
\notag\end{align} 

We want to obtain a power expansion in $c_2$, so it is OK to scale $|\,\tilde I^{(1)}_k(c_1)\,\rangle \to \al^{k/2} |\,\tilde I^{(1)}_k(c_1)\,\rangle$. 
As we have a finite $ \mu_z+ 2 \De_2+ 2 \al_2(\beta' - Q) = 2 (\al'' - \beta')(\beta'-Q)$ then the Ward identities for the rescaled vectors have a neat limit to the Ward identities 
we are after 
\begin{align} 
L_0 |\,\tilde I^{(1)}_k(c_1)\,\rangle &= \left(\De_{\beta'} + k+ c_1 \partial_{c_1} \right)  |\,\tilde I^{(1)}_k(c_1)\,\rangle\cr 
L_1|\,\tilde I^{(1)}_k(c_1)\,\rangle  &= - 2 c_1 (\beta' - Q) |\,\tilde I^{(1)}_k(c_1)\,\rangle + (\partial_{c_1}+ \nu_1/c_1)|\,\tilde I^{(1)}_{k-2}(c_1)\,\rangle \cr 
L_2 |\,\tilde I^{(1)}_k(c_1)\,\rangle &= -c_1^2 |\,\tilde I^{(1)}_k(c_1)\,\rangle  + (3Q - 2 \al'') |\,\tilde I^{(1)}_{k-2}(c_1)\,\rangle \cr
 L_3 |\,\tilde I^{(1)}_k(c_1)\,\rangle  &= - 2 c_1 |\,\tilde I^{(1)}_{k-2}(c_1)\,\rangle \cr
 L_4 |\,\tilde I^{(1)}_k(c_1)\,\rangle  &= -|\,\tilde I^{(1)}_{k-4}(c_1)\,\rangle \cr
 L_n |\,\tilde I^{(1)}_k(c_1)\,\rangle  &= 0 \qquad n>4
\end{align} 

Thus, using the assumption of uniqueness, through the collision limit we have verifies the existence of a solution to this recursion, and of $\Psi^{1,2}$ as a formal power series. 

Finally, we should look at the prefactor, remembering from section $2$ that we expect $z^{2 \al_2 \al'} e^{- \frac{2 \al_2(c_1 -z \al_2)}{z}} |\,\tilde IR_1\,\rangle$ to have a good collision limit
\begin{align}
& z^{\mu_z+2 \al_a \al'}c_1^{\nu_1}(c_1 - \al_2z)^{\mu_1-\nu_1} \,e^{(\beta'-\al') \frac{2(c_1 - \al_2z)}{z}}e^{- \frac{2 \al_2 (c_1 - \al_2z)}{z}} = \cr
& c_1^{\nu_1} z^{\mu_z+\mu_1+2 \al_2 \al'-\nu_1}\al_2^{\mu_1-\nu_1}(1-\frac{c_1}{\al_2z})^{\mu_1-\nu_1}  \,e^{(\beta'-\al'') \frac{2(c_1 - \al_2z)}{z}} \sim \cr
& c_1^{\nu_1} c_2^{\nu_2} \al_2^{\mu_1-\nu_1-\nu_2} e^{(\al''-\beta')\frac{c_1^2}{c_2}} \, e^{-2\al_2 (\beta'-\al'') }
\end{align}
Hence we find the desired rank $2$ irregular vector ansatz, with a spurious power prefactor $\al_2^{2 \al_2 (\beta' -\al'') - 1/2 \De_{\beta'-\al''}}$ and an exponential $e^{- 2 \al_2(\beta'-\al'')}$.
 
 \subsubsection{Normalization}
 The effect of the second collision limit on the normalization of $\Psi^{(1)}$ is simple: 
 \begin{align}
C^{(1) \De_2}_{\beta',\al''-\al_2}= 
(\mu_0)^{\frac{1}{b}(\be'-\al'')}\times \frac{\up_0\up(2\al_2)}
{\up(\al''-\be')\up(\al''-\be'+Q-2\al_2)
}2^{2\De_z- 2\De_{\beta'-\al'}}
\,,
\notag\end{align}
has a single divergent factor, which we can decompose into 
\begin{equation}
(2 \al_2)^{\De_{\beta'-\al''}- 4 \al_2 (\be'-\al'')} e^{ 4 \al_2 (\be'-\al'')}
\end{equation}
This fully cancels the spurious prefactors we found in the collision limit. 

We are left with the normalization for $\Psi^{1,2}$
 \begin{align}
C^{1,2}_{\beta',\al''}= 
(\mu_0)^{\frac{1}{b}(\be'-\al'')}\times \frac{\up_0}
{\up(\al''-\be')
} 2^{-\De_{\beta'-\al''}}
\,,
\notag\end{align}

 \subsubsection{Generalizations}
 There is no obvious reason for which this limit procedure cannot be iterated.
 Starting from 
 \begin{equation}
 \Psi_{\gamma',\beta'}^{(1)\De_1}(w)  \Psi_{\beta',\al'}^{(1)\De_2}(z) |I_1 \rangle
 \end{equation}
 expanded in powers of $z$ and pulling the Virasoro generators and $c_1$ derivatives through  $ \Psi_{\gamma',\beta'}^{(1)\De_1}(w)$
 we can write it as a sum of descendants of the formal module $|IR_1\rangle$ representing 
 an irregular puncture at the origin and a regular puncture at $w$. Then a collision limit 
 $|IR_1\rangle \to |I_2\rangle$ sending $w \to 0$ to produce an irregular puncture of rank two 
 should give us the formal series for 
  \begin{equation}
\Psi_{\beta'',\al''}^{(2)\De_2}(z) |I_2 \rangle
 \end{equation}
expanded in positive powers of $z$. Then a further limit $z \to 0$ should give us the 
formal power series expansion of 
$\Psi^{2,3}_{\beta'',\al'''} |I_3 \rangle$,
and so on.

\subsection{Existence of the collision limit $(n=0)\ra(n=1)$ from null vector
equations}

We want to study the behavior of $\CG^{(0)}(y;z_1,z_2)$ 
in the limit $z_1\ra 0$, $\al_1\ra\infty$, $\al_3\ra\infty$ 
such that $\al':=\al_1+\al_3$ and $c_1':=\al_1z_1$ are kept fixed.
In order to get a well-defined limit we will furthermore send
the intermediate representation label $\beta\ra\infty$ such that
$\beta':=\beta+\al_1$ stays finite. 
We will use the notation $\lim_{(0)\ra(1)}$ for the limit 
defined in this way.

The key observation is that the differential operator $\CD$ 
has a finite limit $\CD^{(1)}$ 
in the collision limit $\lim_{(0)\ra(1)}$. However, it is
not straightforward to analyze the behavior of $\CG^{}(y;z_1,z_2)$ 
in the collision limit using the factorization \rf{Gfactor} due to the 
divergence of $\mu'$ in the limit under considertation. This divergence
yields divergent behavior both in the prefactor $\CG_s$ and in the
power series $\CG_r(y/z_1,z_2/z_1)$ appearing
in \rf{Gfactor}. We are going to show that the divergence in 
$\CG_r(y/z_1,z_2/z_1)$
can be factored out, and that it essentially\footnote{There is going to be a remaining divergent piece that cancels in physical correlation functions.}  
cancels the divergence in $\CG_s$. 

\subsubsection{Existence of $\lim_{(0)\ra(1)}\CG_r(y/z_1,z_2/z_1)$}

The trick is to replace the factorization \rf{Gfactor} by
\begin{equation}\label{Gfactor'}
\CG^{}(y;z_1,z_2)\,=\,\CG_s'(y;z_1,z_2)
\CG_r'\left({y}/{z_1},
{z_2}/{z_1}\right)\,,
\end{equation}
where $\CG_s'(y;z_2,z_1)$ is the function defined as 
\begin{equation}
\CG_s'(y;z_1,z_2)\,:=\,G_0\;z_1^{\la}
\left(\frac{z_2}{z_1}\right)^{\mu}
\left(\frac{y}{z_1}\right)^{\nu}\left(1-\frac{z_2}{z_1}\right)^{\rho}\,.
\end{equation}
The exponent $\rho$ has to be chosen appropriately.
The function 
$\CG_r'\equiv \big(1-\frac{z_2}{z_1}\big)^{-\rho}\CG_r$ in \rf{Gfactor'}
is a convergent power series in $z_2/z_1$, $y/z_1$, allowing us
to interpret the relation between 
$\CG_r(y/z_1,z_2/z_1)$ and $\CG_r'(y/z_1,z_2/z_1)$ as a relation between
power series.

In order to cancel the divergence from
$\mu$ in $\CG_r^{(0)}(y/z_1,z_2/z_1)$, we will choose $\rho$ such that
\begin{equation}\label{rhochoice}
\mu':=\,\mu+\rho
\end{equation}
stays finite.
Indeed, we then have 
\begin{equation}
\CG_s'(y;z_1,z_2)\,=\,e^{\pm\pi \rho}
G_0\,z_1^{\la}\,\left(\frac{z_2}{z_1}\right)^{\mu'}
\left(\frac{y}{z_1}\right)^{\nu}
\left(1-\frac{z_1}{z_2}\right)^{\rho}\,.
\end{equation}
This function has a simple behavior in the collision limit
\begin{equation}\label{g(0)-(1)}
\CG_s'(y;z_1,z_2)\sim e^{\pm\pi \rho} G_0\,\al_1^{\la-\mu'-\nu}
(c_1')^{\la}\,\left(\frac{z_2}{c_1'}\right)^{\mu'}
\left(\frac{y}{c_1'}\right)^{\nu}e^{2\si_2\frac{c_1'}{z_2}}\,
\big(1+\CO(\al_1^{-1})\big)\,.
\end{equation}
On the right hand side of \rf{g(0)-(1)} we identify
the function that was denoted $\CG_s^{(1)}(y;c_1',z_2)$ 
in Subsection \ref{n=1PDE} provided that we 
adopt the choice for $\mu'$ given in \rf{mu''def}. 
The prefactor $e^{\pm\pi \rho}$ drops out when one forms
physical correlation functions by combining holomorphic and
anti-holomorphic conformal blocks. Choosing the normalization
constant $G_0$ as $\sqrt{C(\al_0,\al_1,\be)C(Q-\be,\al_2,\al_3)}$,
we find as in  Subsection \ref{sec:existence} that 
the constant prefactor $G_0\,\al_1^{\la-\mu'-\nu}$ stays finite in 
this limit.
We may finally conclude that 
$\lim_{(0)\ra (1)}{\CD_{\CG_s}^{}}$ exists, which implies that
$\lim_{(0)\ra (1)}\CG_r(y/z_1,z_2/z_1)$ exists, 
as we wanted to show.

\subsubsection{Asymptotics of lowest order terms}

We had previously seen that the differential equation \rf{PDE0'} 
determines the higher order terms of the expansion in powers of $z_2/z_1$ 
in terms of the lowest order term 
$\CG_0(y;z_1)$, for which we had found the expression
\rf{hypsol0}. It is therefore interesting to understand what happens 
in the limit of interest 
to $\CG_0(y,z_1)$. To this end we shall employ the the integral 
representation 
\begin{equation}\label{Euler}
F(A,1-C+A;1-B+A;w)=
\frac{\Ga(1-B+A)}{\Ga(A)\Ga(1-B)}
\int_0^1dt\;t^{A-1}(1-t)^{-B}(1-tw)^{C-A-1}\,.
\end{equation}
In the  limit in question we have $w\ra\infty$, $B\ra \infty$ 
such that $w/B\ra u:=y/2bc_1'$. In order to study this limit we may
use the substitution $\tau:=-tz$ to rewrite the integral \rf{Euler}
as
\begin{equation}\label{Euler'}
\begin{aligned} 
F &(A,1-C+A ;1-B+A;w)\,=\,\\
&=\,\frac{\Ga(1-B+A)}{\Ga(A)\Ga(1-B)}(-w)^{-A}
\int_0^{-z}d\tau\;\tau^{A-1}(1+\tau)^{C-A-1}
\left(1+\frac{\tau}{w}\right)^{-B}\,.
\end{aligned}\end{equation}
The behavior of the prefactor in \rf{Euler} is found by using 
Stirling's formula:
\begin{equation}
\frac{\Ga(1-B+A)}{\Ga(1-B)}\,\underset{B\ra \infty}{\sim}\,(-B)^A\,,
\end{equation}
while the behavior of the integrand in \rf{Euler'} follows from
\begin{equation}
\left(1+\frac{\tau}{w}\right)^{-B}\,=\,
\left(1+\frac{\tau}{uB}\right)^{-B}\,\underset{B\ra \infty}{\sim}\,
e^{-\frac{\tau}{u}}\,.
\end{equation}
By combining these ingredients we are lead to the conclusion that
\begin{equation}
\lim_{\substack{B,w\ra\infty\\ u=w/B \;{\rm fixed}}} 
F(A,1-C+A;1-B+A;w)\,=\,u^{-A}\,\Psi(A;C;1/u)\,,
\end{equation}
where the function $\Psi(A;C;z)$ is defined by the integral representation
\rf{PsiEuler}.
We may thereby conclude that the following limit exists:
\begin{equation}
\lim_{(0)\ra(1)}{\al_1}^{b(\al_2+\al_3-\be)}
\CG_0^{(0)}(y;z_1)\,=\,\CG_0^{(1)}(y;c_1')\,.
\end{equation}

\subsection{Existence of the collision limit $(n=1)\ra(n=2)$ 
from null vector
equations}


We want to study the behavior of $\CG^{(1)}(y;c_1',z_2)$ 
in the limit $z_2\ra 0$, $c_1'\ra\infty$, $\al_2\ra\infty$, $\al'\ra\infty$ 
such that 
\begin{equation}\label{n=2lim}
\al'':=\al_2+\al',\quad c_1:=c_1'+z_2\al_2,\;\;
{\rm and} \quad d_2:=\sqrt{\al_2}\,z_2\,,
\end{equation} 
are kept fixed.
We will use the notation $\lim_{(1)\ra(2)}$ for the limit 
defined in this way.

The existence of this limit will again be based on  
the fact that
the differential operator $\CD^{(1)}$ turns into  $\CD^{(2)}$. 
Our goal will be to rewrite $\CG^{(1)}\big(y;c_1',z_2\big)$ in the
form 
\begin{equation}
\label{Gfactor(1)''}
\CG^{(1)}\big(y;c_1',z_2\big)\,=\,
\tilde\CG_s^{(1)}\big(y;c_1,d_2\big)\,
\tilde\CG_r^{(1)}\big({y}/{2b c_1},{d_2}/{c_1}\big)\,,
\end{equation}
where $\tilde\CG_r^{(1)}\big({y}/{2b c_1},{d_2}/{c_1}\big)$ is a power series
in the indicated variables. If $\tilde\CG_s^{(1)}\big(y;c_1,d_2\big)$
turns out to have a finite collision limit, the 
existence of a limit for 
$\big(\tilde{\CG}_s^{(1)}\big)^{-1}
\cdot\CD^{(1)}\cdot\tilde\CG_s^{(1)}$ will imply the existence
of the limits for the expansion coefficients of 
$\tilde\CG_r^{(1)}\big({y}/{2b c_1},{d_2}/{c_1}\big)$.
In order to take the limit $\lim_{(1)\ra(2)}$ we will need to rearrange
the series expansion of $\CG^{(1)}(y;c_1',z_2)$ in two steps:


\noindent{\it Step 1:}
First, let us 
invert the relations \rf{n=2lim},
\begin{equation}\label{n=2lim-inv}
\al'=\al''-\al_2,\quad c_1'=c_1-{\sqrt{\al_2}}d_2,\;\;
{\rm and} \quad z_2=\frac{d_2}{\sqrt{\al_2}}\,,
\end{equation} 
and rewrite the
formal series solution $\CG^{(1)}(y;c_1',z_2)$
of $\CD^{(1)}\CG^{(1)}(y;c_1',z_2)=0$ as a formal expansion in powers of
$d_2/c_1$ which is denoted
as $\CG^{(1)}(y;c_1,d_2)$. 
This expansion may be constructed by rewriting
the differential equations \rf{PDE-G-n=1}
in terms of the variables $d_2$ and $c_1$. We find 
\begin{subequations}\label{PDE-G-n=1a}
\begin{align}
&0\,=\,\bigg[\,\frac{1}{b^2}y^3\frac{\pa^2}{\pa y^2}
+\frac{2}{b}(c_1+y(\al-b))y\frac{\pa}{\pa y}+\kappa y\\
&\hspace{1.5cm}-\frac{d_2}{\sqrt{\al_2}}
\left( c_1\frac{\pa}{\pa c_1}+
\frac{1}{b^2}y^2\frac{\pa^2}{\pa y^2}
+\left(\frac{2}{b}c_1-y\right)\frac{\pa}{\pa y}
\right)+d_2^2\left( 
\frac{\pa}{\pa c_1}+\frac{2}{b}
\frac{\pa}{\pa y}\right)
\bigg]\CG(y;c_1,d_2)\,,\notag\\
&0\,=\,\bigg[\,y\frac{\pa}{\pa y}+{c_1}\frac{\pa}{\pa c_1}
+{d_2}\frac{\pa}{\pa d_2}-\kappa\,\bigg]\CG(y;c_1,d_2)\,.
\label{scaling1a}\end{align}
\end{subequations}
Observe that the differential equations \rf{PDE-G-n=1a} become
the differential equations satisfied by 
$\CG^{(2)}(y;c_1,c_2)$ with $c_2=d_2^2$ in the limit 
$\lim_{(1)\ra(2)}$. 

\noindent
{\it Step 2:}
Let us finally modify the factorization \rf{Gfactor(1)} into
\rf{Gfactor(1)''},
where 
$\tilde\CG_s^{(1)}(y;c_1,d_2)$ is the function defined as 
\begin{align}\label{CG0'(1)}
& \tilde\CG_s^{(1)}\big(y;c_1,d_2\big): =
G_0^{(1)}\,e^{2(\be'-\al'-\al_2)\frac{c_1'}{z_2}}
\,(c_1')^{\mu_0+\mu_1}\,z_2^{\mu_2'}\,
\left(\frac{y}{2bc_1'}\right)^{\nu}
\left(\frac{c_1}{c_1'}\right)^{\mu_0+\nu_1-\nu}\\
& \quad =G_0^{(1)}\,e^{2(\be'-\al'')\frac{c_1'}{z_2}}\,
c_1^{\mu_0+\nu_1}\,z_2^{\mu_1+\mu_2'-\nu_1}(-\al_2)^{\mu_1-\nu_1}
\left(1-\frac{c_1}{\al_2z_2}\right)^{\mu_1-\nu_1}
\left(\frac{y}{2bc_1}\right)^{\nu}
\,,
\notag\end{align}
where $\nu_1$ is given in \rf{eq:rk1exp}.
Note that 
\begin{equation}\label{CGr-CGr'}
\tilde\CG_r^{(1)}\big({y}/{2b c_1},{d_2}/{c_1}\big)\,=\,
\left(\frac{c_1'}{c_1}\right)^{\mu_0+\nu_1-\nu}
\CG_r^{(1)}\big({y}/{2b c_1},{d_2}/{c_1}\big)\,.
\end{equation}
Before taking the limit $\lim_{(1)\ra(2)}$ we may use the
expression for $c_1'$ in \rf{n=2lim-inv} and expand the factor
$
(c_1'/c_1)^{\mu_0+\nu_1-\nu}
$
appearing in the relation \rf{CGr-CGr'}
as a power series in $d_2/c_1$. The relation \rf{CGr-CGr'} may therefore
be understood as a relation between formal power series in $d_2/c_1$.

Having reorganized the formal series expansion of $\CG(y;c_1,d_2)$ in this
way finally allows us to take the limit $\lim_{(1)\ra(2)}$.
\begin{equation}\label{limaux}\begin{aligned}
\left(1-\frac{c_1}{\al_2z_2}\right)^{\mu_2'-\nu_1}=
e^{-2\si_2\sqrt{\al_2}\frac{c_1}{d_2}}\,e^{-\si_2\frac{c_1^2}{c_2}}\,
\Big(1+\CO\big(\al_2^{-\frac{1}{2}}\big)\Big)\,.
\end{aligned}\end{equation}
Using this and keeping in mind that $\mu_1+\mu_2'=\nu_1+\nu_2$
we find that  $\tilde\CG_s^{(1)}\big(y;c_1,d_2\big)$ behaves in 
the collision limit as 
\begin{equation}
\tilde\CG_s^{(1)}\big(y;c_1,d_2\big) \sim 
e^{\pm \pi i(\mu_1-\nu_1)}
G_0^{(2)}
c_1^{\mu_0+\nu_1}\,c_2^{\nu_2}\,
e^{-\si_2\frac{c_1^2}{c_2}}
\left(\frac{y}{2bc_1}\right)^{\nu}
\,,
\end{equation}
where $G_0^{(2)}:=G_0^{(1)}e^{-2\si_2\al_2}\al^{\mu_1-\nu_1-\nu_2}$.
This combination has a finite limit as was observed in Subsection
\ref{sec:existence}.

Combining these observations with the fact that $\CD^{(1)}$ turns into 
$\CD^{(2)}$ in this limit, we may argue as before in the 
case of the limit $\lim_{(0)\ra(1)}$ that the 
formal expansion of $\tilde\CG_r^{(1)}(y;c_1,d_2)$ in powers of 
$d_2$ approaches the expansion of $\CG_r^{(2)}(y;c_1,c_2)$, order 
by order in $c_2$.

\end{document}